\documentclass[a4paper,12pt]{article}

\usepackage[cp1251]{inputenc}
\usepackage[T1,T2A]{fontenc}
\usepackage[english,russian]{babel}

\usepackage{fullpage}
\usepackage{amsmath,amssymb}
\usepackage{graphicx}

\def\Re{\mathop{\mathrm{Re}}\nolimits}
\def\Im{\mathop{\mathrm{Im}}\nolimits}
\def\sign{\mathop{\mathrm{sign}}\nolimits}
\def\const{\mathop{\mathrm{const}}\nolimits}

\def\Sp{\mathop{\mathrm{Sp}}\nolimits}

\begin{document}
\title{Современное модельное описание магнетизма}

\author{В.Ю.~Ирхин}


\date{ }

\maketitle

\begin{abstract}
Проведено сопоставление основополагающих работ С. В. Вонсовского
по многоэлектронным полярной и $s-d(f)$ обменной моделям с
последующим развитием теории магнетизма переходных и
редкоземельных металлов, а также их соединений. Особый упор
делается на вывод различных многоэлектронных моделей (Гейзенберга,
Хаббарда, Андерсона), а также соотношения и внутреннюю связь между
ними. Среди тем, рассмотренных в обзоре, - многоэлектронные
подходы при описании систем $d$- и $f$-электронов, атомное
представление Х-операторов, проблема сильного
коллективизированного магнетизма и формирования локальных
моментов, роль неквазичастичных (некогерентных состояний).
Обсуждается применение этих концепций к
сильнокоррелированным системам, в частности, к полуметаллическим
ферромагнетикам и решеткам Кондо.
\end{abstract}

1. Введение

2. Полярная модель и модель Хаббарда.

2.1 Атомное представление и метод многоэлектронных операторов

2.2 Электронный спектр в модели Хаббарда и переход металл-изолятор

2.3 Ферромагнетизм сильно коррелированных $d$--систем

3. $s-d(f)$-обменная модель и модель Андерсона

3.1 Электронные состояния в $s-d$ обменной модели

3.2 $s$---$d$ обменная модель с узкими зонами и $t$---$J$ модель

3.3 Сопротивление магнитных переходных металлов

3.4 $s-f$-обменная модель и свойства редкоземельных металлов

3.5 Эффект Кондо

3.6 Свойства аномальных $f$-соединений

Заключение

\section{Введение}

Проблема двойственной природы электронных состояний в кристалле, проявляющих как зонные, так и атомные черты, – до сих пор одна из центральных в физике твердого тела.
Особенно существенна эта проблема для описания поведения $d$--электронов. В частности, именно в переходных металлах, их сплавах и соединениях наблюдается столь важное явление сильного магнетизма, обусловленное формированием локальных магнитных моментов вследствие межэлектронного взаимодействия.

Уже в 20-30-е годы ХХ века были достигнуты первые успехи теории металлов в рамках новой квантовой механики – после открытия статистики Ферми. В рамках приближения свободных электронов, а затем одноэлектронной зонной теории --- в работах Паули, Блоха, Вильсона, Пайерлса, Зоммерфельда -- было дано объяснение парамагнетизма, поведения теплоемкости и кинетических свойств \cite{1}.
Однако для описания ферромагнетизма и ряда других явлений (например, перехода металл-изолятор) эти представления оказались недостаточными. С другой стороны, попытки использовать для магнитных металлов модель Дирака-Гейзенберга, основанную на атомной картине локализованных спинов, также не дали хороших результатов (в частности, она была не в состоянии объяснить дробные значения магнитных моментов). Таким образом, потребовался определенный синтез модели Гейзенберга и одноэлектронной зонной модели.

В 1934 г. была предложена полярная модель Шубина и Вонсовского \cite{662}, а в 1946 г. --- $s-d$--обменная модель \cite{1946,265}. Обе эти модели сыграли исключительно важную роль в теоретическом описании $d$-- и $f$--металлов и их соединений.

Работы Шубина и Вонсовского по полярной модели \cite{662} были опубликованы в престижном английском журнале Proc. Roy. Soc. и (в более подробном изложении) в харьковском журнале Phys. Zs. UdSSR, выходившем на немецком языке; их русский перевод можно найти в книге \cite{Shubin}. В этих статьях была намечена программа на много лет вперед, которая целиком не выполнена до сих пор: построение систематической теории металлов, позволяющей рассматривать их электрические и магнитные свойства одновременно, и выбор подходящей системы приближений.

Настоящий обзор посвящен эволюции идей многоэлектронных моделей, которые были заложены и развиты в работах С.В. Вонсовского, его коллег и учеников.

Мы ограничимся модельными аспектами теории металлов, хотя в
настоящее время широко используются как первопринципные зонные
расчеты, так и попытки их комбинации с многоэлектронными моделями
(что позволяет значительно улучшить учет корреляционных эффектов).
Изложение будет придерживаться классических теоретико-полевых
методов (преимущественно метода двухвременных запаздывающих
функций Грина, в том числе для многоэлектронных операторов). Во
время написания первых работ Шубина и Вонсовского этих наглядных
аналитических методов, основанных на представлении вторичного квантования,
еще не было (под рукой был только громоздкая
техника слэтеровских детерминантов), и их последующее применение
позволило существенно продвинуться в понимании многоэлектронных
эффектов. Следует также отметить, что, несмотря на имеющиеся
принципиальные трудности (в особенности так называемая “проблема
знака”, обусловленная фермиевской статистикой), в последнее время
достигнуты существенные успехи в прямых численных расчетах
многоэлектронных систем квантовым методом Монте--Карло (см., напр., \cite{mc}). Они,
однако, пока далеко не достаточны, чтобы заменить аналитические
модельные подходы.

Данная работа в значительной мере продолжает и дополняет фундаментальный обзор \cite{81}, где сопоставляются локализованные и делокализованные аспекты поведения электронов в переходных металлах и их сильнокоррелированных соединениях.
Теоретическое изложение по возможности сопровождается примерами реальных физических систем. Более подробно физические свойства переходных металлов и систем с сильными корреляциями на их основе рассмотрены в книге \cite{II}.


В разделе 2 приводится формулировка полярной модели Шубина-Вонсовского и ее частного случая – модели Хаббарда. С использованием формализма углового момента проведено рассмотрение вырожденных атомных состояний, которое существенно в случае $d$-- и $f$--электронов. Обсуждаются атомное представление, спектр электронных состояний, переход металл-изолятор и ферромагнетизм в системах сильнокоррелированных электронов.

В разделе 3 рассмотрена $s-d(f)$-обменная модель Вонсовского, ее обобщения, частные случаи и применения к различным физическим ситуациям. Более подробно обсуждаются редкоземельные металлы, сильные полуметаллические ферромагнетики, решетки Кондо.

\section{Полярная модель и модель Хаббарда}

В работах \cite{662} Шубин и Вонсовский  поставили своей целью одновременно описать широкий круг явлений в твердом теле, включая магнетизм и электропроводность.
Полярная модель была предложена ими как синтез гомеополярной модели Гейзенберга, описывающей систему локализованных моментов, и подхода Слэтера для описания многоэлектронной системы металла. В кристалле, где на атом приходится один электрон (или в простейшем примере молекулы водорода, рассмотренном Гайтлером и Лондоном), это означает учет полярных состояний -- двоек и дырок, т.е. дважды занятых и пустых узлов. В исходной формулировке модели были учтены перескоки электронов с узла на узел и все типы межэлектронного взаимодействия

Дальнейшее развитие полярная модель получила в работах Боголюбова \cite{651}, который вывел ее гамильтониан через последовательное разложение по интегралу перекрытия атомных волновых функций в представлении вторичного квантования. В простейшем случае невырожденной зоны его можно записать в виде
\begin{equation}
{\mathcal H} =\sum_{\nu _1\neq \nu _2,\sigma }t_{\nu _1\nu _2}c_{\nu _1\sigma
}^{\dagger }c_{\nu _2\sigma }^{} +\frac 12\sum_{\nu _i\sigma _1\sigma _2}I_{\nu _1\nu _2\nu _3\nu
_4}c_{\nu _1\sigma _1}^{\dagger }c_{\nu _2\sigma _2}^{\dagger }c_{\nu
_4\sigma _2}^{ }c_{\nu _3\sigma _1}^{ }
\label{eq:G0}
\end{equation}

Здесь $t_{\nu _1\nu _2}$ и $I_{\nu _1\nu _2\nu _3\nu_4}$ -- матричные элементы одноэлектронного переноса и межэлектронного взаимодействия. В частности, $V_{\nu _1\nu _2}=I(\nu_1\nu_2\nu_1\nu_2)$ -- кулоновское взаимодействие на разных узлах (ответственное, например, за зарядовое упорядочение), $J_{\nu _1\nu _2}=-I(\nu_1\nu_2\nu_2\nu_1)$ -- "<прямое"> обменное взаимодействие (этот член получается перестановкой (обменом) спиновых индексов).

Новый импульс многоэлектронной теории кристалла придали идеи Хаббарда \cite{28,29,30,31}, выделившего в своей модели наиболее существенную часть кулоновского взаимодействия -- сильное отталкивание электронов на одном узле $U=I(\nu\nu\nu\nu)$.
В~случае невырожденной зоны ее гамильтониан запишется как
\begin{equation}
\mathcal{H}=\sum_{\mathbf{k}\sigma
}t_{\mathbf{k}}c_{\mathbf{k}\sigma }^{\dagger }c_{\mathbf{k}\sigma
}+ U\sum_ic_{i\uparrow }^{\dagger
}c_{i\uparrow }c_{i\downarrow }^{\dagger }c_{i\downarrow }, \label{eq:G.1}
\end{equation}
где $ t_{\mathbf{k}}$ -- зонный спектр. Модель Хаббарда широко
использовалась для рассмотрения ферромагнетизма
коллективизированных электронов, перехода металл-изолятор и других физических явлений. Несмотря на очевидную простоту, эта модель содержит очень богатую
физику и ее строгое исследование является весьма трудной
проблемой.

Поскольку в случае сильных корреляций теория возмущений не работает,
Хаббард использовал метод двухвременных запаздывающих функций Грина, разработанный Боголюбовым и Тябликовым. Предложенная им схема расцепления на разных узлах позволила  получить формальный переход от зонной к атомной картине.
Стартуя с атомного предела, Хаббард нашел интерполяционное решение,
описывающее как атомный, так и зонный пределы для
$s$"~состояний~\cite{28}; затем он рассмотрел простую модель
вырожденных зон~\cite{29}.
В то же время интерполяционное описание оказалось в значительной мере иллюзорным.
(в частности, корреляционное расщепление в спектре сохраняется при сколь угодно малых $U$, неудовлетворительно описываются ферромагнитные решения).

В третьей работе Хаббарда \cite{30} рассмотрено улучшенное
расцепление -- одноузельное приближение, аналогичное теории
неупорядоченных сплавов, позволяющее учесть одноузельные
корреляции и поправки на резонансное уширение и рассеяние. Оно
позволило, в частности, получить переход металл-изолятор. Однако и
ему присущи недостатки -- переоценка затухания, отсутствие
фермижикостного поведения.

В~работе~\cite{31} Хаббард предложил общий формализм многоэлектронных
$X$"~операторов (атомное представление), который позволяет
учесть внутриатомные взаимодействия в нулевом приближении (этот метод детально обсуждается в обзоре \cite{654} и монографии \cite{II}).

В отсутствие стандартного малого параметра стандартные диаграммные
подходы  здесь оказались неприменимыми, а успех нестандартных
диаграммных техник \cite{633} -- весьма ограниченным в силу
неоднозначности их правил.

В работах \cite{699,694,695} методом уравнений движения было
развито разложение по обратному координационному числу $1/z$,
которое позволило последовательно учесть вклады спиновых и
зарядовых флуктуаций, а также фермиевских возбуждений, однако оно
также встретилось с рядом трудностей в случае парамагнитного
состояния.


Второе дыхание модель Хаббарда получила после открытия
высокотемпературных сверхпроводников, поскольку позволяла описать
движение носителей тока в медь-кислородных плоскостях.

Для описания электронных состояний в CuO${}_2$-плоскостях
перовскитов могут быть использованы и более сложные многозонные
модели, например так называемая модель Эмери:
\begin{equation}
\mathcal{H}=\sum_{\mathbf{k}\sigma }\left[ \varepsilon
p_{\mathbf{k}\sigma }^{\dagger }p_{\mathbf{k}\sigma }+\Delta
d_{\mathbf{k}\sigma }^{\dagger }d_{\mathbf{k}\sigma
}+V_{\mathbf{k}}(p_{\mathbf{k}\sigma }^{\dagger
}d_{\mathbf{k}\sigma }+d_{\mathbf{k}\sigma }^{\dagger
}p_{\mathbf{k}\sigma })\right]
+U\sum_id_{i\uparrow }^{\dagger }d_{i\uparrow }d_{i\downarrow
}^{\dagger }d_{i\downarrow },
 \label{eq:6.108}
\end{equation}
где $\varepsilon $ и $\Delta $~"--- положения $p$- и $d$"~уровней
для Cu- и O"~ионов соответственно. $\mathbf{k}$"~зависимость
матричных элементов $p$---$d$~гибридизации для квадратной решетки
имеет вид
\begin{equation}
V_{\mathbf{k}}=2V_{pd}\left( \sin ^2k_x+\sin ^2k_y\right) ^{1/2}.
 \label{eq:6.109}
\end{equation}
При $|V_{pd}|\ll \varepsilon -\Delta $
гамильтониан~(\ref{eq:6.108}) приводится каноническим
преобразованием~\cite{619} к модели Хаббарда с сильным кулоновским
отталкиванием и эффективными интегралами перескока Cu"--~Cu
\begin{equation}
t_{\text{eff}}=V_{pd}^2/(\varepsilon -\Delta ).
 \label{eq:6.110}
\end{equation}

\subsection{Атомное представление и метод многоэлектронных операторов}

Вывод и анализ уравнений полярной модели для случая $s$-зоны был
дан в оригинальных работах \cite{662} и далее в статьях и обзорах
\cite{663,81}. Здесь мы обсудим более общий случай вырожденных
электронных состояний на узле, поскольку такое вырождение важно
для переходных металлов и их соединений. Однако вначале, как и в
работах Шубина и Вонсовского \cite{662}, рассмотрим
многоэлектронные (MЭ) волновые функции и процедуру вторичного
квантования для систем с сильными межузельными кулоновскими
корреляциями.

При переходе к стандартному представлению вторичного квантования
MЭ волновые функции кристалла $\Psi (x_1\ldots
x_N)$ ($x=\{\mathbf{r}_is_i\}$, $s_i$~"--- спиновые координаты)
выбираются в виде линейных комбинаций слэтеровских определителей.
Последние составляются из одноэлектронных волновых функций $\psi
_\lambda (x)$ ($\lambda =\{{\nu \gamma \}}$, $\nu $~"--- индексы
ячеек в решетке, а $\gamma $~"--- одноэлектронные наборы квантовых
чисел):
\begin{equation}
\Psi (x_1\ldots x_N)=\sum_{\lambda _1\ldots \lambda _N}c(\lambda
_1\ldots \lambda _N)\Psi _{\lambda _1\ldots \lambda _N}(x_1\ldots
x_N),
 \label{eq:A.1}
\end{equation}
где
\begin{equation}
\Psi _{\lambda _1\ldots \lambda _N}(x_1\ldots
x_N)=(N!)^{-1/2}\sum_P(-1)^P P\prod _i\psi _{\lambda _i}(x_i),
 \label{eq:A.2}
\end{equation}
а $P$ пробегает всевозможные перестановки $x_i$.
Разложение~(\ref{eq:A.1}) справедливо при условии, что система
функций $\psi _\lambda $ полная~\cite{651}. Представление
вторичного квантования вводится путем использования
одноэлектронных чисел заполнения $n_\lambda $ в качестве новых
переменных:
\begin{equation}
\Psi (x_1\ldots x_N)=\sum_{\{n_\lambda \}}c(\ldots n_\lambda
\ldots )\Psi _{\{n_\lambda \}}(x_1\ldots x_N).
 \label{eq:A.3}
\end{equation}
Тогда величина $c(\ldots n_\lambda \ldots )$ играет роль новой волновой
функции. Одноэлектронные операторы рождения и уничтожения Ферми
определяются следующим образом:
\[
a_\lambda c(\ldots n_\lambda \ldots )=(-1)^{\eta _\lambda
}n_\lambda c(\ldots n_\lambda -1\ldots ),
\]
\begin{equation}
a_\lambda ^{\dagger }c(\ldots n_\lambda \ldots )=(-1)^{\eta
_\lambda }(1-n_\lambda )c(\ldots n_\lambda +1\ldots ),
 \label{eq:A.4}
\end{equation}
причем
$\eta _\lambda =\sum_{\lambda ^{\prime }>\lambda }n_{\lambda
^{\prime }}$, $a_\lambda ^{\dagger }a_\lambda^{ } =\hat{n}_\lambda$.

Теперь попробуем обобщить этот метод, вводя для одноузельной (атомной) задачи
квантовые числа электронных групп.

C физической точки зрения ясно, что межэлектронные корреляции
наиболее важны для электронов одной и той~же атомной оболочки
(эквивалентных электронов). Современная теория атомных спектров
базируется на формализме Рака для угловых моментов (см., напр.,
\cite{20}). Эта мощная математическая методика (отметим, что она
может быть обобщена на случай кристаллического поля, расщепляющего
атомные термы \cite{2020}) вводит представление многоэлектронных
квантовых чисел $\Gamma =\{{SL\mu M\}}$ вместо одноэлектронных
$\gamma =\{{lm\sigma \}}$, причем
$\mathbf{S}=\sum_i\mathbf{s}_i,\qquad
\mathbf{L}=\sum_i\mathbf{l}_i$ суть полные спиновый и орбитальный
угловые моменты, а $\mu $ и $M$~"--- их проекции. Тогда
многочисленные возможные комбинации наборов $\gamma $ для частично
занятой оболочки заменяются наборами $\Gamma $. Общее количество
МЭ~состояний то~же самое, но энергетическое вырождение снято, так
что в большинстве физических задач можно сохранить только самый
низкий МЭ~терм. Согласно правилам Хунда, он соответствует
максимальным~$L$ и $S$. В~рамках такого подхода проблема
электростатического взаимодействия в системе сводится к вычислению
нескольких интегралов Слэтера~$F^{(p)}$, которые могут быть
рассчитаны с использованием атомных волновых функций~\cite{33} или
определены из экспериментальных данных.

Объединяя электроны на каждом узле в решетке ($\Lambda =\{{\nu \Gamma }\}$),
получим
\begin{equation}
\Psi (x_1\ldots x_N)=\sum_{\{N_\lambda \}}c(\ldots N_\lambda
\ldots )\Psi _{\{N_\lambda \}}(x_1\ldots x_N).
 \label{eq:A.5}
\end{equation}
В~случае конфигурации эквивалентных электронов $l^n$ с одинаковым
орбитальным квантовым числом MЭ волновая функция электронной
группы определяется следующим рекуррентным соотношением
(см.~\cite{20}):
\begin{equation}
\Psi _{\Gamma _n}(x_1\ldots x_N)=\sum_{\Gamma _{n-1},\gamma
}G_{\Gamma _{n-1}}^{\Gamma _n}C_{\Gamma _{n-1},\gamma }^{\Gamma
_n}\Psi _{\Gamma _{n-1}}(x_1\ldots x_{n-1})\psi _\gamma (x_n),
 \label{eq:A.6}
\end{equation}
где ${C}$~"--- коэффициенты Клебша"--~Гордана. В~случае $LS$"~связи используем обозначения
\begin{equation}
C_{\Gamma _{n-1},\gamma }^{\Gamma _n}\equiv
C_{L_{n-1}M_{n-1},lm}^{L_nM_n}C_{S_{n-1}\mu _{n-1},\frac 12\sigma
}^{S_n\mu _n},
 \label{eq:A.7}
\end{equation}
где суммирование по $\gamma =\{lm{\sigma }\}$ (опускаем для
краткости главное квантовое число) стоит вместо суммирования по
одноэлектронным орбитальным проекциям $m$ и спиновым проекциям
$\sigma $, но не~по~$l$. Величины
$G_{\Gamma _{n-1}}^{\Gamma _n}
\equiv G_{S_{n-1}L_{n-1}\alpha_{n-1}}^{S_nL_n\alpha _n}$
называются генеалогическими коэффициентами ($\alpha $~"---
дополнительные квантовые числа, которые отличают различные
состояния с совпадающими $S$, $L$, например, число "<сеньорити">,
введенное Рака). Они не~зависят от проекций момента импульса, а величины
$\left( G_{\Gamma _{n-1}}^{\Gamma _n}\right) ^2$ имеют смысл
вкладов терма $\Gamma _{n-1}$ в формирование терма $\Gamma _n$.

Если добавленный электрон принадлежит другой оболочке, можно
записать
\begin{equation}
\Psi _{\Gamma _n}(x_1\ldots x_n)=n^{-1/2}\sum_{i,\Gamma
_{n-1},\gamma }(-1)^{n-i}C_{\Gamma _{n-1},\gamma }^{\Gamma
_n} \Psi _{\Gamma _{n-1}}(x_1\ldots x_{i-1},x_{i-1}\ldots
x_{n-1})\psi _\gamma (x_i)
 \label{eq:A.10}
\end{equation}
так что, в отличие от случая эквивалентных электронов, здесь дополнительная
антисимметризация необходима. Следует иметь в виду, что
такое представление MЭ функций и операторов, которые описывают
несколько электронных оболочек, работает в теории твердого тела
лишь при условии, что взаимодействие между оболочками велико по
сравнению с зонными энергиями.

Волновая функция всего кристалла~(\ref{eq:A.5}) может быть теперь
получена как антисимметризованное произведение MЭ функций для
электронных групп.
По аналогии с~(\ref{eq:A.6}), (\ref{eq:A.10}) можно ввести МЭ
операторы рождения для электронных групп~\cite{652}. Для
эквивалентных электронов и при добавлении электрона из другой оболочки соответственно имеем
\begin{equation}
A_{\Gamma _n}^{\dagger }=n^{-1/2}\sum_{\Gamma _{n-1},\gamma
}G_{\Gamma _{n-1}}^{\Gamma _n}C_{\Gamma _{n-1},\gamma }^{\Gamma
_n}a_\gamma ^{\dagger }A_{\Gamma _{n-1}}^{\dagger },.
A_{\Gamma _n}^{\dagger }=\sum_{\Gamma _{n-1},\gamma }C_{\Gamma
_{n-1},\gamma }^{\Gamma _n}a_\gamma ^{\dagger }A_{\Gamma
_{n-1}}^{\dagger }.
 \label{eq:A.11}
\end{equation}
Антисимметрия функций $|\Gamma _n\rangle =A_{\Gamma _n}^{\dagger
}|0\rangle$ обеспечивается антикоммутацией ферми-операторов.
Используя соотношения ортогональности для коэффициентов
Клебша"--~Гордана и генеалогических коэффициентов, легко получить
$\langle 0|A_{\Gamma ^{\prime }}A_\Gamma ^{\dagger }|0\rangle
=\delta _{\Gamma \Gamma ^{\prime }}$. Однако при $m<n$ имеем
$A_{\Gamma _m^{\prime }}A_{\Gamma _n}^{\dagger }|0\rangle \neq 0$.
Поэтому операторы~(\ref{eq:A.11}), (\ref{eq:A.11}) удобны только
для работы с конфигурациями с фиксированным числом электронов
(скажем, в гомеополярной модели Гайтлера"--~Лондона). Для
рассмотрения проблемы с перемещением электронов между оболочками
или узлами удобно определить новые MЭ операторы рождения, которые
содержат проекционные множители, введенные в~\cite{653,654}:
\begin{equation}
\tilde{A}_\Gamma ^{\dagger }=A_\Gamma ^{\dagger }\prod_\gamma
(1-\hat{n}_\gamma ).
 \label{eq:A.16}
\end{equation}
Формально произведение в~(\ref{eq:A.16}) идет по всем допустимым
одноэлектронным состояниям $\gamma $. Однако в силу тождества
$a_\gamma ^{\dagger }\hat{n}_\gamma =0$ достаточно сохранить
только те $\gamma $, которые не~входят в соответствующие
произведения операторов в $A_\Gamma $.
Теперь получаем
\begin{equation}
\tilde{A}_\Gamma \tilde{A}_\Gamma ^{\dagger }=\delta _{\Gamma
\Gamma ^{\prime }}\prod_\gamma (1-\hat{n}_\gamma ); \quad
\tilde{A}_\Gamma \tilde{A}_{\Gamma ^{\prime }} =\tilde{A}_{\Gamma
^{\prime }}^{\dagger }\tilde{A}_\Gamma ^{\dagger }=0 \quad
(|\Gamma \rangle \neq 0)
 \label{eq:A.18}
\end{equation}
Таким образом, можно прийти к
представлению MЭ чисел заполнения $N_\Gamma $ на данном узле:
\[
\tilde{A}_\Gamma |\Gamma ^{\prime }\rangle =\delta _{\Gamma \Gamma
^{\prime }}|0\rangle ,\qquad \tilde{A}_\Gamma ^{\dagger }|\Gamma
^{\prime }\rangle =\delta _{\Gamma ^{\prime }0}|\Gamma \rangle,
\]
\begin{equation}
\tilde{A}_\Gamma ^{\dagger }\tilde{A}_\Gamma =\hat{N}_\Gamma
,\qquad \hat{N}_\Gamma |\Gamma \rangle =\delta _{\Gamma \Gamma
^{\prime }}|\Gamma \rangle,\qquad \sum_\Gamma \hat{N}_\Gamma =1.
 \label{eq:A.19}
\end{equation}

Подчеркнем, что, вводя MЭ операторы, которые зависят от всех
одноэлектронных квантовых чисел (как занятых, так и свободных
состояний), мы делаем следующий шаг в квантово-полевом описании
после обычного вторичного квантования. В принципе оно возможно и
за пределами одноатомной задачи. Здесь снова (как и при обсуждении
полярной модели) полезно рассмотрение простых модельных систем
типа молекулы водорода, а также сравнение с приближениями типа
Хартри-Фока (в точном смысле, как оно используется в
атомной теории). Вообще говоря, следует использовать
многоэлектронные волновые функции, которые не
сводятся к слэтеровским детерминантам и не факторизуются на
одноэлектронные.

При практических вычислениях удобно перейти от
МЭ~операторов рождения и уничтожения к $X$"~операторам Хаббарда
$X(\Gamma ,\Gamma ^{\prime })=\tilde{A}_\Gamma ^{\dagger}\tilde{A}_{\Gamma ^{\prime }}$,
которые переводят состояние $\Gamma ^{\prime }$ в состояние $\Gamma $.
Такие операторы впервые были предложены в~\cite{31}
аксиоматическим способом как обобщенные проекционные операторы:
\begin{equation}
X(\Gamma ,\Gamma ^{\prime })=|\Gamma \rangle \langle \Gamma
^{\prime }|, \quad \sum_\Gamma X(\Gamma ,\Gamma )=1,
 \label{eq:A.22}
\end{equation}
где $|\Gamma \rangle $ --- точные собственные состояния
гамильтониана. При этом произвольный оператор~$\hat{O}$,
действующий на электроны на данном узле~$i$, выражается через $X$"~операторы так:
\begin{equation}
\hat{O}=\sum_{\Gamma \Gamma ^{\prime }}\langle \Gamma |\hat{O}|\Gamma
^{\prime }\rangle X(\Gamma ,\Gamma ^{\prime }).
 \label{eq:A.29}
\end{equation}

Использование введенных выше операторов электронных конфигураций
позволяет получить явные выражения для $X$"~операторов через
одноэлектронные операторы:
\begin{equation}
X(\Gamma ,0)=\tilde{A}_\Gamma ^{\dagger },\qquad X(\Gamma ,\Gamma
)=\hat{N}_\Gamma .
 \label{eq:A.23}
\end{equation}
Например, рассмотрим простейший случай $s$"~-электронов, где
$\gamma =\sigma =\pm (\uparrow ,\downarrow )$, $\Gamma =0,\
\sigma ,\ 2$ и $|0\rangle $~"--- свободное состояние (дырка),
а~$|2\rangle $~"--- дважды занятое синглетное состояние на узле.
Тогда
\[
X(0,0)=(1-\hat{n}_{\uparrow })(1-\hat{n}_{\downarrow }),\qquad
X(2,2)=\hat{n}_{\uparrow }\hat{n}_{\downarrow },\qquad
X(2,0)=a_\uparrow  ^{\dagger }a_\downarrow  ^{\dagger },
\]
\[
X(\sigma ,\sigma )=\hat{n}_\sigma (1-\hat{n}_{-\sigma }),\qquad
X(\sigma ,-\sigma )=a_\sigma ^{\dagger }a_{-\sigma },
\]
\begin{equation}
X(\sigma ,0)=a_\sigma ^{\dagger }(1-\hat{n}_{-\sigma }),\qquad
X(2,\sigma )=-\sigma a_{-\sigma }^{\dagger }\hat{n}_\sigma .
 \label{eq:A.24}
\end{equation}

Как следует из~(\ref{eq:A.18}), правила умножения,
постулированные Хаббардом~\cite{31}, имеют вид
\begin{equation}
X(\Gamma ,\Gamma ^{\prime })X(\Gamma ^{\prime \prime },\Gamma
^{\prime \prime \prime })=\delta _{\Gamma ^{\prime }\Gamma
^{\prime \prime }}X(\Gamma ,\Gamma ^{\prime \prime \prime }).
 \label{eq:A.25}
\end{equation}

$X$"~операторы можно подразделить на операторы бозе- и ферми-типа:
они соответственно меняют количество электронов на узле на четное
и нечетное число, а на разных узлах решетки коммутируют и
антикоммутируют.
На одном узле $X$"~операторы обладают значительно более сложной
алгеброй (точнее говоря, супералгеброй) коммутационных и
антикоммутационных соотношений, чем фермиевские и бозевские
операторы ~\cite{697}. Имеются попытки использовать
соответствующие динамические симметрии за пределами атомной
задачи~\cite{697a}.

Из~(\ref{eq:A.11}) получаем представление
\begin{equation}
a_\gamma ^{\dagger }=\sum_nn^{1/2}\sum_{\Gamma _n\Gamma
_{n-1}}G_{\Gamma _{n-1}}^{\Gamma _n}C_{\Gamma _{n-1},\gamma
}^{\Gamma _n}X(\Gamma _n,\Gamma _{n-1}).
 \label{eq:A.31}
\end{equation}
В~частности, для $s$"~-электронов
\begin{equation}
a_\sigma ^{\dagger }=X(\sigma ,0)+\sigma X(2,-\sigma ).
 \label{eq:A.32}
\end{equation}

В работах Шубина и Вонсовского было использовано квазиклассическое
приближение. Оно по существу состоит в замене Х-операторов
с-числовыми функциями, определяющими амплитуду вероятности
пребывания узла в состоянии однократно занятого узла, двойки или
дырки:
\begin{equation}
X_{i}(+,0)\rightarrow \varphi _{i}^{\ast }\Psi _{i},~X_{i}(2,-)\rightarrow
\Phi _{i}^{\ast }\psi _{i},~X_{i}(2,0)\rightarrow \Phi _{i}^{\ast }\Psi _{i}
\label{quasicl}
\end{equation}%
с дополнительным условием
\begin{equation*}
|\varphi _{i}|^{2}+|\psi _{i}|^{2}+|\Phi _{i}|^{2}+|\Psi _{i}|^{2}=1
\end{equation*}%
Это соответствует определению полной энергии системы из вариационного принципа с волновой функцией
\begin{equation}
\phi =\prod_{i}(\varphi _{i}^{\ast }X_{i}(+,0)+\psi _{i}^{\ast
}X_{i}(-,0)+\Phi _{i}^{\ast }X_{i}(2,0~)+\Psi _{i}^{\ast })|0\rangle
\label{dops}
\end{equation}
Она смешивает возбуждения бозевского и фермиевского типа, а потому
не удовлетворяет принципу Паули. Тем не менее, квазиклассическое
приближение позволяет грубо описать переход металл-изолятор, что и
было проделано впоследствии рядом авторов (см. обзор \cite{81}).
Так, в работе Карона и Пратта \cite{caron}  было даже рассмотрено
среднее поле для фермионов, так что фактически на с-числа
заменялись не Х-операторы, а обычные фермиевские операторы.


Предвосхищая работы Хаббарда, уравнения квазиклассического
приближения \cite{662} дают изменение ширины энергетической
полосы, а также относительного расположения различных полос в
зависимости от числа двоек и магнитного момента (соответствующие
результаты в представлении Х-операторов рассмотрены ниже). В
работах \cite{662}, также впервые, по существу введено атомное
представление, в дальнейшем детально разработанное в работах
\cite{31,697,653,654}.

В связи с теорией двумерных высокотемпературных сверхпроводников (ВТСП)
Андерсон \cite{633a} выдвинул идею разделения спиновых и зарядовых
степеней свободы электрона, используя для Х-операторов
представление вспомогательных ("auxiliary", "slave") бозевских и
фермиевских операторов
\begin{equation}
c_{i\sigma }^{\dagger
}=X_i(\sigma ,0)+\sigma X_i(2,-\sigma )=s_{i\sigma }^{\dagger
}e_i+\sigma d_i^{\dagger }s_{i-\sigma }.
 \label{eq:6.131}
\end{equation}
Здесь $s_{i\sigma }^{\dagger }$~"--- операторы рождения для
нейтральных фермионов (спинонов), $e_i^{\dagger }$, $d_i^{\dagger
}$~"--- операторы рождения для заряженных бесспиновых бозонов.
Физический смысл таких возбуждений можно объяснить
следующим образом. Рассмотрим решетку с одним
электроном на узле с сильным хаббардовским отталкиванием, так что
каждый узел нейтрален. В~основном состоянии
 резонирующих валентных связей~(РВС, у Андерсона RVB) каждый узел
принимает участие в одной связи. Когда связь нарушается,
появляются два неспаренных узла, которые обладают спинами,
равными~$1/2$. Соответствующие возбуждения (спиноны) не заряжены.
Вместе с тем, пустой узел (дырка) в системе несет заряд, но
не~спин.

Для полузаполненной зоны присутствуют только спинонные возбуждения
с кинетической энергией порядка~$|J|$. При допировании системы
дырками возникают носители заряда, которые описываются операторами
холонов~$e_i^{\dagger }$. В~простейшей бесщелевой версии
гамильтониан системы для квадратной решетки может быть представлен
в виде
\begin{equation}
\mathcal{H}=\sum_{\mathbf{k}}(4t\phi _{\mathbf{k}}-\zeta
)e_{\mathbf{k}}^{\dagger }e_{\mathbf{k}}+4\sum_{\mathbf{k}}(\Delta
+t\delta )\phi _{\mathbf{k}}(s_{\mathbf{k}\sigma }^{\dagger
}s_{-\mathbf{k}-\sigma }^{\dagger }+s_{\mathbf{k}\sigma
}s_{-\mathbf{k}-\sigma })+\ldots ,
 \label{eq:6.133}
\end{equation}
где $\phi _{\mathbf{k}}=(1/2)(\cos k_x+\cos k_y)$,
$\Delta $ -- параметр порядка состояния РВС, который
определен аномальными средними спинонных операторов, причем
$\delta =\langle e^{\dagger }e\rangle$~"--- концентрация
дырок, $\zeta$ -- химический потенциал.
Таким образом, возникает состояние спиновой жидкости с
подавленным дальним магнитным порядком. При этом в чисто спиновых
(недопированных) системах вследствие существования
фермиевской поверхности спинонов появляются малый энергетический
масштаб $J$ и большой линейный член в удельной теплоемкости
с $\gamma \sim 1/|J|$ (некоторые экспериментальные данные
указывают на присутствие $T$"~линейного члена в непроводящей фазе
медь-кислородных систем).

Позже были развиты более сложные варианты теории РВС, которые
используют топологическое рассмотрение и аналогии с дробным
квантовым эффектом Холла (см., напр., \cite{633}). Эти идеи
привели к довольно необычным и красивым результатам. Например,
показано, что спиноны могут подчиняться дробной статистике, т.~е.
волновая функция системы приобретает комплексный коэффициент при
перестановке двух квазичастиц.

Исходя из конкретной физической задачи и ситуации, используются различные представления для операторов Хаббарда. В работе \cite{Kotliar} было предложено представление четырех бозонов $p_{i \sigma}$, $e_i$,  $d_i$, которые осуществляют проектирование на однократно занятые состояния, дырки и двойки соответственно.
В  результате гамильтониан Хаббарда принимает вид

\begin{equation}
\mathcal{H}=\sum_{ij \sigma}t _{ij}f_{i \sigma}^{\dagger }f_{j \sigma}z_{i \sigma}^{\dagger }z_{j \sigma} + U\sum_{i}d _{i}^{\dagger }d _{i}, \, z_{i \sigma} =e_{i }^{\dagger } p_{i \sigma} +p_{i - \sigma}^{\dagger }d_{i},
 \label{eq:133}
\end{equation}
причем накладываются дополнительные ограничения
\[
\sum_{\sigma}p_{i \sigma}^{\dagger }p_{i \sigma}
+d _{i}^{\dagger }d _{i}+d _{i}^{\dagger }d _{i}=1, \,
f_{i \sigma}^{\dagger }f_{i \sigma}=p_{i \sigma}^{\dagger }p_{i \sigma}+d _{i}^{\dagger }d _{i}.
\]
Это представление позволило качественно воспроизвести ряд прежних результатов (например, как и квазиклассическое приближение (\ref{quasicl}), получить описание перехода металл--изолятор по Гутцвиллеру).

В работе \cite{Wang} было предложено представление, содержащее два
сорта фермиевских операторов -- холонов $ e_{i}$ и дублонов
$d_{i}$, которые соответствуют дыркам и двойкам:
\begin{eqnarray}
X_{i}(+,0) &=&e_{i}(1-d_{i}^{\dagger
}d_{i}^{{}})(1/2+s_{i}^{z}),~X_{i}(-,0)=e_{i}(1-d_{i}^{\dagger
}d_{i}^{{}})s_{i}^{-},  \nonumber \\
X(2,-) &=&d_{i}^{\dagger }(1-d_{i}^{\dagger
}d_{i}^{{}})s_{i}^{+},~X(2,+)=d_{i}^{\dagger }(1-d_{i}^{\dagger
}d_{i}^{{}})(1/2+s_{i}^{z})  \label{ed}
\end{eqnarray}%
При этом операторы физических спинов связаны с псевдоспиновыми
операторами $ s_{i}^{\alpha}$ соотношением
$\mathbf{S}_{i}=\mathbf{s}_{i}(1-d_{i}^{\dagger
}d_{i}^{{}}-e_{i}^{\dagger}e_{i}^{{}})$. Ранее рассматривались
различные частные случаи этого представления, соответствующие
пределу больших $U$ ($t-J$ модели, см. обзор \cite{Izyumov1}).
В дальнейшем были предложены также суперсимметричные представления \cite{Pepin}.
Отметим, что в теоретико-полевых подходах оказалось полезным обобщение обычной модели Хаббарда на $N$  "<цветов">,  которое позволяет выполнить $1/N$-разложение по обратной кратности вырождения.

При увеличении силы корреляций в МЭ системах происходит смена статистики элементарных возбуждений с зонной на атомную, которая проявляется как переход металл-изолятор (формирование хаббардовских подзон). Формальное  описание такого перехода  оказывается  крайне сложным.
Для решения этой проблемы, в частности,  предлагалось смешивание
базиса из обычных одноэлектронных и многоэлектронных Х-операторов, а также
использовалось конструкция башни симметрии \cite{697a}.

Операторы спинового и углового момента в чисто локализованных системах также могут быть представлены через $X$"~операторы.  Учитывая выражения для матричных элементов углового момента, находим
для циклических компонент вектора $\mathbf{I}=\mathbf{S},\
\mathbf{L},\ \mathbf{J}$
\begin{align}
I^{+} =\sum_M\gamma _I(M)X(M+1,M), I^z =\sum_MMX(M,M), \gamma _I(M)=[(I-M)(I+M+1)]^{1/2}.
\label{eq:B.8}
\end{align}

Использование операторов Хаббарда дает возможность простым
способом получить главные результаты теории гейзенберговских
магнетиков. В~частности, этот формализм позволяет учесть сильную
одноионную магнитную анизотропию в нулевом
приближении~\cite{674,II}. Гамильтониан модели
Гейзенберга с произвольной одноузельной анизотропией имеет вид
\begin{equation}
\mathcal{H}=\sum_{ij}J_{ij}\mathbf{S}_i\mathbf{S}_j
+\mathcal{H}_{\text{a}}
 \label{eq:F.1}
\end{equation}
В представлении Х-операторов гамильтониан анизотропии принимает
диагональный вид. В случае анизотропии типа легкая ось имеем
\begin{equation}
\mathcal{H}_{\text{a}}=-\sum_i[\varphi (S_i^z)+ HS_i^z]  =-\sum_{iM}[\varphi (M)+HM]X_i(M,M),
 \label{eq:F.2}
\end{equation}
где $H$ -- магнитное поле. Удобно ввести коммутаторные функции Грина
\begin{equation}
G_{\mathbf{q}}(\omega ) =\langle \!\langle
S_{\mathbf{q}}^{+}|S_{-\mathbf{q}}^{-}\rangle \!\rangle _\omega
,\qquad G_{\mathbf{q}M}(\omega )=\langle \!\langle
X_{\mathbf{q}}(M+1,M)|S_{-\mathbf{q}}^{-}\rangle \!\rangle _\omega
.
\label{eq:F.6}
\end{equation}
Запишем уравнение движения, в котором выполним простейшее
расцепление, соответствующее расцеплению Тябликова на различных
узлах решетки
\begin{multline}
(\omega -H-\varphi (M+1)+\varphi (M)+2J_0\langle S^z\rangle
)G_{\mathbf{q}M}(\omega )=
\\
=\gamma _S(M)(N_{M+1}-N_M)[1+J_{\mathbf{q}}G_{\mathbf{q}}(\omega
)],
 \label{eq:F.7}
\end{multline}
где $N_M=\langle X(M,M)\rangle$. После суммирования по $M$ получаем
\begin{equation}
G_{\mathbf{q}}(\omega )=\frac{\Phi _S(\omega
)}{1-J_{\mathbf{q}}\Phi _S(\omega )}, \qquad
\Phi _S(\omega )=\sum_M\frac{\gamma _S^2(M)(N_{M+1}-N_M)}{\omega
-H-\varphi (M+1)+\varphi (M)+2J_0\langle S^z\rangle }.
 \label{eq:F.8}
\end{equation}
Спектр возбуждений
определяется полюсом (\ref{eq:F.8}) и содержит $2S$ ветвей.
Выражения~(\ref{eq:F.7}), (\ref{eq:F.8})
позволяют вычислить числа заполнения $N_M$ и получить
самосогласованное уравнение для намагниченности. Еще более интересными оказываются
результаты в случае анизотропии типа легкая плоскость и кубических кристаллов \cite{674}.

Современные исследования квантовых двумерных антиферромагнетиков с
локализованными спинами (см., напр., \cite{Sachdev}) демонстрируют
очень красивую физику; при этом  фазовая диаграмма является очень
богатой --- она включает как магнитоупорядоченные фазы, так и состояние
спиновой жидкости. Интересно, что эта проблема сводится к двумерному бозе-газу в магнитном поле \cite{633}.

\subsection{Электронный спектр в модели Хаббарда и переход металл-изолятор}

Обсудим теперь полный гамильтониан многоэлектронной системы кристалла.
Для перехода к представлению вторичного квантования в случае
вырожденных электронных зон можно использовать приближение сильной
связи, предполагая, что зоны происходят из атомных волновых
функций
\begin{equation}
\varphi _{lm\sigma }(x)=\varphi _{lm}(\mathbf{r})\chi _\sigma
(s)=R_l(r)Y_{lm}(\hat{\mathbf{r}})\chi _\sigma (s)
 \label{eq:C.2}
\end{equation}
где $s$~"--- спиновая координата, $R_l$~"--- радиальная волновая
функция, $Y$~"--- сферическая гармоника,
$\hat{\mathbf{r}}=(\theta,\phi )$, $l$ и $m$ -- орбитальное и
магнитное квантовые числа. Тогда гамильтониан примет вид
(\ref{eq:G.1}) с заменой $\nu_i \rightarrow {\nu_i l_i m_i}$.
Следует, однако, иметь в виду, что атомные функции
не~удовлетворяют условию ортогональности для различных узлов $\nu
$, которое должно выполняться для процедуры вторичного
квантования.

Проще всего использовать процедуру
ортогонализации, предложенную Боголюбовым~\cite{651}. С~точностью
до первого порядка по перекрытию атомных функций
ортогонализованные функции имеют вид
\begin{equation}
\psi _{\nu lm}(\mathbf{r})=\varphi _{\nu lm}(\mathbf{r})-\frac
12\sum_{\nu ^{\prime }\neq \nu }\sum_{l^{\prime }m^{\prime
}}\varphi _{\nu ^{\prime }l^{\prime }m^{\prime }}(\mathbf{r})\int
\varphi _{\nu ^{\prime }l^{\prime }m^{\prime
}}^{*}(\mathbf{r}^{\prime })\varphi _{\nu
lm}(\mathbf{r})\,d\mathbf{r}^{\prime }.
 \label{eq:C.3}
\end{equation}

Запишем гамильтониан  в многоэлектронном
представлении, учитывая одноузельное кулоновского отталкивание
и межузельный перенос электронов (что соответствует модели
Хаббарда):
\begin{multline}
\mathcal{H} =\sum_{\nu \Gamma }E_\Gamma X_\nu (\Gamma ,\Gamma )+
\\
+\sum_{\nu _1\neq \nu _2}\sum_{\Gamma _n\Gamma _{n-1}\Gamma
_{n^{\prime }}\Gamma _{n^{\prime }-1}}B_{\nu _1\nu _2}(\Gamma
_n\Gamma _{n-1},\Gamma _{n^{\prime }}\Gamma _{n^{\prime }-1})
X_{\nu _1}(\Gamma _n\Gamma _{n-1})X_{\nu _2}(\Gamma _{n^{\prime
}}\Gamma _{n^{\prime }-1}),
 \label{eq:C.21}
\end{multline}
В пренебрежении зависимостью энергии терма от МЭ квантовых чисел имеем
\begin{equation}
E_\Gamma =\frac 12n(n-1)F^{(0)}(ll).
 \label{eq:C.20}
\end{equation}
где $ F^{(p)}$ -- интегралы Слэтера. Учет таких вкладов с $p=2,\
4,\ \ldots $ дают зависимость энергии термов от MЭ квантовых чисел
$S$, $L$ в соответствии с правилом Хунда.

Многоэлектронные интегралы переноса содержат вклад, связанный с матричными
элементами электростатического взаимодействия для
$\nu _1\neq \nu _3$, $\nu _2=\nu _4$ (ср. (\ref{eq:G0})).
В частности, для $s$"~зон (в модели (\ref{eq:G0})) имеем
\begin{multline}
\mathcal{H} =U\sum_\nu X_\nu (2,2)+
\sum_{\nu _1\nu _2\sigma }\{t _{\nu _1\nu _2}^{(00)}X_{\nu
_1}(\sigma ,0)X_{\nu _2}(0,\sigma )+t _{\nu _1\nu
_2}^{(22)}X_{\nu _1}(2,\sigma )X_{\nu _2}(\sigma ,2)+
\\
+\sigma t _{\nu _1\nu _2}^{(02)}[X_{\nu _1}(\sigma ,0)X_{\nu
_2}(-\sigma ,2)+X_{\nu _1}(2,-\sigma )X_{\nu _2}(0,\sigma )]\},
 \label{eq:C.23}
\end{multline}
где $U=I_{\nu \nu \nu \nu }=F^{(0)}(00)$~"--- параметр Хаббарда,
\begin{align}
t _{\nu _1\nu _2}^{(00)} =t _{\nu _1\nu _2}, \quad
t _{\nu _1\nu _2}^{(22)} =t _{\nu _1\nu _2}+2I_{\nu _1\nu
_1\nu _2\nu _1},
\notag \\
t _{\nu _1\nu _2}^{(02)} =t _{\nu _1\nu _2}^{(20)}=t
_{\nu _1\nu _2}+I_{\nu _1\nu _1\nu _2\nu _1}
 \label{eq:C.24}
\end{align}
суть интегралы переноса для дырок и двоек и интеграл рождения
дырок и двоек; в соответствии с~(\ref{eq:C.3})
\begin{equation}
I_{\nu _1\nu _1\nu _2\nu _1}=\widetilde{I}_{\nu _1\nu _1\nu _2\nu
_1}-\frac U2\int \varphi _{\nu _2}(\mathbf{r})\varphi _{\nu
1}(\mathbf{r})\,d\mathbf{r},
 \label{eq:C.25}
\end{equation}
где интегралы $\widetilde{I}$ вычисляются для атомных функций $\varphi $.

Зависимость интегралов переноса от
атомных MЭ термов может быть менее тривиальной, если использовать при
решении атомной проблемы более сложные подходы, чем в
разделе~2.1. Например, общее приближение
Хартри"--~Фока (см.~\cite{20}) дает радиальные одноэлектронные
функции, которые явно зависят от атомных термов. В~некоторых
вариационных подходах многоэлектронной атомной теории
(см.~\cite{664}) MЭ волновые функции не~факторизуются на
одноэлектронные. Поэтому интегралы переноса должны быть вычислены
с использованием MЭ волновых функций, как обсуждалось выше.
В~частности, для $s$"~зон интегралы~(\ref{eq:C.24}) могут
отличаться даже в пренебрежении межатомным кулоновским
взаимодействием и неортогональностью. Кроме этого, может
потребоваться многоэлектронный подход, который принимает во
внимание взаимодействие различных электронных оболочек.

Приведем еще выражение для параметра прямого обмена в случае
невырожденной зоны \cite{727} (в общем случае матричные элементы обменного и
кулоновского взаимодействия могут быть выражены через коэффициенты
Клебша-Гордана \cite{660,II}):
\begin{equation}
\ J_{\nu _1\nu _2} = - I_{\nu _1\nu _2\nu _2\nu _1}
=  \widetilde{J}_{\nu _1\nu _2}+2\gamma _{\nu _1\nu _2}L_{\nu _1\nu _2}-%
\frac 12(U+Q_{\nu _1\nu _2})\gamma _{\nu _1\nu _2}^2  \label{J}
\end{equation}
где
\begin{equation}
\widetilde{J}_{\nu _1\nu _2}=-\widetilde{I}_{\nu _1\nu _2\nu _2\nu
_1},\,L_{\nu _1\nu _2}=\widetilde{I}_{\nu _1\nu _1\nu _2\nu _1},\,
\gamma _{\nu _1\nu _2}=\int d{\bf r}\varphi _{\nu _2}^{*}({\bf r})\varphi _{\nu _1}({\bf r}).
\end{equation}

Кроме "<потенциального"> обмена типа~(\ref{J}) рассмотрим
"<кинетическое"> обменноe взаимодействие, которое появляется во
втором порядке теории возмущений по переносу. В приближении
(\ref{eq:C.20}) имеем
\begin{equation}
\tilde{\mathcal{H}}=\sum_{\nu _1\nu _2}\left\{
\frac{\overline{t }_{\nu _1\nu
_2}^2(ll0)}{F^{(0)}(ll)}\right\}
\{n_1n_2+4(\mathbf{S}_1\mathbf{S}_2)-(2l+1)(n_1+n_2)\}.
 \label{eq:D.27}
\end{equation}
Этот механизм приводит к антиферромагнитному
взаимодействию, так как выигрыш в кинетической энергии достигается
при антипараллельной ориентации спинов электронов. Численные
расчеты (см., напр., \cite{265}) показывают, что при
реалистических межатомных расстояниях данный вклад, как правило,
преобладает над ферромагнитным потенциальным обменом. Таким
образом, модель локализованных спинов не~объясняет ферромагнетизм
металлов группы железа.

Теперь запишем гамильтониан модели Хаббарда с~сильными
корреляциями в более простом виде
\begin{equation}
\mathcal{H}=\sum_{\mathbf{k}m\sigma
}t_{\mathbf{k}}a_{\mathbf{k}lm\sigma }^{\dagger
}a_{\mathbf{k}lm\sigma }+\sum_{i\Gamma }E_\Gamma X_i(\Gamma,\Gamma ),
 \label{eq:H.1}
\end{equation}
Здесь мы не~учитываем зависимость интегралов переноса от $m$,
т.~е. пренебрегаем эффектами кристаллического поля. Такое
приближение позволяет в простейшем случае рассмотреть эффекты
многоэлектронной термовой структуры в спектре. Разумеется, оно
является не слишком реальным, поскольку орбитальные моменты в
твердом теле обычно в значительной степени заморожены
~\cite{II,654}.

Рассмотрим одноэлектронную функцию Грина. Согласно~(\ref{eq:A.31}),
\begin{equation}
G_{\mathbf{k}\gamma }(E)=\langle \!\langle a_{\mathbf{k}\gamma
}|a_{\mathbf{k}\gamma }^{\dagger }\rangle \!\rangle
_E=\sum_{n\Gamma _n\Gamma _{n-1}}n^{1/2}G_{\Gamma _{n-1}}^{\Gamma
_n}C_{\Gamma _{n-1},\gamma }^{\Gamma _n}\langle \!\langle
X_{\mathbf{k}}(\Gamma _{n-1},\Gamma _n)|a_{\mathbf{k}\gamma
}^{\dagger }\rangle \!\rangle _E.
 \label{eq:H.3}
\end{equation}
В~уравнении движения для функции Грина в правой
части~(\ref{eq:H.3}) выполним простейшее расцепление, которое
соответствует расцеплению на разных узлах решетки
"<Хаббард"~I">~\cite{28,29,31}.
В результате получим \cite{653}
\begin{equation}
G_{\mathbf{k}\gamma }(E)=\frac{\Phi _\gamma
(E)}{1-t_{\mathbf{k}}\Phi _\gamma (E)}, \,
\Phi _\gamma (E)=\sum_{n\Gamma _n\Gamma _{n-1}}n\left( G_{\Gamma
_{n-1}}^{\Gamma _n}C_{\Gamma _{n-1},\gamma }^{\Gamma _n}\right)
^2\frac{N_{\Gamma _n}+N_{\Gamma _{n-1}}}{E-E_{\Gamma _n}+E_{\Gamma
_{n-1}}}.
 \label{eq:H.4}
\end{equation}
Отметим, что выражения~(\ref{eq:H.4}) имеют
структуру, которая напоминает~(\ref{eq:F.8}) и
легко обобщается с
учетом одноузельного кристаллического поля (см. также~\cite{29}).
Используя для $E_\Gamma $ приближение~(\ref{eq:C.20}), можно
просуммировать генеалогические коэффициенты в~(\ref{eq:H.4}).
Тогда зависимость от МЭ квантовых чисел $L$, $S$ исчезает, что
соответствует приближению~\cite{29}.

В~отсутствие магнитного и орбитального упорядочения числа
заполнения $N_\Gamma $ в~(\ref{eq:H.4}) не~зависят от проекции
спина и мы имеем
\begin{equation}
\Phi _\gamma (E)=\sum_{n\Gamma _n\Gamma _{n-1}}\frac
n{2[l]}([S_{n-1}][L_{n-1}])^{-1}\left( G_{\Gamma _{n-1}}^{\Gamma
_n}\right) ^2\frac{N_{\Gamma _n}+N_{\Gamma _{n-1}}}{E-E_{\Gamma
_n}+E_{\Gamma _{n-1}}}
 \label{eq:H.6}
\end{equation}
($[A]=2A+1$). Спектр возбуждений определяется уравнением
$1-t_{\mathbf{k}}\Phi _\gamma (E)=0$.
Таким образом, межузельный электронный перенос ведет к размытию
каждого перехода между атомными уровнями в хаббардовскую подзону.
Эти подзоны разделены корреляционными щелями. В~частности, для
$s$"~зоны мы получаем спектр, который содержит в ферромагнитной
фазе четыре подзоны:
\begin{equation}
E_{\mathbf{k}\sigma }^{1,2}=\frac 12\left\{ t_{\mathbf{k}}+U\mp
\left[ (t_{\mathbf{k}}-U)^2+4t_{\mathbf{k}}U(N_{-\sigma
}+N_2)\right] ^{1/2}\right\}.
 \label{eq:H.8}
\end{equation}
Выражение~(\ref{eq:H.8})может быть также переписано через
одноэлектронные числа заполнения, поскольку $N_\sigma
+N_2=n_\sigma ,\qquad N_\sigma +N_0=1-n_{-\sigma }$. В отличие от
приближения Хартри--Фока--Стонера $ E_{\mathbf{k}\sigma}=
t_{\mathbf{k}}+U n_{-\sigma} $, зависимость спектра от
чисел заполнения не~сводится к постоянному сдвигу подзон. Спектр
приближения "<Хаббард"~I"> имеет наиболее простой вид в случае больших~$U$,
когда
\begin{equation}
E_{\mathbf{k}\sigma }^1=(1-n_{-\sigma })t_{\mathbf{k}},\qquad
E_{\mathbf{k}\sigma }^2=t_{\mathbf{k}}n_{-\sigma }+U.
 \label{eq:H.10}
\end{equation}
Можно предположить, что в действительности некоторые подзоны плохо
определены из-за большого затухания.

Для иллюстрации рассмотрим простой пример насыщенного хаббардовского ферромагнетика с малой концентрацией носителей тока (двоек) $c$, где  вычисление дает~\cite{338}
\begin{equation}
G_{\mathbf{k}\downarrow }(E)=\left\{ E-t_{\mathbf{k}}+(1-c)\left[\sum_{\mathbf{q}} \frac{n_{\mathbf{k+q}}}{E-t_{\mathbf{k}}+\omega _ {\bf q}}\right]
^{-1}\right\} ^{-1}.
 \label{eq:J.23}
\end{equation}
($\omega _ {\bf q}$ -- частота магнонов).
Таким образом, некоторые энергетические знаменатели заменяются
резольвентами и соответствующие состояния имеют неквазичастичную
природу. При малых значениях~$c$ функция Грина~(\ref{eq:J.23}) не~имеет
полюсов ниже уровня Ферми. Однако с увеличением $c$ функция Грина приобретает
спин-поляронный полюс ниже~$E_{\text{F}}$ и насыщенный
ферромагнетизм разрушается~\cite{332}.

Выражение~(\ref{eq:J.23})
можно сравнить с соответствующим результатом для парамагнитной
фазы в приближении "<Хаббард"~-III"> (ср.~(\ref{eq:H.16}))
\begin{equation}
G_{\mathbf{k}}(E)=\left\{ E-t
_{\mathbf{k}}+\frac{1-c}2\left[\sum_{\mathbf{q}}
G_{\mathbf{q}} (E)\right] ^{-1}\right\} ^{-1}.
 \label{eq:J.24}
\end{equation}
В~отличие от~(\ref{eq:J.23}), уравнение~(\ref{eq:J.24})
не~содержит фермиевские функции, так что некогерентные
(неквазичастичные) состояния не~исчезают на~$E_{\text{F}}$.
Подобная ситуация всегда имеет место в приближении
"<Хаббард"~III">~\cite{30,694,695}, где затухание на уровне Ферми
конечно (см.~(\ref{eq:H.17}), (\ref{eq:J.24})).

Флуктуационные поправки к электронным функциям Грина приближения
"<Хаббард"~I"> были получены в работах~\cite{337,338,730,729} в рамках
формального разложения по~$1/z$ ($z$~"--- число ближайших соседей). Они
выражаются через одночастичные числа заполнения и спиновые и
зарядовые корреляционные функции
\begin{eqnarray}
\chi _{\mathbf{q}}^{-\sigma \sigma } &=&\langle S_{-\mathbf{q}}^{-\sigma }S_{%
\mathbf{q}}^{\sigma }\rangle =\langle X_{-\mathbf{q}}^{-\sigma \sigma }X_{%
\mathbf{q}}^{\sigma -\sigma }\rangle ,\quad \kappa _{\mathbf{q}%
}^{{}}=\langle X_{\mathbf{-q}}^{20}X_{\mathbf{q}}^{02}\rangle ,  \nonumber \\
\chi _{\mathbf{q}}^{zz} &=&\langle \delta (X_{-\mathbf{q}}^{00}+X_{-%
\mathbf{q}}^{\sigma \sigma })\delta (X_{\mathbf{q}}^{00}+X_{\mathbf{q}%
}^{\sigma \sigma })\rangle , \quad \delta A=A-\langle A\rangle
\end{eqnarray}%
Формальную проблему, связанную с нарушения аналитических свойств в
этом разложении \cite{694,695}, удалось решить переходом к
локаторному представлению функций Грина \cite{729}.
Соответствующий результат имеет вид
\begin{equation}
G_{\mathbf{k}\sigma }(E)=\frac{1}{F_{\mathbf{k}\sigma }(E)-t_{\mathbf{k}}}%
,\quad F_{\mathbf{k}\sigma }(E)=\frac{b_{\mathbf{k}\sigma }(E)}{a_{\mathbf{k}%
\sigma }(E)},  \label{locator}
\end{equation}%
\begin{eqnarray}
a_{\mathbf{k}\sigma }(E) &=&\frac{N_{0}+N_{\sigma }}{E}+\frac{N_{-\sigma
}+N_{2}}{E-U}+\left( \frac{1}{E}-\frac{1}{E-U}\right) \sum_{\mathbf{q}}t_{%
\mathbf{q}}  \nonumber \\
&&\times \left( \frac{-En_{\mathbf{q}-\sigma }+U(\langle c_{\mathbf{q}%
-\sigma }^{\dagger }X_{\mathbf{q}}^{0-\sigma }\rangle -\chi _{\mathbf{k+q}%
}^{-\sigma \sigma })}{E^{2}-E(t_{\mathbf{q}}+U)+Ut_{\mathbf{q}%
}(N_{0}+N_{-\sigma })}+\frac{En_{\mathbf{q}-\sigma }+U(\sigma \langle X_{-%
\mathbf{q}}^{2\sigma }c_{\mathbf{q}-\sigma }\rangle +\kappa _{\mathbf{k+q}%
}^{{}})}{E^{2}+E(t_{\mathbf{q}}-U)-Ut_{\mathbf{q}}(N_{\sigma }+N_{2})}%
\right. -{}  \nonumber \\
&&{}-\left. \frac{U\chi _{\mathbf{k+q}}^{zz}}{E^{2}-E(t_{\mathbf{q}}+U)+Ut_{%
\mathbf{q}}(N_{0}+N_{\sigma })}\right) {}  \label{ak}
\end{eqnarray}%
\begin{eqnarray}
b_{\mathbf{k}\sigma }(E) &=&1-\left( \frac{1}{E}-\frac{1}{E-U}\right) \sum_{\mathbf{q}}t_{\mathbf{q}}^{2}\left( \frac{En_{\mathbf{q}-\sigma }-U\langle
c_{\mathbf{q}-\sigma }^{\dagger }X_{\mathbf{q}}^{0-\sigma }\rangle }{%
E^{2}-E(t_{\mathbf{q}}+U)+Ut_{\mathbf{q}}(N_{0}+N_{-\sigma })}\right. +{}
\nonumber \\
&&{}+\left. \frac{En_{\mathbf{q}-\sigma }+\sigma U\langle X_{-\mathbf{q}%
}^{2\sigma }c_{\mathbf{q}-\sigma }\rangle }{E^{2}+E(t_{\mathbf{q}}-U)-Ut_{%
\mathbf{q}}(N_{\sigma }+N_{2})}\right) .  \label{bk}
\end{eqnarray}
Эти выражения могут быть использованы для анализа электронного спектра и различных
фазовых переходов в модели Хаббарда; при этом корреляционные функции должны
находиться самосогласованно. К сожалению, такое исследование пока до конца
не выполнено,  поскольку оно сталкивается с вычислительными трудностями.

Для парамагнитной фазы щель в спектре~(\ref{eq:H.8}) сохраняется
при сколь угодно малых $U$. Чтобы описать переход
металл"--~изолятор, который имеет место при $U\sim W$ ($W$~"---
ширина зоны), требуются более сложные самосогласованные
приближения для электронных функций Грина. Первое описание такого
типа было предложено Хаббардом~\cite{30}; более простое
приближение использовалось Зайцевым~\cite{697}.

Выражение "<Хаббард"~III"> для одноэлектронной функции Грина в
случае наполовину заполненной зоны может быть представлено в
виде~\cite{695}
\begin{equation}
G_{\mathbf{k}}(E)=[E-t_{\mathbf{k}}-\Sigma (E)]^{-1},
 \label{eq:H.16}
\end{equation}
причем электронная собственная энергия определяется самосогласованно
через точную резольвенту:
\begin{equation}
\Sigma (E) =\frac{U^2}{16\Psi }R(E)\Biggm/ \left[ 1+\Sigma
(E)R(E)+ER(E)\left( \frac 1{4\Psi }-1\right) \right], R(E)=\sum_{\mathbf{k}}G_{\mathbf{k}}(E),
 \label{eq:H.17}
\end{equation}
где $\Psi =3/4$ -- нормированный квадрат полного спина на узле. Выражение~(\ref{eq:H.17}) выполняется также для
классической ($S\rightarrow \infty $) $s$---$d$~обменной модели,
если мы положим $\Psi =1/4$,
$U\rightarrow 2|IS|$. Тогда формула~(\ref{eq:H.17}) упрощается и
совпадает с результатом приближения когерентного потенциала~(CPA)
в теории неупорядоченных сплавов.

Недостатки приближений~\cite{30,697} (нарушение аналитических
свойств функций Грина, несамосогласованное описание
термодинамических свойств) обсуждаются в работах~\cite{694,695} с
точки зрения $1/z$"~разложения. Правильное описание перехода
Мотта"--~Хаббарда является важной физической проблемой.
В~последнее время здесь широко используется приближение
бесконечной размерности пространства~$d$, которое строится в
рамках динамической теории среднего поля (DMFT) ~\cite{705}. При
этом исходная проблема сводится к эффективной однопримесной модели
Андерсона с нетривиальной динамикой, где остается взаимодействие
электронов только на одном узле, погруженном в термостат свободных
фермионов, параметры которого определяются самосогласованным
образом. Это приближение позволяет получить трехпиковую структуру
плотности состояний, включая кондовский пик на уровне Ферми.
Оно может быть формально обобщен на произвольные решетки. Такой
подход может давать два фазовых перехода: при $U>U_{\text{c}1}$
нарушается фермижидкостная картина, а
при~$U>U_{\text{c}2}>U_{\text{c}1}$ система переходит в
изоляторное состояние.

\begin{figure}[tbp]
\includegraphics[width=0.65\columnwidth]{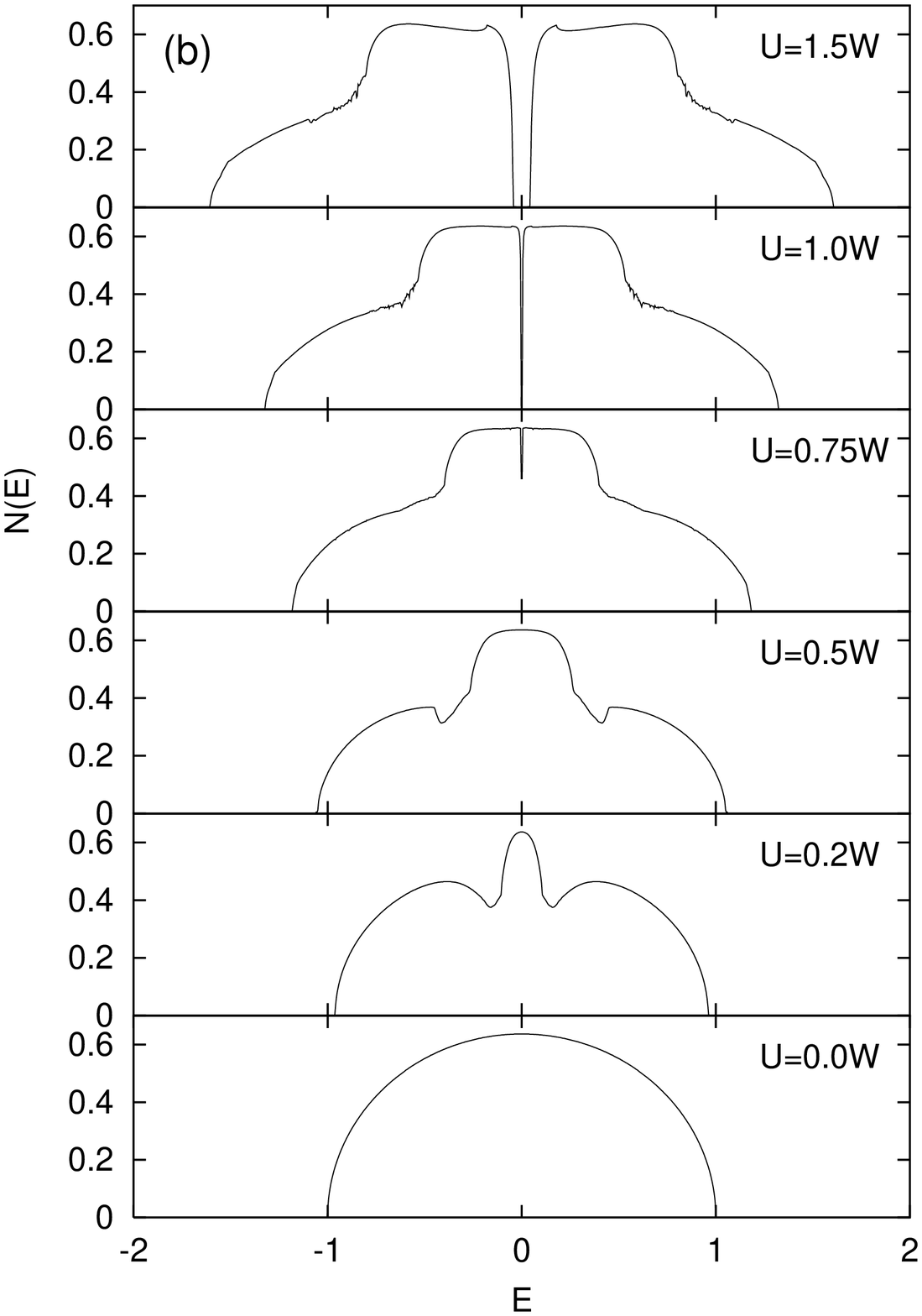}
\caption{Плотность состояний в модели Хаббарда для затравочной
полуэллиптической зоны при различных значениях $U$ \cite{730} ($W$ -- ширина зоны).}
\label{fig:H.1}
\end{figure}

Учет фермиевских возбуждений в рамках $1/z$"~разложения был
выполнен в работе~\cite{730}. Использование локаторного
представления для функции Грина (\ref{locator}) позволило получить
правильные аналитические свойства и воспроизвести трехпиковую
структуру. Используя (\ref{ak}), (\ref{bk}) и пренебрегая
импульсной зависимостью корреляционных функций, самосогласованный
результат для локатора~$F(E)=b(E)/a(E)$ можно записать в виде
\begin{align}
a(E)=&1+\frac
34\frac{U^2}{E^2}\sum_{\mathbf{q}}t_{\mathbf{q}}G_{\mathbf{q}}(E)
+\frac{2U}E\sum_{\mathbf{q}}t_{\mathbf{q}}G_{\mathbf{q}}(E)
n_{\mathbf{q}}, \nonumber
\\
b(E)=&F_0(E)+\frac{2U}E\sum_{\mathbf{q}}t_{\mathbf{q}}^2
G_{\mathbf{q}}(E)n_{\mathbf{q}},
\label{eq:H.177}
\end{align}
где $n_{\mathbf{q}}$~"--- точные числа заполнения. Соответствующая
эволюция электронного спектра в зависимости от параметра
взаимодействия показана на рис.~\ref{fig:H.1}.

Альтернативная картина моттовского перехода, которая была
развита в рамках полярной модели, -- связывание носителей
тока противоположного знака (двоек и дырок) в бестоковые
элементарные возбуждения -- экситоны особого типа;
в основном состоянии происходит их конденсация \cite{exciton,692}.
Важно отметить, что  причиной этих явлений является
не дальнодействующее кулоновское взаимодействие (как в полупроводниках),
а флуктуации полярности. Сами экситоны  в вариационном приближении типа
Хартри-Фока  соответствуют антиферромагнитным спиновым волнам, а щель имеет
слэтеровскую природу \cite{692}.
Обобщение этого приближения  (например, вариационное рассмотрение  состояния, в котором формируются экситоны с разными квазиимпульсами) сталкивается с математическими трудностями и является нерешенной задачей.

Рассмотрим  электронный  спектр антиферромагнетика
Хаббарда с сильными
корреляциями. Появление АФМ~упорядочения приводит к расщеплению
затравочной электронной зоны на две слэтеровские подзоны
\begin{equation}
E_{\mathbf{k}}^{\alpha ,\beta }=\theta _{\mathbf{k}}\mp
\left( \tau
_{\mathbf{k}}^2+U^2\overline{S}^2\right) ^{1/2}, \,
\theta _{\mathbf{k}},\tau _{\mathbf{k}}=\frac
12(t_{\mathbf{k}+\mathbf{Q}/2}\pm t_{\mathbf{k}-\mathbf{Q}/2}).
 \label{eq:G.98}
\end{equation}
Величина
\[
\overline{S}=\sum_{\mathbf{k}}\langle c_{\mathbf{k}\uparrow }^{\dagger
}c_{\mathbf{k}+\mathbf{Q}\downarrow }\rangle,
\]
которая определяет АФМ~расщепление, может быть получена из
самосогласованного уравнения. Если кулоновское взаимодействие
достаточно сильно, вся энергетическая зона расщеплена, так что
щель возникает во всех направлениях. В~частности, в случае одного
электрона на атом происходит переход металл"--~изолятор. Если для
данного вектора~$\mathbf{Q}$ выполняется условие "<нестинга">
$t_{\mathbf{k}}-E_{\text{F}}=E_{\text{F}}-t_{\mathbf{k}+\mathbf{Q}}$,
то щель в спектре сохраняется при произвольно малом $U$.

В рамках этой картины выражение для спектра спиновых волн
и спин-волновые поправки к спектру (\ref{eq:G.98})
были получены в работе~\cite{693}. В отличие от
стонеровского ферромагнетика, в антиферромагнитной фазе удается
получить интерполяцию между пределами слабой и сильной связи.


\subsection{Ферромагнетизм сильно коррелированных $d$"~систем}

Природа магнетизма металлических систем $d$--электронов и
магнитных изоляторов, описываемых моделью Гейзенберга, существенно
отличается. Хотя в обоих случаях для магнитной восприимчивости
выполняется закон Кюри-Вейсса, он не обязательно связан с
существованием локализованных моментов. В частности, совсем иную
природу закон Кюри-Вейсса имеет в слабых зонных магнетиках, где
развитые спиновые флуктуации имеются лишь в узкой области
$q$--пространства. В то же время в сильных зонных магнетиках
флуктуации имеются в широкой области волновых векторов и приводят
к образованию локальных магнитных моментов (ЛММ), т.е.  к
неоднородности спиновой плотности в реальном пространстве.


Для стандартных подходов в теории магнетизма коллективизированных
электронов (зонные расчеты, спин-флуктуационные теории) наиболее
трудны системы, в которых сильные межэлектронные корреляции
приводят к радикальной перестройке электронного спектра ---
формированию хаббардовских подзон. Это оксиды и сульфиды
переходных металлов с большой энергетической щелью, например MeO
(Me${}={}$Ni, Co, Mn), NiS${}_2$~\cite{25}, базовые системы для
медь-оксидных высокотемпературных сверхпроводников
La${}_2$CuO${}_4$ и YBa${}_2$Cu${}_3$O${}_6$. При низких
температурах последние антиферромагнитны, так что щель можно было
бы трактовать как слэтеровскую, т.~е. связанную с АФМ
упорядочением. Однако щель сохраняется и в парамагнитной области и
имеет поэтому природу Мотта"--~Хаббарда. Расщепление Хаббарда
возникает и в некоторых металлических ферромагнетиках, в
частности, в твердом растворе Fe${}_{1-x}$Co${}_x$S${}_2$, имеющем
структуру пирита, CrO${}_2$~\cite{25}. Спонтанное спиновое
расщепление выше точки Кюри, которое в некоторых экспериментах
наблюдается в металлах группы железа, также может быть
интерпретировано с точки зрения хаббардовских подзон.

Хаббардовское расщепление противоречит картине ферми-жидкости --- в частности,
теореме Латтинджера о сохранении объема под поверхностью Ферми:
каждая из двух подзон, возникающих из зоны свободных электронов,
содержит одно электронное состояние на спин. Таким образом,
энергетический спектр коллективизированной электронной системы с
ЛММ, в противоположность слабым коллективизированным магнетикам,
существенно отличается от энергетического спектра нормальной
ферми-жидкости.

Проблема описания ЛММ и вывод закона Кюри-Вейсса --- ключевой
момент теории сильного ферромагнетизма коллективизированных
электронов. В~теориях спиновых флуктуаций~\cite{26} ЛММ вводится
по существу со стороны (например статическое приближение в
интеграле по траекториям, которое соответствует замене
трансляционно-инвариантной системы разупорядоченной системой со
случайными магнитными полями).

С другой стороны, "<многоэлектронная"> (атомная) картина описывает ЛММ
естественным образом. Роль сильных электронных корреляций в
формировании ЛММ может быть качественно показана следующим
способом. Если внутриузельное отталкивание Хаббарда $U$ достаточно
велико, то электронный спектр содержит хаббардовские подзоны
однократно и двукратно занятых состояний. При $n<1$ число пар мало
и стремится к нулю при $U\rightarrow \infty $. Тогда однократно
занятые состояния составляют ЛММ, а пустые узлы (дырки) являются
носителями тока.

Картина ферромагнетизм в модели Хаббарда с сильными корреляциями
существенно отличается от стонеровской. Поэтому условие
существования ферромагнетизма не~должно совпасть с критерием
Стонера, который соответствует неустойчивости немагнитной
ферми-жидкости относительно малой спиновой поляризации. Еще один
принципиальный момент: обычное расцепление Хартри"--~Фока в
одноэлектронном представлении не~описывают образование локальных
моментов, поскольку не~учитывает корректно формирования "<двоек">,
т.~е. дважды занятых состояний на узле. Эта трудность является
принципиальной -- она указывает на смену статистики от зонной к
атомной при увеличении взаимодействия. Неадекватность
одноэлектронного подхода в случае больших $U$ может быть показана
рассмотрением случая малых электронных
концентраций~$n$~\cite{353}. Разложение электронной функции Грина
по числам заполнения носителей тока. в~одноэлектронном
представлении дает число двоек, которое ведет себя в этом пределе
как $1/U$. Тогда теорема Геллмана"--~Фейнмана $N_2=\partial
\mathcal{E}/\partial U$ дает расходимость энергии основного
состояния~$\mathcal{E}$:
\begin{equation}
\mathcal{E}(U)-\mathcal{E}(0)=\int\limits_0^\infty N_2(U)\,dU\sim
\ln{} U.
 \label{eq:H.22}
\end{equation}
Эта расходимость указывает на формирование хаббардовских подзон и
неправильность одноэлектронной картины при больших~$U$. С~другой
стороны, вычисление в представлении операторов Хаббарда~\cite{353} дает правильную
асимптотику $N_2\sim 1/U^2$. Таким образом, возникает неперестановочность
предельного перехода $U \rightarrow \infty$  с разложением по другим малым параметрам.
Эта проблема пока остается
неразрешенной. Следует отметить, что такой недостаток не~играет
большой роли для насыщенного ферромагнитного состояния, где
образование двоек запрещено и описания на языках слэтеровских и
хаббардовских подзон совпадают, поскольку, как видно из
(\ref{eq:H.10}), спектры в обоих подходах тождественны.

В~простейшем приближении "<Хаббард"~I"> (см.~(\ref{eq:H.10})) магнитное
упорядочение есть результат сужения и расширения спиновых подзон (а
не~постоянного спинового расщепления, как в теории Стонера).
Однако оказалось, что эта картина Хаббарда также не дает удовлетворительных результатов.
В частности, Хаббард~\cite{28} не~нашел магнитных решений для
простых затравочных плотностей состояний (хотя ситуация может
измениться в случае вырожденных $d$"~зон~\cite{696}). Метод
$X$"~операторов проясняет причину этой неудачи:
соответствующие выражения для функций Грина при больших $U$
\begin{equation}
\langle \!\langle X_{\mathbf{k}}(\sigma 0)|X_{-\mathbf{k}}(0\sigma
)\rangle \!\rangle _E =\frac{c+N_\sigma }{E- t_{\mathbf{k} }(c+N_\sigma )}.
 \label{eq:H.11}
\end{equation}
($\varepsilon _{\mathbf{k}}=-t_{\mathbf{k}}$) нарушают
кинематические соотношения~(\ref{eq:A.25}), так как при $\langle
S^z\rangle \neq 0$ невозможно одновременно удовлетворить тождества
\begin{equation}
\sum_{\mathbf{k}}\langle X_{-\mathbf{k}}(0\sigma) X_{\mathbf{k}}(\sigma 0)\rangle
=\langle X(00)\rangle =N_0=c
 \label{eq:H.13}
\end{equation}
для обеих проекций спина $\sigma $.

Более успешным оказывается критерий неустойчивости насыщенного
ферромагнетизма, связанный с появлением спин-поляронного полюса
для функции Грина со спином вниз ниже
уровня Ферми \cite{332,729} (ср. (\ref{eq:J.23})).
В~отличие от приближения "<Хаббард"~I"> и
"<Хаббард"~III">, использование выражений для функций Грина
(\ref{locator})-(\ref{bk}), которые содержат фермиевские функции
распределения, позволяет качественно правильно получить магнитную
фазовую диаграмму и описать насыщенное и ненасыщенное
ферромагнитное состояние~\cite{729}. Первая критическая
концентрация носителей тока, соответствующая неустойчивости
насыщенного ферромагнетизма, составляет для различных решеток
около~$30\,\%$ --- значение, которое было ранее получено
различными методами.

Строгое исследование ферромагнетизма в модели Хаббарда
с~$U\rightarrow \infty $ выполнил Нагаока~\cite{349}. Он доказал,
что основное состояние для простой кубической и oцк-решетки в
приближении ближайших соседей с числом электронов
$N_{\text{e}}=N+1$ ($N$~"--- число узлов в решетке) обладает
максимально возможным полным спином, т.~е. является насыщенным
ферромагнетиком. То~же верно для гцк- и гпу-решеток с интегралом
переноса $t<0$, $N_{\text{e}}=N+1$, или $t>0$, $N_{\text{e}}=N-1$.
(Для других комбинаций знаков основное состояние является более
сложным из-за расходимости плотности состояний на границе зоны.)
Физический смысл теоремы Нагаока довольно прост. Для
$N_{\text{e}}=N$, $U=\infty $ каждый узел однократно занят и
движение электронов невозможно, так что энергия системы не~зависит
от спиновой конфигурации. При введении избыточного электрона (или
дырки) его кинетическая энергия будет минимальна при однородном
ферромагнитном спиновом упорядочении, поскольку оно
не~препятствует их движению. Нужно, однако, отметить, что
доказательство теоремы Нагаока использует нетривиальные
топологические соображения. В~частности, оно не~работает в
одномерном случае, когда зависимость кинетической энергии от
спиновой конфигурации отсутствует, поскольку нет замкнутых
траекторий~\cite{349}.

В~случае полузаполненной зоны ($N_{\text{e}}=N$), $|t|\ll U$
основное состояние антиферромагнитно из-за кинетического обменного
взаимодействия Андерсона порядка $t^2/U$ (см. (\ref{eq:D.27})).
Оно
возникает из-за увеличения кинетической энергии при виртуальных
переходах электрона на соседние узлы, которые возможны при
условии, что электрон на данном узле имеет противоположное
направление спина.
В~системах с конечным $U$ и $N_{\text{e}}\neq
N$ имеет место конкуренция между ферро- и антиферромагнитным
упорядочением. Из
вычисления энергии спиновой волны~\cite{349}, следует, что
ферромагнетизм сохраняется при условии
\begin{equation}
|t|/U<\kappa |N_{\text{e}}-N)|/N,
 \label{eq:4.95}
\end{equation}
где константа $\kappa \sim 1$ зависит от структуры кристаллической
решетки. В~то~же время антиферромагнетизм устойчив только
при $N_{\text{e}}=N$. В~ранних работах предполагалось, что в
промежуточной области формируются скошенные магнитные
структуры~\cite{350}. Однако численные расчеты~\cite{351,Igoshev}
показывают, что энергетически более выгодно фазовое расслоение на
антиферромагнитные и ферромагнитные области.
Такое явление, по-видимому, наблюдается в некоторых
сильнолегированных магнитных полупроводниках, причем все носители тока локализуются только в ферромагнитных областях~\cite{352}.

В отличие от обычного критерия Стонера, приближение Хартри-Фока с
учетом неколлинеарных антиферромагнитных флуктуаций \cite{Igoshev}
дает удовлетворительное согласие с пределом сильных корреляций.


\begin{figure}[tbp]
\includegraphics[width=0.65\columnwidth]{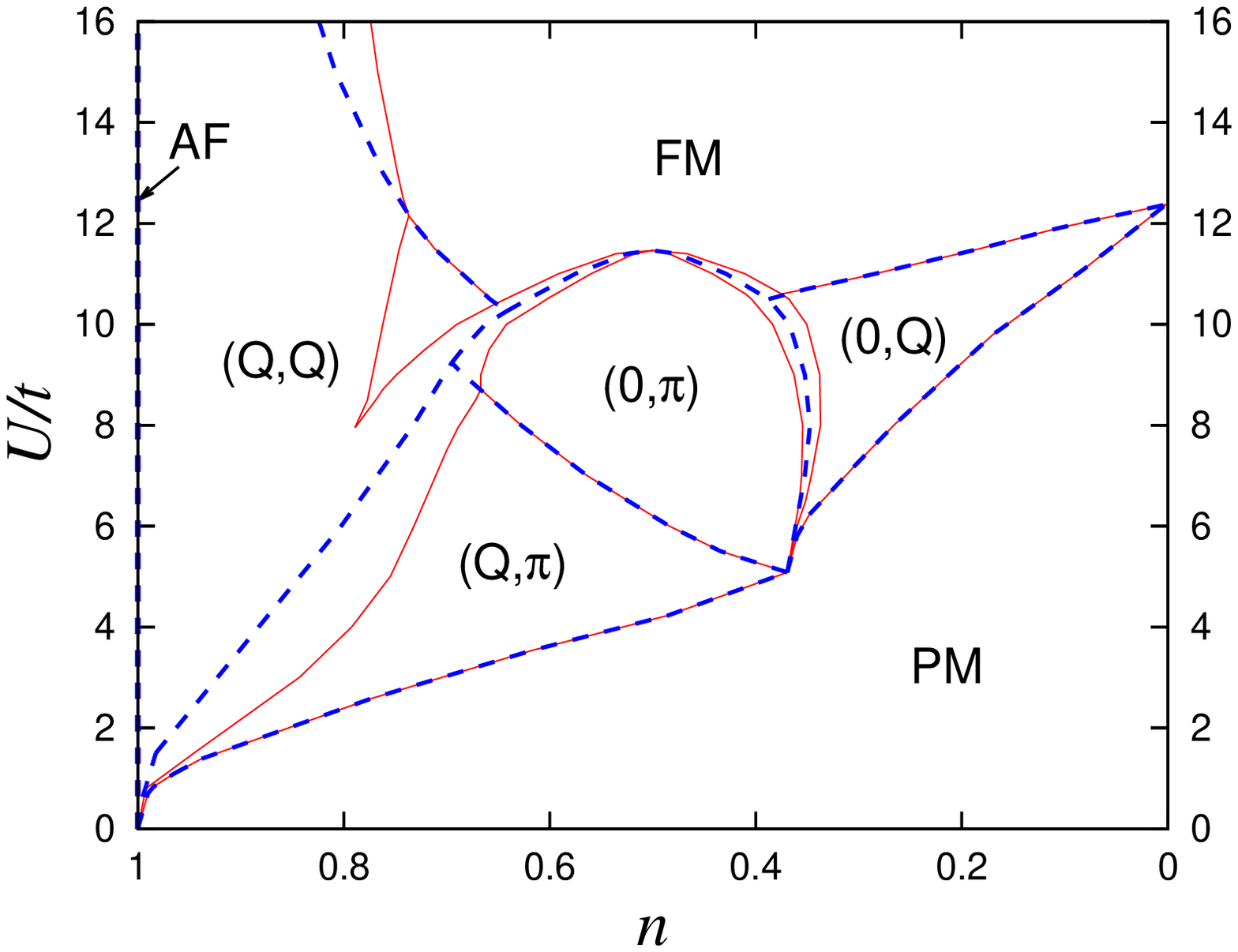}
\caption{Фазовая диаграмма основного состояния двумерной модели
Хаббарда в приближении ближайших соседей \cite{Igoshev};
сплошные линии --- границы фаз,
пунктирные линии --- без учета фазового расслоения. Фазы
обозначены в соответствии с их волновым вектором, FM, AF и PM ---
ферромагнитная, антиферромагнитная и парамагнитная фазы}
\label{fig:1a}
\end{figure}

\begin{figure}[tbp]
\includegraphics[width=0.65\columnwidth]{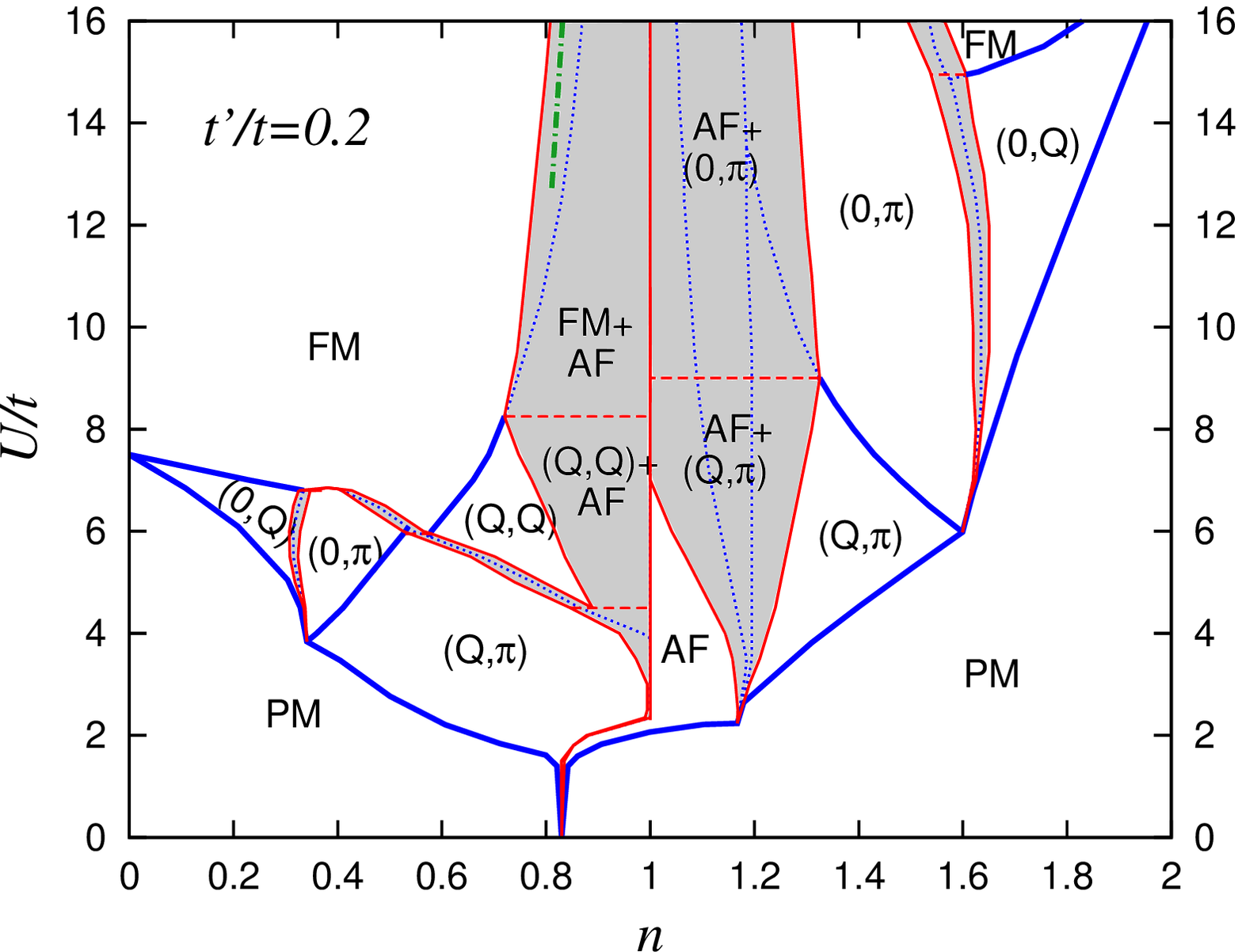}
\caption{Фазовая диаграмма основного состояния квадратной решетки
с $t'/t=0.2$ \cite{Igoshev},  закрашенные области --- области
фазового расслоения,  пунктирные линии --- границы фаз без учета
расслоения, штрих-пунктирная линия -- граница расслоения в пределе
большого взаимодействия $U$ по Вишеру \cite{351}. } \label{fig:3a}
\end{figure}

Фазовая диаграмма модели Хаббарда для квадратной решетки в
приближение среднего поля показана на рис. \ref{fig:1a}.
Видно, что ферромагнетизм возникает только при очень больших, даже
нереалистических значениях $U$. Ситуация существенно меняется при
наличии особенностей ван-Хова вблизи уровня Ферми. На
рис. \ref{fig:3a} видно, что введение даже небольшого переноса
вторых соседей $t'$ приводит к резкой асимметрии фазовой диаграммы
относительно половинного заполнения зоны.  В области $n<1$, в
которой химический потенциал пересекает особенность ван-Хова,
ферромагнитное состояние возникает для умеренных $U$. Аналогичной
ситуации можно ожидать для трехмерных решеток.

Для фазовой диаграммы может иметь существенное значение еще одно
обстоятельство: кинетическое антиферромагнитное обменное
взаимодействие (\ref{eq:D.27}) содержит вклады, обусловленные
поправками неортогональности (второй член в~(\ref{eq:C.25})).
Учитывая (\ref{J}), получаем выражение для эффективного обменного
параметра
\begin{equation}
J_{\text{eff}} =J+2\left( t^{(02)}\right) ^2/U
=\widetilde{J}-2\gamma t+2(t+L)^2/U  \label{Jef}
\end{equation}
(члены порядка $\gamma ^2U$ в выражение для $J_{\text{eff}}$
сокращаются). Таким образом, антиферромагнитное взаимодействие
остается конечным даже в пределе $ U\rightarrow \infty $
~\cite{727}. Первые три члена в (\ref{Jef}) совпадают с
соответствующим результатом для двухузельной проблемы (молекулы
водорода) \cite{656} и дают антиферромагнитное обменное
взаимодействие. В пределе больших $U$ главный вклад в
$J_{\text{eff}}$ принимает вид
\begin{equation}
J_{\text{eff}}\simeq 2\gamma |t|
\end{equation}
где предполагается, что $t$ отрицательно. Мы видим, что в случае
$N_e<N$ отношение $J_{\text{eff}}$ к ширине зоны пропорционально
параметру перекрытия (а не $|t|/U\ll \gamma $, как в обычном
рассмотрении).

По аналогии с рассмотрением Нагаока \cite{349} можно записать
критерий ферромагнетизма $2\kappa c|t^{(\lambda \lambda
)}|>J_{\text{eff}}$, где $\lambda =0$ для $N_e<N$ и $\lambda =2$
для $N_e>N.$ В предположении $\gamma \gg |t|/U\gg \gamma ^2$ этот
критерий принимает вид
\begin{equation}
\kappa c>\left\{
\begin{tabular}{ll}
$J_{\text{eff}}/(2|t|)\simeq \gamma ,$ & $N_e<N$ \\
$J_{\text{eff}}/(\gamma U)\simeq 2|t|/U,$ & $N_e>N$%
\end{tabular}
\right.  \label{new1}
\end{equation}
и существенно видоизменяется для $N_e<N$. Таким образом,
неортогональность приводит к появлению косвенного
антиферромагнитного взаимодействия в узких зонах, причем
ферромагнитный обмен, обусловленный движением носителей тока,
более сильно подавляется в случае ``дырочной'' проводимости
($N_e<N$), чем ``электронной''. Это обстоятельство может иметь значение
для фазовой диаграммы меднооксидных перовскитов и аналогичных систем.

Для рассмотрения проблемы ферромагнитного упорядочения в узких
зонах с экспериментальной точки зрения подробнее обсудим систему
Fe${}_{1-x}$Co${}_x$S${}_2$. Ее электронная структура довольно
проста: все электроны, ответственные и за проводимость, и за
магнетизм, принадлежат одной и той~же узкой
$e_{\text{g}}$"~полосе. CoS$_2$~"--- ферромагнитный металл с
сильными корреляциями, FeS$_2$~"--- диамагнитный изолятор.

Экспериментальные исследования магнитных свойств
Fe${}_{1-x}$Co${}_x$S${}_2$ выполнены в~\cite{347}. Наиболее
интересная особенность~"--- возникновение ферромагнетизма при
удивительно малых электронных концентрациях $n=x<0{,}05$.
Магнитный момент равняется $1\mu _{\text{B}}$ в широком
концентрационном интервале $0{,}15<n<0{,}95$, причем магнитное
состояние не насыщено при $n<0{,}15$. Однако, в отличие от обычных
слабых коллективизированных ферромагнетиков, нет никаких
признаков обменного усиления восприимчивости Паули выше
$T_{\text{C}}$, а закон Кюри"--~Вейсса хорошо выполняется при
произвольных электронных концентрациях, причем константа Кюри
пропорциональна $n$. Такое поведение нельзя объяснить в рамках
одноэлектронных подходов типа Стонера, что отражает важную роль локальных
магнитных моментов (ЛММ).

В работе~\cite{353} был использован
подход на основе спиновой функции Грина.
Ее вычисление  в магнитном поле $H$ дает
\begin{multline}
G_{\mathbf{q}}(\omega ) =\left( 2\langle S^z\rangle
+\sum_{\mathbf{k}}\frac{(\varepsilon
_{\mathbf{k}-\mathbf{q}}-\varepsilon
_{\mathbf{k}})(n_{\mathbf{k}\uparrow
}-n_{\mathbf{k}-\mathbf{q}\downarrow })}{\omega
-H-E_{\mathbf{k}\uparrow }+E_{\mathbf{k}-\mathbf{q}\downarrow
}}\right)\times
\\
\times \left( \omega -H-\sum_{\mathbf{k}}\frac{(\varepsilon
_{\mathbf{k}-\mathbf{q}}-\varepsilon _{\mathbf{k}})(\varepsilon
_{\mathbf{k}}n_{\mathbf{k}\uparrow }-\varepsilon
_{\mathbf{k}-\mathbf{q}}n_{\mathbf{k}-\mathbf{q}\downarrow
})}{\omega -H-E_{\mathbf{k}\uparrow
}+E_{\mathbf{k}-\mathbf{q}\downarrow }}\right) ^{-1},
 \label{eq:4.97}
\end{multline}
где $E_{\mathbf{k}\sigma }=\varepsilon_{\mathbf{k} }(c+N_\sigma )$
-- энергии в приближении "<Хаббард"~I">, $n_{\mathbf{k}\sigma }$ ---
соответствующие числа заполнения, $\varepsilon _{\mathbf{k}}=-t
_{\mathbf{k}}$.

При вычислении статической магнитной восприимчивости $\chi $
необходимо тщательно рассмотреть пределы $H\rightarrow 0$, $\omega
\rightarrow 0$, $q\rightarrow 0$ из-за неэргодичности
ферромагнитного основного состояния. Действительно, простая
подстановка $H=\omega =q=0$  дает только
восприимчивость Паули. Чтобы избежать потери вклада Кюри"--~Вейсса от
локальных моментов, в работе~\cite{353} был использован метод Тябликова
для модели Гейзенберга. Применяя спектральное представление
для функций Грина, находим уравнение для намагниченности
\begin{equation}
\langle S^z\rangle =\frac{1-c}2+\frac 1\pi \int\limits_{-\infty
}^\infty N_{\text{B}}(\omega )\Im
\sum_{\mathbf{q}}G_{\mathbf{q}}(\omega )\,d\omega .
 \label{eq:4.98}
\end{equation}
При малых концентрациях дырок $c\ll 1$ возникает
насыщенный ферромагнетизм с
\begin{equation}
\langle S^z\rangle
=\frac{1-c}2-\sum_{\mathbf{p}}N_{\text{B}}(\omega _{\mathbf{p}}).
 \label{eq:4.99}
\end{equation}
Уравнение~(\ref{eq:4.98}) может быть упрощено при условии $\langle
S^z\rangle \ll 1$, которое имеет место и в парамагнитной области
($\langle S^z\rangle =\chi H$, $H\rightarrow 0$), и при $n\ll 1$
при произвольных температурах. Разложение знаменателя и
числителя~(\ref{eq:4.97}) по $\langle S^z\rangle $ и $H$ имеет вид
\begin{equation}
G_{\mathbf{q}}(\omega )=\frac{\omega A_{\mathbf{q}\omega }+\langle
S^z\rangle B_{\mathbf{q}\omega }+HC_{\mathbf{q}\omega }}{\omega
-\langle S^z\rangle D_{\mathbf{q}\omega }-HP_{\mathbf{q}\omega }}.
 \label{eq:4.100}
\end{equation}
Здесь величина
\begin{multline}
D_{\mathbf{q}0}
=-\sum_{\mathbf{k}}\frac{8E_{\mathbf{k}}}{(1+c)^2}\left(
E_{\mathbf{k}}\frac{\partial f_{\mathbf{k}}}{\partial
E_{\mathbf{k}}}-E_{\mathbf{k}+\mathbf{q}}\frac{f_{\mathbf{k}-\mathbf{q}}
-f_{\mathbf{k}}} {\omega
+E_{\mathbf{k}-\mathbf{q}}-E_{\mathbf{k}}}\right)\times
\\
\times \left( 1+\frac 2{1+c}\sum_{\mathbf{k}}
\frac{E_{\mathbf{k}-\mathbf{q}}f_{\mathbf{k}-\mathbf{q}}
-E_{\mathbf{k}}f_{\mathbf{k}}} {\omega
+E_{\mathbf{k}-\mathbf{q}}-E_{\mathbf{k}}}\right) ^{-1}
 \label{eq:4.102}
\end{multline}
($f_{\mathbf{k}}=f(\varepsilon_{\mathbf{k}})$)
описывает эффективное обменное взаимодействие вследствие движения
носителей тока. Грубо говоря, последнее отличается от взаимодействия
РККИ заменой
\mbox{$s$---$d(f)$}~обменного параметра на интеграл переноса.
Такая замена характерна для предела узкой зоны.

При $T>T _{\text{C}}$, полагая в~(\ref{eq:4.100}) $\langle
S^z\rangle =\chi H$, получаем уравнение для парамагнитной
восприимчивости:
\begin{equation}
\frac{1-c}2+\frac 1\pi \int\limits_0^\infty \coth{} \frac \omega
{2T}\Im{} \sum_{\mathbf{q}}A_{\mathbf{q}\omega }\,d\omega =
T\sum_{\mathbf{q}}\left( \frac{\chi
B_{\mathbf{q}0}+C_{\mathbf{q}0}}{\chi
D_{\mathbf{q}0}+P_{\mathbf{q}0}}+A_{\mathbf{q}0}\right).
 \label{eq:4.104}
\end{equation}
Температура Кюри определяется условием $\chi (T_{\text{C}})=\infty$.
При малых $n$ имеем
\begin{equation}
T_{\text{C}}=\frac{S_0}2\left( \sum
_{\mathbf{q}}D_{\mathbf{q}0}^{-1}\right) ^{-1}.
 \label{eq:4.105}
\end{equation}
Разлагая правую часть~(\ref{eq:4.104}) по $\chi $ при
$T_{\text{C}}\ll T\ll E_{\text{F}}$, получаем
\begin{equation}
\chi =-\frac 14\sum_{\mathbf{k}}\frac{\partial f(\varepsilon
_{\mathbf{k}})}{\partial \varepsilon _{\mathbf{k}}}+\frac
C{T-\theta },
 \label{eq:4.106}
\end{equation}
где первый член соответствует вкладу Паули, второй~"--- вкладу
Кюри"--~Вейсса от ЛММ, где
\begin{equation}
C=\frac 12S_0,\qquad \theta
=\sum_{\mathbf{q}}D_{\mathbf{q}0}>T_{\text{C}}
 \label{eq:4.107}
\end{equation}
суть константа Кюри и парамагнитная температура Кюри.
Таким образом, "<многоэлектронный"> подход обеспечивает простой
вывод закона Кюри"--~Вейсса для коллективизированных магнетиков.

Еще один из факторов, важных для температурной зависимости
магнитной восприимчивости, -- особенности плотности
состояний \cite{81,II}. Такие особенности могут
быть связаны с наличием вырожденных гибридизованных зон. По мнению
Маттиса \cite{656}, именно в вырождении лежит ключ к объяснению
ферромагнетизма.

Рассмотренные концепции могут быть применены к общей теории
металлического магнетизма. Очевидно, предположение о сильном (по
сравнению с полной шириной зоны) межэлектронном отталкивании в
целом несправедливо для переходных $d$"~металлов. Однако оно может
быть верным для некоторых электронных групп около уровня Ферми.
Эта идея использовалась в~\cite{288} для обсуждения
ферромагнетизма металлов группы железа. Узкозонная модель Хаббарда
применялась для описания группы "<магнитных"> состояний, которые
формируют узкий пик плотности состояний, обусловленный
"<гигантскими"> особенностями ван~Хова \cite{81}.
Корреляции для этих состояний должны быть сильными из-за малой
ширины пиков $\Gamma \simeq 0{,}1$~эВ. Остальные $s$-, $p$-,
$d$"~электроны формируют широкие зоны и слабо гибридизованы с
"<магнитными"> электронными пиками. Состояния на пике ответственны
за формирование ЛММ и другие магнитные свойства в Fe и Ni. Данная
модель позволяет простым образом объяснить низкое (по сравнению с
уровнем Ферми) значение температуры Кюри, которая, как следует из
упомянутых соображений, порядка~$\Gamma $.

В~ферромагнитной фазе расщепление пиков со спином вверх и вниз
$\Delta \simeq 1-2\ \text{эВ}\gg \Gamma $, и структуры обоих пиков
подобны. Так как более низкий пик заполнен, ситуация для
"<магнитных"> электронов оказывается близкой к насыщенному
ферромагнетизму, т.~е. к полуметаллическому состоянию в обычной
модели Хаббарда с большим~$U$.


Предельный случай "<сильного"> ферромагнетизма
с большим значением спинового расщепления представляет собой
противоположность слабым коллективизированным ферромагнетикам.
Еще в~теории Стонера рассматривалось
решение, в котором спиновое расщепление превышает уровень Ферми и
одна спиновая подзона пуста или полностью заполнена. Считалось,
что такая ситуация (для дырочного случая) соответствует
ферромагнитному никелю. Однако современные зонные расчеты в
рамках метода функционала спиновой плотности опровергли
это предположение (плотность состояний на уровне Ферми со спином
вверх оказалось малой, но конечной).

В~то~же время зонные расчеты привели к открытию реальных
магнетиков, которые подобны сильным cтонеровским ферромагнетикам.
Расчеты де~Гроота и~др. зонной структуры для сплава Гейслера
NiMnSb, PtMnSb со структурой C${}_{\text{1b}}$ (MgAgAs)
показывают, что уровень Ферми для одной из проекций спина
находится в энергетической щели. Поскольку эти системы ведут себя
как изоляторы лишь для одного значения~$\sigma $, они были названы
"<полуметаллическими ферромагнетиками">~(ПМФ). Позже подобная
картина была получена для CoMnSb, ферримагнитного FeMnSb,
антиферромагнитного CrMnSb. Вычисления зонной структуры для
большой группы ферро- и антиферромагнитных сплавов Гейслера из
другого ряда T${}_2$MnZ (T${}={}$Co, Ni, Cu, Pb) со структурой
$L2_1$ показывают, что состояние, близкое к ПМФ ($N_{\downarrow}(E_F)$
практически равно нулю), имеет место в системах Co${}_2$MnZ с
Z${}={}$Al, Sn и Z${}={}$Ga, Si, Ge. Кроме того, полуметаллическое
состояние обнаружено в зонных расчетах для CrO${}_2$ (структура
рутила), Fe${}_3$O${}_4$, CrAs, VAs (структура цинковой обманки),
двойных перовскитах типа Sr${}_2$FeMoO$_6$ и др. (см. обзоры
\cite{318,RMP,RMP1}). Удивительным образом концепция сильного
полуметаллического ферромагнетизма оказалась применимой для
некоторых $sp$-систем, включая гипероксиды типа RbO$_2$ \cite{RMP}.

Ситуация резко различных состояний для спина вверх и для спина
вниз, которая реализована в ПМФ, интересна для общей теории
коллективизированного магнетизма~\cite{318}. Схема формирования
"<полуметаллического"> состояния в сплавах Гейслера может быть
описана следующим образом. В~пренебрежении
гибридизацией атомных состояний T и~Z $d$"~зона марганца для
упомянутых структур характеризуется широкой энергетической щелью
между связующими и антисвязующими состояниями. Из-за сильного
внутриатомного (хундовского) обмена для ионов марганца в
ферромагнитном состоянии подзоны со спинами вверх и вниз
значительно раздвинуты. Одна из спиновых подзон близко подходит к
$p$"~зоне лиганда, и поэтому соответствующая щель частично или
полностью размыта $p$---$d$~гибридизацией. Энергетическая щель в
другой подзоне сохраняется и может совпадать при известных
условиях с уровнем Ферми, что дает ПМФ-состояние. Для структуры
C${}_{\text{1b}}$ имеем истинную щель, а для структуры $L2_1$~"---
глубокую псевдощель.

Интерес к полуметаллическим ферромагнетикам (ПМФ) вызван в первую
очередь их уникальными магнитооптическими свойствами~\cite{318},
которые тесно связаны с электронной структурой около уровня Ферми.
В дальнейшем интерес к полуметаллическому ферромагнетизму
значительно увеличился в связи с открытием гигантского
магнитосопротивления в ферромагнитных манганитах, которые,
вероятно, близки к ПМФ~\cite{RMP}.

Кроме того, ПМФ важны в связи с проблемой получения большого
магнитного момента насыщения, поскольку их электронный
спектр благоприятен для максимальной спиновой поляризации
(дальнейшее увеличение спинового расщепления в полуметаллическом
состоянии не~приведет к увеличению магнитного момента).
Электронная структура, напоминающая полуметаллическую (глубокий
минимум плотности состояний на $E_{\text{F}}$ для $\sigma =\downarrow $),
обнаружена в системе сплавов Fe"--~Co~\cite{319} и в системах
R${}_2$Fe${}_{17}$, R${}_2$Fe${}_{14}$B с рекордным значением
$M_0$~\cite{320}. Такой минимум типичен для систем с
выраженными локальными магнитными моментами и возникает также в
чистом железе.

С~теоретической точки зрения ПМФ характеризуются отсутствием
распада спиновых волн на электрон-дырочные пары с
антипараллельными спинами (возбуждения Стонера). Таким образом,
магноны существуют во всей зоне Бриллюэна, как и в
гейзенберговских ферромагнетиках и вырожденных ферромагнитных
полупроводниках. В~отличие от обычных коллективизированных
ферромагнетиков, эффекты электрон-магнонного взаимодействия
не~замаскированы возбуждениями Стонера и могут изучаться в чистой
форме. Особенно интересны эффекты, связанные с некогерентными
(неквазичастичными) состояниями, которые будут подробнее
рассмотрены в следующей главе.

\section{$s-d(f)$-обменная модель и модель Андерсона}

Вторая многоэлектронная модель, связанная с именем С.В.
Вонсовского и сыгравшая огромную роль в физике твердого тела, -- это $s-d$-обменная модель. Сам С.В. обычно называл ее моделью
Шубина-Вонсовского, -- в память о своем учителе и друге, с которым
обсуждались идеи, легшие в основу модели (сохранилась
незаконченная совместная работа "<Общие свойства системы внутренних
и внешних электронов в переходных металлах"> 1936-37 г., которая
впервые была опубликована в книге~\cite{Shubin}). Возможно,
определенную роль в формулировке идей $s$---$d$~модели сыграл и
Л.Д. Ландау.

В отличие от полярной модели и модели Хаббарда, в $s-d$-обменной
модели существование локализованных магнитных моментов не
выводится, а постулируется. При этом подсистемы, ответственные за
магнетизм и электропроводность разделены: коллективизированные
"<$s$"~-электроны"> определяют кинетические свойства, а
локализованные "<$d$"~-электроны"> дают главный вклад в магнитный
момент.

Исходно модель $s$---$d$~обмена была предложена для переходных
$d$"~металлов, в особенности для объяснения их электрического
сопротивления~\cite{1946,265,Turov53}. Поскольку
$d$--электроны (как и 5$f$-электроны в актинидах) в
действительности не описываются локализованной моделью
Гейзенберга, а образуют энергетические зоны, предположение о их
локализации едва~ли может быть обосновано количественно. Тем не
менее, полуфеноменологическая картина двух подсистем, связанных
обменным взаимодействием, очень часто оказывается полезной при
качественном анализе. Как предположил Гудинаф \cite{Goodenough},
$d$-электроны с $e_g$ и $t_{2g}$ симметрией могут обладать
различной степенью локализации: первые проявляют почти
локализованное, а вторые -- коллективизированное поведение. В
дальнейшем такая двухзонная модель неоднократно уточнялась
\cite{288,81,katanin}.

С другой стороны, $4f$-электроны в кристаллах хорошо локализованы
и описываются атомным подходом, а потому  $s$---$f$~обменная
модель обеспечивает количественное описание магнетизма в
большинстве редкоземельных металлов, а также в их соединениях с
хорошо локализованными $f$"~состояниями (впрочем, и здесь бывают
исключения -- в частности, церий, системы с промежуточной
валентностью).

Судьба $s$--$d$~модели оказалась очень интересной и непростой, в
некотором смысле даже драматической. В истории физики она
выступала в разнообразных ипостасях.

Важнейшим успехом теории магнетизм металлов, невозможным без
$s$--$d$~модели, было открытие косвенного обменного взаимодействия
Рудермана"--~Киттеля"--~Касуя"--~Иосида~(РККИ) через электроны
проводимости. Оно возникает во втором порядке теории возмущений по
$s$---$f$~обменному параметру. Таким образом, не только
локализованные моменты воздействуют на движение носителей тока -- возникает и обратное влияние. Вначале (в 1954 г.) этот механизм
был предложен Рудерманом и Киттелем для взаимодействия между
ядерными спинами, а затем применен японскими учеными к
локализованным моментам в металлической матрице. С.В. Вонсовский в
сотрудничестве с учениками (в частности, с А.А. Бердышевым) также
проводил исследования в теории косвенного обмена \cite{Karpenko},
однако вовремя опубликовать свои результаты они не успели. Будучи
дальнодействующим, РККИ-взаимодействие является основным
механизмом обмена между $4f$"~оболочками в редких землях и их
проводящих соединениях.

В последнее время в зарубежной физической литературе
$s$--$d$~модель часто именуется моделью Кондо. Такой исторический
"<перекос"> связан с исключительной важностью эффекта Кондо
\cite{552} -- пожалуй, самого красивого явления в физике твердого
тела. Здесь следует отметить, что сам Кондо в своем пионерском
исследовании аномалий магнитного $s$---$d(f)$~рассеяния исходил из
гамильтониана в представлении вторичного квантования, который был
предложен Вонсовским и Туровым \cite{Turov53}. Неоправданно широко
применяется и термин "<решетка Кондо"> (он используется даже для
обычных ферро- и антиферромагнитых металлических соединений). В
действительности область применения $s$--$d$~модели выходит далеко
за пределы систем, где реально наблюдается "<кондовское"> подавление
магнитных моментов.

Удивительно плодотворной $s$--$d$~модель оказалась для магнитных
полупроводников; здесь особенно значительна заслуга Э.Л. Нагаева,
систематически развившего их теорию с использованием
квазиклассического разложения  \cite{352}. В дальнейшем близкие
представления были использованы для теоретического описания
манганитов с гигантским магнитосопротивлением на основе LaMnO$_3$
\cite{nagaev}.

В западных работах $s$--$d$~модель нередко называют моделью
Вонсовского-Зинера -- в честь американского ученого К. Зинера,
предложившего в 1951 г. механизм двойного обменного взаимодействия
в узких зонах (в пределе сильной связи между локальными моментами
и носителями тока). В последнее время этот механизм широко
обсуждается в связи с манганитами \cite{skryabin}.

Отметим также, что t-J модель, широко применяемая для описания
носителей тока в меднооксидных сверхпроводниках, эквивалентна
$s$--$d$~модели с большим по модулю отрицательным $s$--$d$~обменом
и спином $S=1/2$ (см. раздел 3.2).


Гамильтониан $s$---$d(f)$~модели в простейшей форме имеет вид
\begin{equation}
\mathcal{H}= \sum_{\mathbf{k}\sigma
}t_{\mathbf{k}}c_{\mathbf{k}\sigma }^{\dagger }c_{\mathbf{k}\sigma
}+\sum_{\mathbf{q}}J_{\mathbf{q}}\mathbf{S}_{-\mathbf{q}}
\mathbf{S}_{\mathbf{q}}-I\sum_{i\sigma \sigma ^{\prime
}}(\mathbf{S}_i\mbox{\boldmath$\sigma $}_{\sigma \sigma ^{\prime
}})c_{i\sigma }^{\dagger }c_{i\sigma ^{\prime }}, \label{eq:G.2}
\end{equation}
где  {\boldmath$\sigma $}~"--- матрицы Паули, $I$~"--- параметр
$s$---$d(f)$~обменного взаимодействия, которое предполагается
контактным (вывод $s$---$d(f)$~модели в более общем случае
рассматривается ниже). Часто (напр., в~редкоземельных металлах)
взаимодействие между локализованными спинами $J_{\mathbf{q}}$
является косвенным РККИ-обменом через электроны проводимости,
причиной которого служит то~же самое $s$---$f$~взаимодействие.
Однако при построении теории возмущений удобно включить его в
нулевой гамильтониан.

В отличие от внутриатомного кулоновского (хаббардовского)
взаимодействия, $s-d(f)$ взаимодействие как правило не является
сильным. Тем не менее, оно приводит к существенным эффектам в
электронном спектре. С микроскопической точки зрения оно может
иметь различную природу. В ряде редкоземельных систем (например, в
магнитных полупроводниках) это внутриатомный хундовский обмен,
который ферромагнитен. Часто обменное взаимодействие является не
настоящим, а эффективным -- обусловленным гибиридизацией между
 $s$---зонами и атомными уровнями $d(f)$-электронов; в этом
случае он антиферромагнитен. Впрочем, как мы увидим ниже, знак
$s$---$d(f)$~обмена не столь существен, как знак взаимодействия
между локализованными моментами в модели Гейзенберга: грубо
говоря, теория возмущений начинается с квадрата
$s$---$d(f)$~обменного параметра. Однако такой важное физическое
явление, как эффект Кондо, возникает только в третьем порядке. При
этом в случае отрицательного (антиферромагнитного) обмена
эффективное (перенормированное) обменное взаимодействие становится
бесконечным, так что магнитное рассеяние приводит к полному
экранированию магнитных моментов \cite{556,558}.

Имея гамильтониан более сложной формы, $s$---$d$~модель
оказывается в некотором отношении более простой, чем модель
Хаббарда, поскольку в ней возможно выполнить квазиклассическое
разложение по малому параметру~$1/(2S)$. Обе модели позволяют
получить полуфеноменологическое описание электрон-магнонного
взаимодействия в ферро- и антиферромагнитных металлах,
удовлетворяющее требованиям симметрии. Как продемонстрировано в
работах \cite{338,693}, при построении теории возмущений по
электрон-магнонному рассеянию результаты в обеих моделях
отличаются, как правило, только заменой $I\rightarrow U$.

Идеи $s$---$d(f)$~модели получили дальнейшее развитие в
современных теоретико-полевых подходах. Например, благодаря
сильному ближнему порядку в двумерном случае возникает
динамическое разделение коллективизированных и магнитных степеней
свободы. Для описания взаимодействия электронов проводимости со
спиновыми флуктуациями парамагнонного типа в коллективизированных
магнетиках, особенно вблизи квантового фазового перехода в
магнитное состояние, была предложена полуфеноменологическая
спин-фермионная модель \cite{spin-ferm1} с действием
\begin{equation}
\mathcal{S}[c,\mathbf{S}] =\sum_{k}(i\nu _{n}-\varepsilon _{\mathbf{k}})c_{k\sigma }^{\dagger
}c_{k\sigma }-\sum_{q}\mathcal{R}_{q}\mathbf{S}_{q}\mathbf{S}_{-q}+I\sum_{kk^{\prime
}\sigma \sigma ^{\prime }}\mathbf{S}_{k-k^{\prime }}\mbox {\boldmath $\sigma
$}_{\sigma \sigma ^{\prime }}c_{k\sigma }^{\dagger }c_{k^{\prime }\sigma
^{\prime }}  \label{sf}
\end{equation}
где поля $c$ и $\mathbf{S}$ соответствуют электронным и спиновым степеням свободы, член с $I$ описывает взаимодействие между ними, $q=(\mathbf{q},i\omega _{n})$, $k=(\mathbf{k}, \nu _{n})$ ($\omega _{n}=2n\pi T$ и $\nu _{n}=(2n+1)\pi T$ -- бозонные и фермионные мацубаровские частоты),  $\mathcal{R}_{q}=\chi_q^{-1}$, $\chi_q$ -- затравочная восприимчивость спиновой подсистемы.

В статическом случае $\mathcal{R}_{q}=\delta
_{n0}\mathcal{R}_{\mathbf{q}}$ мы получаем из (\ref{sf})
гамильтониан, формально совпадающий с гамильтонианом
(\ref{eq:G.2}). Однако в спин-фермионной модели квадрат полного
спина на узле не фиксирован (как и в сферической модели для
локализованных спинов, обобщающей модель Гейзенберга). Сравнение
электронного спектра двумерного парамагнетика с соответствующими
результатами в классической $s$---$d$ модели с сильным корреляциям
проведено в работе \cite{spin-ferm}.

В случае $s$---$d(f)$~обмена гибридизационной природы
$s-d(f)$-модель тесно связана с моделью Андерсона. В~пренебрежении
орбитальным вырождением ее гамильтониан записывается как
\begin{equation}
\mathcal{H}=\sum_{\mathbf{k}\sigma }\left[
t_{\mathbf{k}}c_{\mathbf{k}\sigma }^{\dagger }c_{\mathbf{k}\sigma
}+\Delta f_{\mathbf{k}\sigma }^{\dagger }f_{\mathbf{k}\sigma
}+V(c_{\mathbf{k}\sigma }^{\dagger }f_{\mathbf{k}\sigma
}+f_{\mathbf{k}\sigma }^{\dagger }c_{\mathbf{k}\sigma })\right]
+U\sum_if_{i\uparrow }^{\dagger }f_{i\uparrow }f_{i\downarrow
}^{\dagger }f_{i\downarrow }.
 \label{eq:6.54}
\end{equation}
где $V$~"--- матричный элемент гибридизации; $\Delta $~"---
положение $d$"~уровня, отсчитываемого от $E_{\text{F}}$; $U$~"---
внутриузельное кулоновское взаимодействие. Исходно эта модель (в
случае одной $d$-примеси) была предложена Андерсоном, чтобы
исследовать проблему формирования локального момента при
гибридизации атомного $d$-уровня с зоной проводимости. В пределе
больших $U$ член с кулоновским взаимодействием может быть опущен,
но при этом одноэлектронные операторы для $d(f)$-состояний
заменяются проекционными $X$-операторами.

Периодическая модель Андерсона описывает ситуацию, когда
сильнокоррелированные $d(f)$"~-электроны не~участвуют
непосредственно в зонном движении, но гибридизуются с состояниями
зоны проводимости. Такое положение имеет место для ряда
редкоземельных и актинидных соединений.
Гибридизационная (многоконфигурационная) картина часто полезна и
для обсуждения электронных свойств переходных $d$"~металлов и других
$d$"~-электронных систем. Например, сильная $p$---$d$~гибридизация
имеет место в медь-кислородных высокотемпературных
сверхпроводниках (ср. (\ref{eq:6.108})).

$s-d(f)$-обменная модель является пределом модели Андерсона в
случае, когда этот уровень лежит глубоко под уровнем Ферми
\cite{584}; формально она получается из (\ref{eq:6.54})
каноническим преобразованием, устраняющим член гибридизации, так
что
\begin{equation}
I=V^2\left( \frac 1\Delta -\frac 1{\Delta +U}\right),
 \label{eq:6.9}
\end{equation}

Случай промежуточной валентности соответствует ситуации, когда,
наоборот, ширина $f$"~пика, обусловленная гибридизацией, $\Gamma
=\pi V^2\rho $ ($\rho $ -- плотность состояний электронов
проводимости на уровне Ферми), мала по сравнению с расстоянием
$|\Delta |=|\varepsilon _{\text{f}}-E_{\text{F}}|$. Ряд
редкоземельных элементов (Ce, Sm, Eu, Tm, Yb и, возможно, Pr)
не~обладают устойчивой валентностью, но меняют ее в различных
соединениях \cite{512,584,545}. В~некоторых соединениях данные элементы могут
находиться в так называемом смешанном валентном состоянии, где на
атом приходится нецелое число $f$"~-электронов. Такая ситуация
возникает, если конфигурации $4f^n(5d6s)^m$ и
$4f^{n-1}(5d6s)^{m+1}$ почти вырождены, так что сильны
межконфигурационные флуктуации. В~металлических системах это
соответствует $f$"~уровню, расположенному около уровня Ферми,
причем $f$"~состояния гибридизованы с состояниями зоны
проводимости.

Состояние промежуточной валентности (ПВ) характеризуется наличием
одной линии в мессбауэровских экспериментах (масштаб времени
измерений~"--- около $10^{-11}$~с), которая имеет промежуточное
положение. Вместе с тем в рентгеновских экспериментах (время~"---
около $10^{-16}$~с) наблюдаются две линии, которые соответствуют
конфигурациям $f^n$ и $f^{n-1}$. Специфической особенностью
перехода в состояние ПВ является также изменение решеточного
параметра к значению, промежуточному между соответствующими
значениями целочисленных валентных состояний. Такие превращения
(например под давлением), как правило, резкие (переходы первого
рода). Помимо этого, ПВ-соединения обладают при низких
температурах значительно увеличенной электронной теплоемкостью и
магнитной восприимчивостью. При высоких температурах
величина~$\chi (T)$ подчиняется закону Кюри"--~Вейсса с
эффективным моментом, промежуточным между значениями для
соответствующих атомных конфигураций.

Состояние решетки Кондо (или с тяжелыми фермионами)
может рассматриваться как предел состояния ПВ
с почти целой валентностью (ее изменение не~превышает нескольких
процентов). В~некотором смысле, ПВ"~системы могут рассматриваться
как решетки Кондо с большими значениями $T_{\text{K}}$~\cite{545},
причем, в отличие от состояния кондо-решетки, в состоянии ПВ
играют важную роль не~только спиновые, но и зарядовые флуктуации.

Рассмотренные модели широко применяются для чистых редкоземельных
металлов и актинидов, а также для объяснений свойств ряда
экзотических соединений -- систем с промежуточной валентностью,
тяжелыми фермионами, "решеток Кондо".

Чтобы описать образование синглетного кондо-состояния в области
сильной связи, можно использовать простой гамильтониан SU$(N)$"~решетки
Андерсона:
\begin{equation}
\mathcal{H}=\sum_{\mathbf{k}m}t_{\mathbf{k}}c_{\mathbf{k}m}^{\dagger
}c_{\mathbf{k}m}+\Delta \sum_{im}X_i(mm)+
V\sum_{\mathbf{k}m}[c_{\mathbf{k}m}^{\dagger
}X_{\mathbf{k}}(0m)+X_{-\mathbf{k}}(m0)c_{\mathbf{k}m}]
 \label{eq:6.38}
\end{equation}
($m=1,\ \ldots,\ N$). Эта модель удобна при описании
межконфигурационных переходов $f^0$---$f^1$ (церий, $J=5/2$) или
$f^{14}$---$f^{13}$ (иттербий, $J=7/2$) и часто применяется в
рамках разложения по $1/N$. Более общая и реалистичная модель
$s$---$f$~гибридизации с включением двух (вообще говоря,
магнитных) конфигураций обсуждается ниже в разделе 3.5.

Модель~(\ref{eq:6.38}) может быть сведена каноническим
преобразованием, исключающим гибридизацию, к так называемой модели
Коблина"--~Шриффера:
\begin{equation}
\mathcal{H}_{\text{CS}}=\sum_{\mathbf{k}m}
t_{\mathbf{k}}c_{\mathbf{k}m}^{\dagger
}c_{\mathbf{k}m}^{}-I\sum_{imm^{\prime }}X_i(mm^{\prime
})c_{i m^{\prime }}^{\dagger }c_{im}^{},
 \label{eq:6.39}
\end{equation}
причем $I=V^2/\Delta $.
Чтобы избежать трудностей вследствие сложных соотношений коммутации для
$X$"~операторов, используют представление~\cite{581}
\begin{equation}
X_i(m0)=f_{im}^{\dagger }b_i^{\dagger },\qquad X_i(m^{\prime
}m)=f_{im^{\prime }}^{\dagger }f_{im}^{},\qquad X_i(00)=b_i^{\dagger
}b_i^{},
 \label{eq:6.40}
\end{equation}
где $f^{\dagger }$~"--- фермиевские операторы, $b^{\dagger }$~"---
вспомогательные бозе-операторы. Это представление удовлетворяет
необходимым коммутационным соотношениям Х-операторов. В~то~же
время, согласно~(\ref{eq:A.22}), нужно требовать выполнения
вспомогательного условия
\begin{equation}
\sum_m X_i(mm)+X_i(00)=\sum_m f_{im}^{\dagger }f_{im}^{}+b_i^{\dagger
}b_i^{}=1.
 \label{eq:6.41}
\end{equation}
Тогда параметр $\langle b_i\rangle $ перенормирует матричные
элементы гибридизации. Помимо применения к реальным системам, такие модели позволяют строить разложение по формальному малому параметру $1/N$.

Для описания зарядовых флуктуаций в системах с промежуточной
валентностью иногда используется так называемая модель
Фаликова-Кимбалла, в которой вводится кулоновское взаимодействие
между локализованными и коллективизированными электронами.
Гамильтониан бесспиновой модели
Фаликова"--~Кимбалла с введением гибридизации имеет вид
\begin{equation}
\mathcal{H}=\sum_{\mathbf{k}}\left[
t_{\mathbf{k}}c_{\mathbf{k}}^{\dagger }c_{\mathbf{k}}+\Delta
f_{\mathbf{k}}^{\dagger }f_{\mathbf{k}}+V(c_{\mathbf{k}}^{\dagger
}f_{\mathbf{k}}+f_{\mathbf{k}}^{\dagger }c_{\mathbf{k}})\right]
+G\sum_if_i^{\dagger }f_ic_i^{\dagger }c_i,
 \label{eq:6.53}
\end{equation}
где $G$~"--- внутриузельный кулоновский $s(d)$---$f$ интеграл; для простоты
пренебрегаем зависимостью гибридизации от~$\mathbf{k}$. Такой
гамильтониан позволяет легко учесть сильное $f$---$f$~отталкивание
на узле (в бесспиновой модели дважды занятые состояния запрещаются
принципом Паули) и удобен при описании валентных фазовых
переходов, причем взаимодействие $G$ важно для многоэлектронных
экситонных эффектов. Эти эффекты могут приводить к существенной
температурной зависимости электронного спектра, в частности,
гибридизационной щели \cite{590}.

Модель Фаликова"--~Кимбалла может быть обобщена включением
кулоновского взаимодействия на различных узлах, которое позволяет
описывать зарядовое упорядочение. При этом однопримесная модель
Фаликова"--~Кимбалла с гибридизацией эквивалентна проблеме
Кондо~\cite{590a}.

\subsection{Электронные состояния в $s-d$ обменной модели}

Спектр состояний электронов проводимости может существенно
меняться благодаря взаимодействию с локализованными моментами,
даже если оно не слишком велико. Рассмотрим одноэлектронную
функцию Грина
\begin{equation}
G_{\mathbf{k}\sigma }(E)=\langle \!\langle c_{\mathbf{k}\sigma }|c_{\mathbf{k}\sigma }^{\dagger }\rangle \!\rangle _E=\left[ E-t_{\mathbf{k}\sigma }-\Sigma _{\mathbf{k}\sigma }(E)\right] ^{-1}, \, t_{\mathbf{k}\sigma} =t_{\mathbf{k}} -\sigma I\langle S^z\rangle
\label{eq:G.30}
\end{equation}
В парамагнитном случае запишем цепочку уравнений движения
\begin{equation}
(E-t_{\mathbf{k}})G_{\mathbf{k\sigma }}^{{}}(E)=1-I\sum_{\mathbf{p}}\Gamma _{%
\mathbf{kp}}^{\mathbf{\sigma }}(E)
\label{Nag1}
\end{equation}%
\begin{equation*}
\Gamma _{\mathbf{kp}}^{\mathbf{\sigma }}(E)=\sum_{\mathbf{\sigma }^{\prime
}}\langle \langle (\mathbf{S}_{\mathbf{p}}^{{}}\sigma _{\mathbf{\sigma
\sigma }^{\prime }})c_{\mathbf{k}-\mathbf{p\sigma }^{\prime }}|c_{\mathbf{%
k\sigma }}^{\dagger }\rangle \rangle _{E}
\end{equation*}%
\begin{equation}
\ (E-t_{\mathbf{k-p}})\Gamma _{\mathbf{kp}}^{\mathbf{\sigma }}(E)=-I(\langle
\mathbf{S}_{\mathbf{p}}^{{}}\mathbf{S}_{-\mathbf{p}}^{{}}\rangle -2m_{%
\mathbf{k-p}}\rangle )G_{\mathbf{k}\sigma }^{{}}(E)-I(1-2n_{\mathbf{k-p}%
})\sum_{\mathbf{q}}\Gamma _{\mathbf{kq}}^{\sigma }(E)
\label{Nag2}
\end{equation}%
Здесь выполнено расцепление типа Нагаока, которое ранее
использовалось для исследования эффекта Кондо в однопримесной
модели (см. \cite{552}), $n_{\mathbf{k}}=\langle
c_{\mathbf{k\sigma }}^{\dagger }c_{\mathbf{k\sigma }}\rangle$,
\begin{equation}
m_{\mathbf{k}}=\sum_{\mathbf{q\sigma }^{\prime }}\langle (\mathbf{S}_{%
\mathbf{q}}^{{}}\sigma _{\mathbf{\sigma \sigma }^{\prime }})c_{\mathbf{%
k\sigma }}^{\dagger }c_{\mathbf{k-q\sigma }^{\prime }}\rangle =-\frac{1}{\pi
}\Im\int dEf(E)\sum_{\mathbf{q}}\Gamma _{\mathbf{kq}}^{\mathbf{\sigma }}(E)
\end{equation}%
Выражая интегральный член в уравнении (\ref{Nag2}) из уравнения
(\ref{Nag1}), мы можем формально решить эту систему уравнений и
получить для собственной энергии
\begin{equation}
\Sigma _{\mathbf{k}}(E)=\frac{2I^{2}P_{\mathbf{k}}(E)}{1-IR(E)}
\label{Nag3}
\end{equation}%
где
\begin{equation}
P_{\mathbf{k}}(E)=\sum_{\mathbf{q}}\frac{\langle \mathbf{S}_{-\mathbf{q}%
}^{{}}\mathbf{S}_{\mathbf{q}}^{{}}\rangle -2m_{\mathbf{k-q}}}{E-t_{\mathbf{%
k-q}}},~R(E)=\sum_{\mathbf{k}}\frac{1-2n_{\mathbf{k}}}{E-t_{\mathbf{k}}}
\end{equation}%
\begin{equation}
\end{equation}%
Как следует из (\ref{Nag3}), в третьем порядке по $s-d$ обмену в мнимой части
собственной энергии возникает кондовский (логарифмический по энергии) вклад от интеграла с фермиевскими функциями $n_{\mathbf{k}}$; в
то же время вещественная часть сингулярного вклада компенсируется
членами с корреляторами $m_{\mathbf{k}}$.


В случае ферромагнетика во втором порядке теории возмущений для собственно-энергетической части получаем
\begin{equation}
\Sigma _{\mathbf{k}\uparrow }(E)=2I^2S
\sum_{\mathbf{q}}\frac{N_{\mathbf{q}}+n_{\mathbf{k}+\mathbf{q}\downarrow
}}{E-t_{\mathbf{k}+\mathbf{q}\downarrow }+\omega _{\mathbf{q}}}, \,
\Sigma _{\mathbf{k}\downarrow }(E)= 2I^2S \sum_{\mathbf{q}}
\frac{1+N_{\mathbf{q}}-n_{\mathbf{k}-\mathbf{q}\uparrow
}}{E-t_{\mathbf{k}-\mathbf{q}\uparrow }-\omega _{\mathbf{q}}},
 \label{eq:G.34}
\end{equation}
где $N_{\mathbf{q}}$ -- бозевские функции распределения магнонов.
В силу вращательной симметрии электрон-магнонного взаимодействия
его амплитуда обращается в нуль при $q \rightarrow 0$, так что
поправки к спектру пропорциональны $T^{5/2}$ (поправки порядка
$T^{3/2}$ от поперечных и продольных вкладов взаимно
компенсируются).

Приведенные выражения позволяют рассмотреть картину плотности состояний с учетом корреляционных
эффектов~\cite{329,orb,RMP,RMP1}. Записывая разложение уравнения Дайсона (\ref{eq:G.30}), получаем
\begin{multline}
N_\sigma (E) =-\frac 1\pi \Im{}
\sum_{\mathbf{k}}G_{\mathbf{k}\sigma }(E)=
\\
=\sum_{\mathbf{k}}\delta (E-t_{\mathbf{k}\sigma
})-\sum_{\mathbf{k}}\delta ^{\prime }(E-t_{\mathbf{k}\sigma
})\Re{} \Sigma _{\mathbf{k}\sigma }(E)-\frac 1\pi
\sum_{\mathbf{k}}\frac{\Im{} \Sigma _{\mathbf{k}\sigma
}(E)}{(E-t_{\mathbf{k}\sigma })^2}.
 \label{eq:4.87}
\end{multline}
Обсудим подробнее случай сильного полуметаллического
ферромагнетика (ПМФ), где  расщепление спиновых подзон превышает энергию Ферми,
так что заполнена только одна из них.
Второй член в правой части~(\ref{eq:4.87})
описывает перенормировку энергии квазичастиц. Третий член, который
возникает из-за разреза собственной энергии $\Sigma
_{\mathbf{k}\sigma }(E)$, описывает некогерентный
(неквазичастичный) вклад вследствие рассеяния магнонов. Видно, что
он не~обращается в нуль в энергетической области, соответствующей
"<чужой"> спиновой подзоне с противоположной проекцией~$-\sigma $.
$T^{3/2}$"~зависимость магнонного вклада в
вычет функции Грина, т.~е. эффективной массы в нижней спиновой
подзоне, и увеличение с температурой хвоста верхней подзоны
приводят к сильным температурным зависимостям парциальных
значений~$N_{\sigma }(E)$ противоположного знака. Соответствующее
поведение спиновой поляризации электронов проводимости $P(T)\simeq
\langle S^z\rangle $  подтверждается экспериментальными данными по
полевой эмиссии из ферромагнитных полупроводников и кинетическим
свойствам полуметаллических сплавов Гейслера~\cite{318,RMP}.

\begin{figure}[tbp]
\includegraphics[width=0.65\columnwidth]{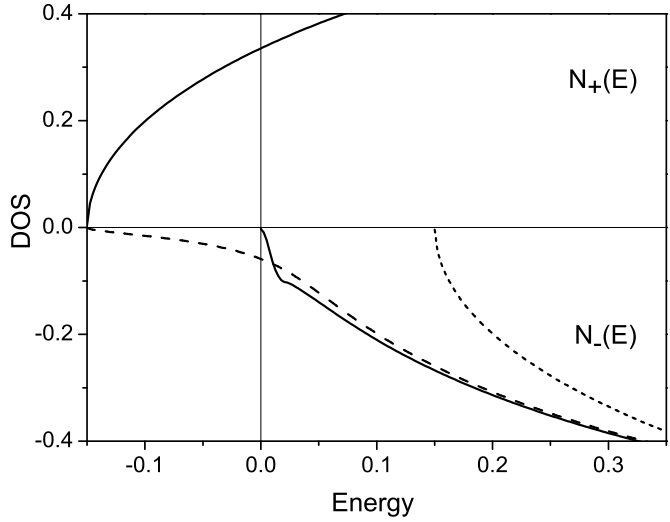}
\caption{Плотность состояний полуметаллического ферромагнетика
с~$I=0.3>0$ для затравочной полуэллиптической зоны с шириной
$W=2$. Неквазичастичные состояния с~$\sigma =\downarrow $ (нижняя
половина рисунка) отсутствуют ниже уровня Ферми. В случае пустой
зоны (пунктир) спин-поляронный хвост состояний со спином вниз
достигает дна полосы; короткий пунктир -- приближение среднего
поля. } \label{fig:1}
\end{figure}
\begin{figure}[tbp]
\includegraphics[width=0.65\columnwidth]{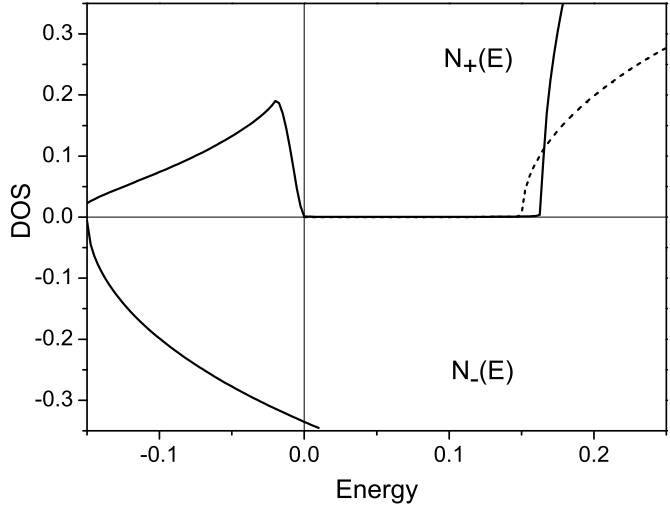}
\caption{Плотность состояний полуметаллического ферромагнетика
с~$I=-0.3<0$ (остальные параметры как на рис.~\ref{fig:1}).
Неквазичастичные состояния с~$\sigma =\uparrow $
возникают ниже уровня Ферми}
\label{fig:2}
\end{figure}

При нулевой температуре картина $N(E)$ около уровня Ферми в ПМФ
(или вырожденных полупроводниках) оказывается также
нетривиальной. Если пренебречь магнонными
частотами в знаменателях~(\ref{eq:G.34}), то парциальная плотность
некогерентных состояний должна появляться скачком выше или ниже
уровня Ферми для $I>0$ и $I<0$ соответственно из-за наличия
функций распределения Ферми. Учет конечности магнонных частот
$\omega _{\mathbf{q}}$  ведет к размытию этих особенностей на
энергетическом интервале $\omega _{\text{max}}\ll E_{\text{F}}$
(рис.~\ref{fig:1}, \ref{fig:2}), причем величина $N_{-\sigma}(E_{\text{F}})$
оказывается равна нулю.

Случай ПМФ, где заполнена только одна спиновая подзона, может быть
исследован более строго. В спин-волновой области здесь возможно
последовательное рассмотрение путем разложения по числам
заполнения электронов и магнонов. С этой целью могут быть
использованы методы уравнений движения для функций
Грина~\cite{329},  производящего функционала \cite{329a} и
разложение оператора эволюции ~\cite{329b}. При этом в каждом
порядке возникают интегральные уравнения, описывающие
электрон-магнонное рассеяния (по структуре они похожи на
интегральные уравнения типа Нагаока ~\cite{349}).

Здесь мы ограничимся простой иллюстрацией. Составляя аналогично
(\ref{Nag2}) интегральные уравнения для функций Грина
$\Gamma
_{\mathbf{kp}}^{\mathbf{\sigma }}(E)=\langle \langle
b_{\mathbf{p} }^{\sigma }c_{\mathbf{k}-\mathbf{p-\sigma
}}|c_{\mathbf{k\sigma }}^{\dagger }\rangle \rangle _{E}$, находим
при $I>0$
\begin{equation}
G_{\mathbf{k\uparrow }}^{{}}(E)=G_{\mathbf{k\uparrow }}^{0}(E)=(E-t_{\mathbf{%
k}}+IS)^{-1}  \label{up}
\end{equation}
\begin{eqnarray}
G_{\mathbf{k\downarrow }}^{{}}(E) &=&\left( \ E-t_{\mathbf{k}}+IS-\frac{2IS}{%
1-IR_{\mathbf{k\uparrow }}(E)}\right) ^{-1},  \label{down} \\
R_{\mathbf{k\uparrow }}(E) &=&\sum_{\mathbf{q}}(1-n_{\mathbf{k-q\uparrow }%
})G_{\mathbf{k-q\uparrow }}(E-\omega _{\mathbf{q}})  \nonumber
\end{eqnarray}
Таким образом, электроны со спином вверх движутся свободно, а
состояния со спином вниз имеют некогерентный характер (функция
Грина имеет разрез, но не имеет полюсов ниже уровня Ферми).

Для $I<0$ результат для спина вниз совпадает с (\ref{down}) (при
$n_{\mathbf{k} \uparrow }=0$). Эта функция Грина дает точное
решение задачи о взаимодействии электрона со спиновой волной и
имеет спин-поляронный полюс $E_{\bf k}^*$, причем в пределе
$I\rightarrow -\infty$ имеем
\[
G_{\mathbf{k\downarrow }}^{-1}(E)=\frac{2S+1}{2S}\varepsilon -t_{\mathbf{k}}, \, \varepsilon = E-I(S+1)
\]
Напротив, функция Грина со спином вверх имеет неполюсную структуру:
\begin{eqnarray}
G_{\mathbf{k\uparrow }}^{{}}(E) &=&\left( \ E-t_{\mathbf{k}}-IS+\frac{2IS}{%
1+IR_{\mathbf{k\downarrow }}(E)}\right) ^{-1},~  \label{up1} \\
R_{\mathbf{k\downarrow }}(E) &=&\sum_{\mathbf{q}}n_{\mathbf{k-q\downarrow }}G_{%
\mathbf{k-q\downarrow }}(E+\omega _{\mathbf{q}})  \nonumber
\end{eqnarray}
В координатном представлении эти  выражения для функций Грина могут быть обобщены
на случай беспорядка  \cite{RMP} и  использованы для описания
пространственно неоднородных систем.

В пределе $I\rightarrow +\infty$ результат (\ref{down}) дает правильный предельный переход
\begin{equation}
G_{\mathbf{k}\downarrow }(E)=\frac{1}{\varepsilon -t_{\mathbf{k}}+2S/R_{\mathbf{k\uparrow }}(E)},~\varepsilon =E+IS
\end{equation}
С другой стороны, выражение (\ref{up1})
в пределе $I\rightarrow -\infty$ дает качественно правильную неполюсную структуру, но все же не обеспечивает согласия с атомным пределом, поскольку при расцеплениях некорректно учитывается сильное внутриатомное $s-d$-взаимодействие.
Таким образом, при больших $|I|$, как и в модели Хаббарда, необходим переход к атомному представлению (см. следующий раздел). Вычисление с его помощью дает вблизи нижнего края зоны (ср. \cite{orb})
\begin{equation}
G_{\mathbf{k}\uparrow }(E)=\frac{2S/(2S+1)}{E-E_{\mathbf{k}}^{\ast
}+(2S-n)/R_{\mathbf{k}\downarrow }^{\ast }(E)}
,~R_{\mathbf{k}\downarrow }^{\ast
}(E)=\sum_{\mathbf{q}}\frac{n_{\mathbf{k-q\uparrow }}^{\ast }}{E-E_{\mathbf{%
k-q}}^{\ast }+\omega _{\mathbf{q}}}  \label{l15}
\end{equation}


Для рассмотрения электронного и магнонного спектров металлического
антиферромагнетика в $s$---$d(f)$~обменной модели перейдем к
локальной системе координат
\[
S_i^x \rightarrow S_i^z\cos{} \mathbf{Q}\mathbf{R}_i-S_i^y\sin{}
\mathbf{Q}\mathbf{R}_i,
\]
\begin{equation}
S_i^y\rightarrow S_i^z\sin{} \mathbf{Q}\mathbf{R}_i+S_i^x\cos{}
\mathbf{Q}\mathbf{R}_i,\qquad S_i^z\rightarrow -S_i^x.
 \label{eq:E.8}
\end{equation}
Тогда
гамильтониан $s$---$d(f)$~обменного взаимодействия примет вид
\begin{multline}
\mathcal{H}_{sd} =-I\sum_{\mathbf{k}\mathbf{q}}
[S_{\mathbf{q}}^x(c_{\mathbf{k}+\mathbf{q}\downarrow }^{\dagger
}c_{\mathbf{k}\downarrow }-c_{\mathbf{k}+\mathbf{q}\uparrow
}^{\dagger }c_{\mathbf{k}\uparrow })
+iS_{\mathbf{q}}^y(c_{\mathbf{k}-\mathbf{Q}\downarrow }^{\dagger
}c_{\mathbf{k}-\mathbf{Q}\uparrow
}-c_{\mathbf{k}+\mathbf{q}\uparrow }^{\dagger
}c_{\mathbf{k}-\mathbf{Q}\downarrow })+
\\
+S_{\mathbf{q}}^z(c_{\mathbf{k}+\mathbf{q}\uparrow }^{\dagger
}c_{\mathbf{k}-\mathbf{Q}\downarrow
}+c_{\mathbf{k}-\mathbf{Q}\downarrow }^{\dagger
}c_{\mathbf{k}-\mathbf{q}\uparrow })].
 \label{eq:G.68}
\end{multline}
В~приближении среднего поля электронный спектр содержит две
расщепленные антиферромагнитные подзоны, которые определяется
выражением (\ref{eq:G.98}) с заменой $U  \rightarrow I$. Переходя
в локальной системе координат к магнонному представлению и
вычисляя электронную собственную энергию во втором порядке по $I$,
получаем
\begin{multline}
\Sigma _{\mathbf{k}}(E)
=\frac{I^2\overline{S}^2}{E-t_{\mathbf{k}-\mathbf{Q}}}
+\frac 12I^2S\sum_{\mathbf{q}}\left\{
(u_{\mathbf{q}}-v_{\mathbf{q}})^2\left[
\frac{1-n_{\mathbf{k}-\mathbf{q}}+N_{\mathbf{q}}}
{E-t_{\mathbf{k}-\mathbf{q}}-\omega
_{\mathbf{q}}}+\frac{n_{\mathbf{k}-\mathbf{q}}+N_{\mathbf{q}}}
{E-t_{\mathbf{k}-\mathbf{q}}+\omega _{\mathbf{q}}}\right] +\right.
\\
\left.+(u_{\mathbf{q}}+v_{\mathbf{q}})^2\left[
\frac{1-n_{\mathbf{k}+\mathbf{q}-\mathbf{Q}}+N_{\mathbf{q}}}
{E-t_{\mathbf{k}+\mathbf{q}-\mathbf{Q}}-\omega
_{\mathbf{q}}}+\frac{n_{\mathbf{k}+\mathbf{q}-\mathbf{Q}}
+N_{\mathbf{q}}}{E-t_{\mathbf{k}+\mathbf{q}-\mathbf{Q}}+\omega
_{\mathbf{q}}}\right] \right\},
 \label{eq:G.69}
\end{multline}
где $\overline{S}$~"--- намагничeнность подрешетки, $u_{\mathbf{q}}$,
$u_{\mathbf{q}}$ --- коэффициенты преобразования Боголюбова.
Вычисление дает $T^2$"~зависимость электронного спектра;
она является следствием линейной дисперсии спектра спиновой волны и
зависимости амплитуды электрон-магнонного взаимодействия вида
$q^{-1}$, которые специфичны для антиферромагнетиков. Поправки к
энергии дна зоны ($t_{\mathbf{k}}=t_{\text{min}}$) из-за
намагниченности подрешетки  и поперечных флуктуаций имеют
противоположные знаки. Вклад от флуктуаций преобладает, что
приводит к "<синему"> сдвигу дна зоны проводимости с уменьшением
температуры, который наблюдается в антиферромагнитных
полупроводниках~\cite{352}, в отличие от "<красного"> сдвига в
ферромагнитных полупроводниках.

Запишем также многоэлектронный вклад третьего порядка в
собственную энергию, который описывает перенормировку
антиферромагнитной щели из-за подобных расходимостей кондовского
типа~\cite{367}:
\begin{equation}
\delta \Sigma _{\mathbf{k}}^{(3)}(E)
=2I^3S^2\sum_{\mathbf{q}}\frac{n_{\mathbf{k}+\mathbf{q}}
(E-t_{\mathbf{k}+\mathbf{q}})}{(E-t_{\mathbf{k}+\mathbf{q}})^2-\omega _{\mathbf{q}}^2}
\left( \frac
1{t_{\mathbf{k}+\mathbf{q}}-t_{\mathbf{k}-\mathbf{Q}+\mathbf{q}}}-\frac
1{E-t_{\mathbf{k}+\mathbf{Q}}}\right).
 \label{eq:G.73}
\end{equation}

Отметим, что все эти вклады могут быть получены из выражения для
парамагнитного случая (\ref{Nag3}), если учесть специфический вид
динамической спиновой корреляционной функции в антиферромагнетике
\begin{equation}
K_{\mathbf{q}}(\omega )= \overline{S}^2\delta
{(\mathbf{q}-\mathbf{Q}})\delta (\omega )+
\overline{S}(u_{\mathbf{q}}-v_{\mathbf{q}})^2
[(1+N_{\mathbf{q}})\delta (\omega +\omega
_{\mathbf{q}})+ N_{\mathbf{q}}\delta (\omega
-\omega _{\mathbf{q}})].
\label{eq:P.22}
\end{equation}

Для двумерных ферро- и антиферромагнетиков дальний магнитный
порядок при $T>0$ отсутствует, однако при низких температурах в
корреляционной функции содержатся почти дельта-функционные вклады,
связанные с сильным ближним порядком в локализованной подсистеме.
Таким образом, в спин-волновой области температур структура
спектра сохраняется \cite{620}.

Как показано в работе \cite{kampf} в рамках $s-d$ обменной модели
ферромагнетика, вблизи поверхности Ферми спектр возбуждений
демонстрирует нефермижидкостное поведение. Спектральная функция
при температурах $T<\Delta _0$ ($\Delta _0$ -- спиновое расщепление в
основном состоянии) имеет двухпиковую структуру, что означает
квазирасщепление поверхности Ферми в парамагнитной фазе в
присутствии сильных ферромагнитных флуктуаций.

\subsection{$s$---$d$~обменная модель с~узкими зонами и~$t$---$J$~модель}

$s$---$d$~обменную модель можно использовать и в пределе сильных
корреляций при рассмотрении переноса электронов в узких
вырожденных зонах. Эта модель соответствует случаю, когда носители
тока не~принадлежат той~же энергетической зоне, где формируются
магнитные моменты. Такая ситуация имеет место в некоторых
магнитных полупроводниках и металлах, например
манганитах~\cite{nagaev}.

В~случае сильного $s$---$d$~обмена
$I$ удобно перейти к атомному представлению~\cite{698,699}.
Подставляя значения коэффициентов Клебша"--~Гордана, отвечающих
сложению моментов $S$ и~$1/2$, находим собственные функции
$\mathcal{H}_{sd}$:
\begin{equation}
|M\rangle \equiv |SM\rangle |0\rangle ,\qquad |M2\rangle \equiv
|SM\rangle |2\rangle,
 \label{eq:I.1}
\end{equation}
\begin{equation}
|\mu \pm \rangle =\left( \frac{S\pm \mu +1/2}{2S+1}\right)
^{1/2}|S,\mu -\frac 12\rangle |\uparrow \rangle \pm \left(
\frac{S\mp \mu +1/2}{2S+1}\right) ^{1/2}|S,\mu +\frac 12\rangle
|\downarrow \rangle,
 \label{eq:I.2}
\end{equation}
где $|\mu \alpha \rangle $~"--- состояния, занятые одним
электроном, с полным спином на узле $S+\alpha /2$ и его
проекцией~$\mu $. Тогда $\mathcal{H}_{sd}$ диагонализуется:
\begin{equation}
\mathcal{H}_{sd}=-IS\sum_{\mu =-S-1/2}^{S+1/2}\sum_iX_i(\mu +,\mu
+)+I(S+1)\sum_{\mu =-S+1/2}^{S-1/2}\sum_iX_i(\mu -,\mu -).
 \label{eq:I.3}
\end{equation}
Одноэлектронные операторы выражаются через $X$"~операторы как
\begin{equation}
c_{i\sigma }^{\dagger }=\sum_{\alpha =\pm}(g_{i\sigma \alpha }^{\dagger
}+h_{i\sigma \alpha }^{\dagger }),
 \label{eq:I.4}
\end{equation}
\begin{align*}
g_{i\sigma \alpha}^{\dagger }=&\sum_M\{[S+\sigma \alpha M+(1+\alpha)/2]/(2S+1)\}^{1/2}
X_i\left( M+\frac \sigma 2,\alpha;M\right),
\\
h_{i\sigma \alpha}^{\dagger }=&\sum_M\{[S+\sigma \alpha M+(1-\alpha)/2/(2S+1)\}^{1/2}
X_i\left( M2;M-\frac \sigma 2,-\alpha \right).
\end{align*}
В~пределе $I\rightarrow \alpha \infty $ для концентрации
электронов проводимости~$n<1$ нужно сохранить в~(\ref{eq:I.4})
только члены, содержащие $g_{i\alpha }$, и опустить гамильтониан
$\mathcal{H}_{sd}$, который дает постоянный сдвиг энергии:
\begin{equation}
\mathcal{H}=\sum_{\mathbf{k}\sigma
}t_{\mathbf{k}}g_{\mathbf{k}\sigma \alpha }^{\dagger
}g_{\mathbf{k}\sigma \alpha }+\mathcal{H}_d,\quad \alpha =\sign{}
I.
 \label{eq:I.5}
\end{equation}
Для $n>1$ мы должны оставить члены, содержащие $h_{i\alpha }$, и
перейти к "<дырочному"> представлению введением новых
локализованных спинов $\tilde{S}=S\pm 1/2$. Тогда гамильтониан
принимает вид~(\ref{eq:I.6}) с заменой~\cite{699}:
\begin{equation}
t_{\mathbf{k}}\rightarrow -t_{\mathbf{k}}(2\tilde{S}+1)/(2S+1).
 \label{eq:I.6}
\end{equation}

При теоретическом рассмотрении сильнокоррелированных соединений,
например, медь-кислородных высокотемпературных сверхпроводников,
широко используется $t$---$J$~модель -- модель Хаббарда для
$s$"~зоны с $U\rightarrow \infty $ и учетом гейзенберговского
обмена. Ее гамильтониан в МЭ представлении имеет вид
\begin{multline}
\mathcal{H}=-\sum_{ij\sigma }t_{ij}X_i(0\sigma )X_j(\sigma 0)-
\\
+\sum_{ij}J_{ij}\left\{ X_i(+-)X_j(-+)+\frac
14[X_i(++)-X_i(--)][X_j(++)-X_j(--)]\right\}.
 \label{eq:I.7}
\end{multline}
Эта модель описывает движение дырок на фоне локальных моментов без образования полярных состояний, однако дополнительно вводится обменное взаимодействие между локальными моментами. Даже в такой упрощенной модели возникает богатая фазовая диаграмма, включающая спиральные магнитные структуры и неоднородные состояния (см. обзор \cite{Izyumov1}).

Вывод антиферромагнитного кинетического обмена  из модели Хаббарда
с большим $U$ каноническим преобразованием дает $J=2t^2/U$. С
другой стороны, иногда удобно считать $J$ независимой переменной.
В~частности, "<суперсимметричный"> случай с~$t=J$ позволяет
использовать нетривиальные математические методы (см., напр.,
\cite{701}).

Легко видеть, что модель~(\ref{eq:I.7})~"--- частный случай
$s$---$d$~обменной модели с $I\rightarrow -\infty $, $S=1/2$,
причем $t_{\mathbf{k}}$ заменяется в~(\ref{eq:I.5}) на
$2t_{\mathbf{k}}$ (множитель~$2$ возникает из-за того, что в модели Хаббарда электроны с противоположными спинами в синглетной двойке эквивалентны).
$s$---$d$~модель с произвольным спином $S$ иногда оказывается
более удобной, так как позволяет использовать при вычислениях, помимо малого
параметра $1/z$ ($z$~"--- число ближайших соседей),
квазиклассический параметр $1/2S$.

Как и в общей модели Хаббарда, в $t$---$J$~модели могут
использоваться различные представления Х-операторов через
вспомогательные фермионы, бозоны и псевдоспины \cite{Izyumov1}.
Используя (\ref{ed}), в случае дырочной проводимости $N_e<N$
можно записать
\begin{equation}
\mathcal{H} =-\sum_{ij}t_{ij}e_{i}^{\dagger }e_{j}^{{}}(1/2+\mathbf{s}_{i}\mathbf{s}%
_{j})+\sum_{ij}J_{ij}(1-e_{i}^{\dagger }e_{i}^{{}})(\mathbf{s}_{i}\mathbf{s}%
_{j}-1/4)(1-e_{j}^{\dagger }e_{j}^{{}})  \label{tjd}
\end{equation}%
Такие представления широко применялись при рассмотрении проблемы магнитного полярона в антиферромагнетике.

В то же время гамильтониан~(\ref{eq:I.7}) или(\ref{eq:I.5}) можно
использовать непосредственно в рамках
$1/z$"~разложения~\cite{694}. В~антиферромагнетике со спиральной
магнитной структурой, соответствующей волновому
вектору~$\mathbf{Q}$, мы должны перейти
в~\mbox{$s$---$d$}~гамильтониане к локальной системе координат с
использованием~(\ref{eq:E.8}). Далее, переходя от операторов
$d_{i\sigma }^{\dagger }$ к МЭ операторам, получаем
вместо~(\ref{eq:I.5})
\begin{equation}
\mathcal{H} =\sum_{\mathbf{k}\sigma }(\theta
_{\mathbf{k}}g_{\mathbf{k}\sigma \alpha }^{\dagger }g_{\mathbf{k}\sigma
\alpha }+\tau _{\mathbf{k}}g_{\mathbf{k}\sigma \alpha
}^{\dagger }g_{\mathbf{k},-\sigma ,\alpha })+\mathcal{H}_d,
 \label{eq:I.14}
\end{equation}
где $\theta _{\mathbf{k}}$ и $\tau _{\mathbf{k}}$ определены в (\ref{eq:G.98}).
Выполняя расцепление "<Хаббард"~I">, для электронного
спектра имеем
\begin{equation}
E_{\mathbf{k}}^{1,2}=P_\alpha \theta _{\mathbf{k}}\pm \left\{
\left( \frac{\overline{S}}{2S+1}\theta _{\mathbf{k}}\right)
^2+\left[ P_\alpha ^2-\left( \frac{\overline{S}}{2S+1}\right)
^2\right] \tau _{\mathbf{k}}\right\} ^{1/2},
 \label{eq:I.15}
\end{equation}
В~приближении
ближайших соседей ($\theta _{\mathbf{q}}=0$) для~$I>0$ зона
при~$T=0$ сужается в $(2S+1)^{1/2}$ раз. В~то~же время
в~рассматриваемом приближении для~$I<0$ (а~также
в~$t$---$J$~модели) электроны не~могут переходить на соседние
узлы, и их движение возможно только благодаря квантовым эффектам
(Нагаевым \cite{352} такие состояния были названы
квазиосцилляторными).

Результат вычисления функции Грина
\[
G_{\mathbf{k}\alpha \sigma }(E)=\langle \!\langle
g_{\mathbf{k}\alpha \sigma }| g_{\mathbf{k}\alpha \sigma
}^{\dagger }\rangle \!\rangle _E,\quad \alpha =\sign{} I,
\]
с учетом спиновых флуктуаций имеет вид~\cite{620}:
\begin{equation}
G_{\mathbf{k}\alpha }(E)=\left[ E\left( \Psi _{\mathbf{k}\alpha
}(E)-\frac{\overline{S}_{\text{eff}}^2t_{\mathbf{k}+\mathbf{Q}}/(2S+1)^2}{E-\Psi
_{\mathbf{k}+\mathbf{Q}}(E)t_{\mathbf{k}+\mathbf{Q}}}\right)
^{-1}-t_{\mathbf{k}}\right] ^{-1},
 \label{eq:6.125}
\end{equation}
где
\begin{equation}
\Psi _{\mathbf{k}\alpha }(E)=P_\alpha
+\sum_{\mathbf{q}\ne\mathbf{Q}}\frac{t_{\mathbf{k}+\mathbf{q}}}{(2S+1)^2}\int
K_{\mathbf{q}}(\omega )\Psi _{\mathbf{k}+\mathbf{q},\alpha
}^{-1}(E)G_{\mathbf{k}+\mathbf{q},\alpha }(E+\omega )\,d\omega ,
 \label{eq:6.126}
\end{equation}
\[
P_{+}=\frac{S+1}{2S+1},\qquad P_{-}=\frac S{2S+1}.
\]
В~пренебрежении спиновыми флуктуациями $\Psi _\alpha =P_\alpha $ и
мы получаем спектр (\ref{eq:I.15}).
Второй член в~(\ref{eq:6.126}) (поправки от спиновых флуктуаций)
ведет к качественным изменениям в спектре около дна зоны. Для
решения системы~(\ref{eq:6.125}), (\ref{eq:6.126}) при $T=0$,
используется так называемое приближение доминирующего полюса
\begin{equation}
G_{\mathbf{k}\alpha }(E)=\Psi _{\mathbf{k}\alpha }\left[
\frac{Z_{\mathbf{k}}}{E-\tilde{E}_{\mathbf{k}}}
+G_{\text{incoh}}(\mathbf{k},E)\right].
 \label{eq:6.129}
\end{equation}
Оценка вычета дает
\begin{equation}
Z^{-1}-1\sim
\begin{cases}
|t/JS|^{1/2},             & D=2, \\ %
S^{-1}\ln{} |t/JS|^{1/2}, & D=3.    %
\end{cases}
 \label{eq:6.130}
\end{equation}
Таким образом, помимо некогренентного вклада, около дна затравочной зоны в двумерной ситуации
формируется узкая квазичастичная зона с шириной порядка~$|J|$.
Данный результат был впервые получен в работе~\cite{625}. Видно,
что сильное спин-электронное взаимодействие в двумерных системах
может приводить к большой электронной массе около дна зоны даже в
случае одного носителя тока. Этот эффект должен рассматриваться
совместно с многоэлектронным эффектом Кондо (разд. 3.5-3.6).

\subsection{Сопротивление магнитных переходных металлов}

Наряду с другими моделями, для теоретического описания
кинетических свойств магнитных металлов удобно использовать
$s$---$d(f)$~обменную модель.

Существование магнитных моментов в переходных элементах приводит к
дополнительным факторам, влияющим на поведение носителей тока во
внешнем электрическом поле. Во-первых, тепловые флуктуации в
системе магнитных моментов дают новый механизм рассеяния
вследствие $s$---$d$~обменного взаимодействия. Во-вторых,
электронный спектр магнитных кристаллов сильно зависит от
самопроизвольной намагниченности (или намагниченности подрешетки в
антиферромагнетиках), а следовательно от температуры.

Рассмотрим рассеяние электронов проводимости на спиновом
беспорядке в рамках $s$---$d$~обменной модели. Результат Касуи для
магнитного сопротивления при высоких температурах в приближении
среднего поля для спина $S=1/2$ имеет вид~\cite{II}
\begin{equation}
\rho _{\text{mag}}=\frac{9\pi
}2\frac{m^{*}}{ne^2}\frac{I^2}{E_{\text{F}}}\left(\frac
14-\overline{S}^2\right).
 \label{eq:5.57}
\end{equation}
В~далекой парамагнитной области для произвольного $S$ имеем
\begin{equation}
\rho _{\text{mag}}=\frac{3\pi
}2\frac{m^{*}}{ne^2}\frac{I^2}{E_{\text{F}}}S(S+1).
 \label{eq:5.58}
\end{equation}
Выражение (\ref{eq:5.57})
довольно хорошо описывает экспериментальную температурную
зависимость сопротивления ферромагнитных металлов около точки
Кюри. Для редкоземельных металлов выражение~(\ref{eq:5.58}) с
заменой $S(S+1)\rightarrow (g-1)^2J(J+1)$
удовлетворительно описывает изменение высокотемпературного
сопротивления на спиновом беспорядке в $4f$"~ряде~\cite{II}.

Обсудим магнитное рассеяние при низких температурах. Переходя к
операторам спиновых волн, получаем
из формулы Кубо
\begin{equation}
\rho =\frac{\pi k_{\text{B}}T}{\langle j_x^2\rangle
}2I^2Se^2\sum_{\mathbf{k}\mathbf{q}}(v_{\mathbf{k}\uparrow
}^x-v_{\mathbf{k}+\mathbf{q}\downarrow
}^x)^2N_{\mathbf{q}}n_{\mathbf{k}\uparrow
}(1-n_{\mathbf{k}+\mathbf{q}\downarrow })\delta
(E_{\mathbf{k}\uparrow }-E_{\mathbf{k}+\mathbf{q}\downarrow
}+\omega _{\mathbf{q}}).
 \label{eq:5.59}
\end{equation}
($v_{\mathbf{k}}^x$ -- операторы скорости, $j_x$ -- оператор тока).
Интегрирование дает для удельного сопротивления
\begin{equation}
\rho =C_1T^2\int\limits_{T_0/T}^\infty \frac{x}{\sinh{}
x}\,dx+C_2T_0T\ln{} \coth{} \frac{T_0}{2T},
 \label{eq:5.60}
\end{equation}
где константы $C_i$ определяются электронным спектром, величина
\begin{equation}
T_0\sim T_{\text{С}} q_0^2\sim (I/E_{\text{F}})^2T_{\text{С}}
 \label{eq:5.61}
\end{equation}
совпадает с границей стонеровского континуума,
$q_0=2|IS|/v_{\text{F}}$~"--- пороговый
вектор для одномагнонных процессов рассеяния. При очень низких
температурах $T<T_0$ одномагнонное сопротивление~(\ref{eq:5.60})
экспоненциально мало, так как законы сохранения квазиимпульса и
энергии не~могут быть выполнены для характерных тепловых
квазиимпульсов магнонов. При $T\gg T_0$ имеем
\begin{equation}
\rho _0(T)\sim T^2N_{\uparrow }(E_{\text{F}})N_{\downarrow
}(E_{\text{F}})
 \label{eq:5.62}
\end{equation}
Таким образом, спин-волновое рассеяние в широком диапазоне
температур приводит к квадратичной температурной зависимости
сопротивления. Отличие от случая электрон-фононного рассеяния
(когда при низких температурах сопротивление пропорционально
$T^5$) объясняется квадратичным законом дисперсии магнонов, так
что их число пропорционально $T^{3/2}$, а не~$T^3$.

Зависимость $T^2$ была установлена Туровым~\cite{425} и
Касуя~\cite{426} и в дальнейшем подтверждена многими авторами.
Однако при очень низких температурах в ферромагнитных переходных
металлах обнаруживаются вклады в сопротивление, которые
пропорциональны $T^{3/2}$ или $T$ (см. обсуждение в
монографии~\cite{265}). Линейные температурные поправки вследствие
релятивистских взаимодействий по-видимому слишком малы, чтобы
объяснить экспериментальные данные. Была сделана
попытка~\cite{288} объяснить $T^{3/2}$"~члены неквазичастичными
вкладами в примесное сопротивление, которые появляются из-за
сильной энергетической зависимости некогерентных состояний около
уровня Ферми (\ref{eq:4.87}). Учитывая, что вблизи $E_F$ $\delta
N(E)\sim |E-E_F|^{3/2}$, получаем поправку к проводимости
\[
\delta \sigma (E)\sim -V^2\int \left( -\frac{\partial
f(E)}{\partial E}\right) \delta N(E)\,dE\sim -T^{3/2}.
\]

Обсудим теперь кинетические свойства полуметаллических
ферромагнетиков (ПМФ). Поскольку вклад в кинетические свойства от
электронов с разными проекциями спина в этих материалах должен
радикально отличаться, в последнее время они вызывают большой
интерес в связи со спинтроникой (спин-зависящей электроникой). Как
было предсказано в \cite{318}, в гетероструктурах, содержащих ПМФ,
следует ожидать гигантского магнитосопротивления. Отметим здесь,
что полуметаллическая зонная структура была обнаружена в системах
с гигантским магнитосопротивлением La$_{1-x}$Sr$_{x}$MnO$_{3}$
(хотя соответствующие экспериментальные данные не вполне
однозначны, см. обзор \cite{RMP}).

При рассмотрении кинетических свойств ПМФ оказываются важными
корреляционные эффекты. Как обсуждалось в разделе 3.1, вследствие
электрон-магнонного рассеяния в энергетической щели появляются
некогерентные состояния, поэтому спиновая поляризация сильно
зависит от температуры. Такое заполнение щели весьма важно для
возможных применений ПМФ в спинтронике, существенно ограничивая
их. Часто используемый подход, основанный на теории Стонера и
дающий лишь температурные поправки от одночастичных возбуждений,
экспоненциально малые при $T\ll \Delta $ ($\Delta $ -- спиновое
расщепление), оказывается в корне неверным. Поскольку
ферромагнитные полупроводники могут рассматриваться как частный
случай ПМФ, некогерентные состояния должны также учитываться в
теории спиновых диодов и транзисторов \cite{RMP}.

Поскольку при нулевой температуре  в ПМФ существуют состояния только с
одной проекцией спина на уровне Ферми, одномагнонные процессы рассеяния
в спин-волновой области температур запрещены и
$T^2$"~член в сопротивлении ~(\ref{eq:5.62}) отсутствует. Это,
по-видимому, подтверждается экспериментальными данными по
удельному сопротивлению сплавов Гейслера TMnSb (T${}={}$Ni, Co,
Pt, Cu, Au) и PtMnSn~\cite{331}. $T^2$"~вклады от одномагнонных
процессов в удельное сопротивление полуметаллических систем
(T${}={}$Ni, Co, Pt) не~выделяются, тогда как зависимости $\rho (T)$
для "<обычного"> ферромагнетика значительно круче.
В~случае ПМФ, так~же как и для обычных
ферромагнетиков, при $T<T_0$ сопротивление определяется
двухмагнонными процессами рассеяния. Они приводят к
$T^{7/2}$"~зависимости сопротивления~\cite{428a}),
которая возникает из-за обращения в нуль
амплитуды электрон-магнонного рассеяния при нулевом волновом
векторе магнона.

Обсудим теперь спин-волновое сопротивление
в редкоземельных металлах, которые являются
ферромагнетиками при низких температурах. Из-за сильной
анизотропии закон дисперсии спиновых волн отличается от случая
$d$"~металлов. Спектр магнонов в редкоземельных элементах содержит
щель порядка $T^{*}\sim 10$~К; в отсутствие анизотропии в базисной
плоскости имеем линейное поведение $\omega _{\mathbf{q}}\sim q$.
Щель приводит к появлению экспоненциального множителя $\exp{}
(-T^{*}/T)$ в магнитном сопротивлении. Для линейного закона
дисперсии возникает зависимость $\rho \sim T^4$  вместо
$T^2$, поскольку каждая степень~$q$ дает при интегрировании
множитель $T/T_{\text{С}}$ (вместо ($T/T_{\text{С}})^{1/2}$ при
$\omega _{\mathbf{q}}\sim q^2$). Последний результат подтвержден
экспериментальной зависимостью $\rho \sim T^{3{,}7}$ для гадолиния
в диапазоне температур $4-20$~К~\cite{430}.

Используя формулу Кубо с гамильтонианом
$s$---$d$~модели~(\ref{eq:G.2}) в спин-волновой области, получаем
для низкотемпературного магнитного удельного сопротивления
антиферромагнитных металлов
\begin{multline}
\rho =\frac{\pi k_{\text{B}}T}{\langle j_x^2\rangle
}2I^2Se^2\sum_{\mathbf{k}\mathbf{q}}(v_{\mathbf{k}}^x
-v_{\mathbf{k}+\mathbf{q}}^x)^2n_{\mathbf{k}}
(1-n_{\mathbf{k}+\mathbf{q}})
[N_{\mathbf{q}}(u_{\mathbf{q}}+v_{\mathbf{q}})^2)\times
\\
\times \delta (E_{\mathbf{k}}-E_{\mathbf{k}+\mathbf{q}}+\omega
_{\mathbf{q}})+N_{\mathbf{q}+\mathbf{Q}}
(u_{\mathbf{q}+\mathbf{Q}}-v_{\mathbf{q}+\mathbf{Q}})^2)\delta
(E_{\mathbf{k}}-E_{\mathbf{k}+\mathbf{q}}+\omega
_{\mathbf{q}+\mathbf{Q}})].
 \label{eq:5.63}
\end{multline}
Сопротивление при очень низких температурах определяется вкладами малых $q$ в~(\ref{eq:5.63}), т.~е. переходами внутри
антиферромагнитных подзон. Из-за
линейного закона дисперсии магнонов такие переходы приводят, как и
в случае электрон-фононного рассеяния, к зависимости
сопротивления $T^5$. Благодаря сингулярности коэффициентов
$uv$"~преобразования вклады от малых
$|\mathbf{q}-\mathbf{Q}|$ (т.~е. межзонные вклады), вообще говоря,
больше. Однако, как и для ферромагнетиков, невозможно удовлетворить
закону сохранения квазиимпульса при $\mathbf{q}\rightarrow
\mathbf{Q}$ из-за антиферромагнитного расщепления, так что эти
вклады экспоненциально малы при
\begin{equation}
T<T_0=\omega (q_0)\sim (|IS|/E_{\text{F}})T_{\text{N}},
 \label{eq:5.64}
\end{equation}
где $q_0=2|IS|/v_{\text{F}}$~"--- пороговое значение
$|\mathbf{q}-\mathbf{Q}|$. (Следует обратить внимание на то, что
граничная температура не~настолько мала, как для
ферромагнетика~(\ref{eq:5.61}).) При более высоких температурах
$T>T_0$ межзонные вклады дают $T^2$"~поведение удельного
сопротивления~\cite{433}. В~двумерном случае эти вклады становятся
линейными по $T$, что дает одни из механизмов объяснения
характерной зависимости $\rho (T)$ в высокотемпературных
сверхпроводниках.

\subsection{$s$---$f$~обменная модель и свойства редкоземельных металлов}

Как отмечалось выше, для редкоземельных металлов, где
$4f$"~-электроны хорошо локализованы, $s$---$f$~модель может быть
основой для количественной теории. В частности, уже простой
гамильтониан (\ref{eq:G.2}), который дает дальнодействующее и
осциллирующее РККИ-взаимодействие между $f$-спинами, позволяет
описать их геликоидальные структуры. Для детального рассмотрения
магнитных и электронных свойств необходима более реалистическая
модель $4f$"~металлов с орбитальными степенями свободы, которая, в
частности, позволяет учесть магнитную анизотропию. Мы обсудим такую модель,
следуя \cite{652,389,15,39} (имеются также работы С.В.
Вонсовского и М.С. Свирского по этой проблеме \cite{3939}).

Для большинства редких земель (исключая Eu и Sm) матричные
элементы межузельного взаимодействия малы по сравнению с
расстояниями между $LSJ$"~мультиплетами, поэтому хорошим
приближением является схема связи Рассела"--~Саундерса. Используя
для простоты представление плоских волн $s$"~типа для электронов
проводимости, находим для $s$---$f$~гамильтониана
\begin{multline}
\mathcal{H}_{sf}=\sum_{\mathbf{k}\mathbf{k}^{\prime }\sigma \sigma
^{\prime }}\sum_{\nu \Gamma _1\Gamma _2\gamma _1\gamma
_2}e^{i(\mathbf{k}-\mathbf{k}^{\prime })\mathbf{R}_\nu }
\left\langle \gamma _1,\mathbf{k}\sigma \left|
\sum_{ic}\frac{e^2}{|\mathbf{r}_i-\mathbf{r}_c|}(1-P_{ic})\right|
\gamma _2,\mathbf{k}^{\prime }\sigma ^{\prime }\right\rangle
\times
\\
\times \langle \Gamma _1|a_{\nu \gamma _1}^{\dagger }a_{\nu \gamma
_2}|\Gamma _2\rangle X_\nu (\Gamma _1,\Gamma
_2)c_{\mathbf{k}\sigma }^{\dagger }c_{\mathbf{k}^{\prime }\sigma
^{\prime }},
 \label{eq:K.3}
\end{multline}
где $c_{\mathbf{k}\sigma }^{\dagger }$~"--- операторы рождения для
электронов проводимости, $\gamma _i=\{lm_i\}$,
$\Gamma_i=\{SLJM_i\}$; $P_{ic}$~"--- операторы перестановки электронов
проводимости и локализованных электронов.
Вычисление матричных элементов кристаллического потенциала с учетом
разложения плоских волн
по сферическим гармоникам дает
ряды по $\lambda $, $\lambda ^{\prime }$ с "<интегралами Слэтера">
\begin{align}
F_{\lambda \lambda ^{\prime }}^{(p)}(\mathbf{k}\mathbf{k}^{\prime
})=&e^2\int
r_1^2r_2^2R_l^2(r_1)\frac{r_{<}^p}{r_{>}^{p+1}}
j_{\lambda ^{\prime }}(k^{\prime }r_1)\,dr_1\,dr_2,
 \label{eq:K.4}
\\
G_{\lambda \lambda ^{\prime }}^{(p)}(\mathbf{k}\mathbf{k}^{\prime
})=&e^2\int r_1^2r_2^2R_l(r_1)j_\lambda
(kr_2)\frac{r_{<}^p}{r_{>}^{p+1}}R_l(r_2)j_{\lambda ^{\prime
}}(k^{\prime }r_1)\,dr_1\,dr_2,
 \label{eq:K.5}
\end{align}
где $l=3$ для $f$"~-электронов. Малым параметром разложения
является $k_{\text{F}}r_f\sim 0{,}2$, где $r_f$ --- радиус
$4f$"~-электронной оболочки. Возникшие матричные элементы могут
быть вычислены с помощью метода неприводимых тензорных операторов
и выражены через матричные элементы полного углового момента $J$
~\cite{II}). Для члена нулевого порядка имеем
\begin{equation}
\mathcal{H}_{sf}(00)=-\frac{4\pi }{2l+1}\sum_{\nu \sigma \sigma
^{\prime }}G_{00}^{(0)}\left[ \frac n2\delta _{\sigma \sigma
^{\prime }}+(g-1)(\mbox{\boldmath$\sigma $}_{\sigma \sigma
^{\prime }}\mathbf{J}_\nu )\right] c_{\nu \sigma }^{\dagger
}c_{\nu \sigma ^{\prime }}.
 \label{eq:K.6}
\end{equation}
где введен фактор Ланде.
\[
g=1+\frac{(\mathbf{L}\mathbf{S})}{J^2}
=1+\frac{J(J+1)-S(S+1)-L(L+1)}{2J(J+1)}.
\]
причем
\begin{equation}
\mathbf{S}=(g-1)\mathbf{J}, \quad \mathbf{L}=(2-g)\mathbf{J},
\label{eq:B.20}
\end{equation}
Члены высшего порядка анизотропны и имеют структуру
\begin{equation}
\mathcal{H}_{sf}^{\text{coul}}=\sum_{\nu
\mathbf{k}\mathbf{k}^{\prime }\sigma \sigma ^{\prime
}}e^{i(\mathbf{k}-\mathbf{k}^{\prime })\mathbf{R}_\nu }
c_{\mathbf{k}\sigma }^{\dagger }c_{\mathbf{k}^{\prime }\sigma
^{\prime }} (B_0+B_1[3\{(\mathbf{k}\mathbf{J}_\nu ),(\mathbf{k}^{\prime
}\mathbf{J}_\nu )\}-2(\mathbf{k}\mathbf{k}^{\prime
})J(J+1)]+\ldots ),
 \label{eq:K.7}
\end{equation}
\begin{multline}
\mathcal{H}_{sf}^{\text{exch}}=\sum_{\nu
\mathbf{k}\mathbf{k}^{\prime }\sigma \sigma ^{\prime }}(A_0\delta
_{\sigma \sigma ^{\prime }}+A_1(\mbox{\boldmath$\sigma $}_{\sigma
\sigma ^{\prime }}\mathbf{J}_\nu
)+iA_2([\mathbf{k}\mathbf{k}^{\prime }]\mathbf{J}_\nu )\delta
_{\sigma \sigma ^{\prime }}+
\\
+A_3\{(\mathbf{k}\mathbf{J}_\nu ),(\mathbf{k}^{\prime
}\mathbf{J}_\nu )\}+A_4[(\mathbf{k}\mbox{\boldmath$\sigma
$}_{\sigma \sigma ^{\prime }})(\mathbf{k}^{\prime }\mathbf{J}_\nu
)+(\mathbf{k}^{\prime }\mbox{\boldmath$\sigma $}_{\sigma \sigma
^{\prime }})(\mathbf{k}\mathbf{J}_\nu )]+
\\
+A_5[(\mathbf{k}\mbox{\boldmath$\sigma $}_{\sigma \sigma ^{\prime
}})(\mathbf{k}\mathbf{J}_\nu ) +(\mathbf{k}^{\prime
}\mbox{\boldmath$\sigma $}_{\sigma \sigma ^{\prime }})(\mathbf{k}
^{\prime }\mathbf{J}_\nu )]+
\\
+A_6[(\mathbf{k}\mathbf{J}_\nu )^2+(\mathbf{k}^{\prime
}\mathbf{J}_\nu )^2]+iA_7\{(\mbox{\boldmath$\sigma $}_{\sigma
\sigma ^{\prime }}\mathbf{J}_\nu ),([\mathbf{k}\mathbf{k}^{\prime
}]\mathbf{J}_\nu )\}+\ldots ),
 \label{eq:K.8}
\end{multline}
где $\{ , \}$~"--- антикоммутатор. Члены с векторными
произведениями $[\mathbf{k},\mathbf{k}^{\prime }]$, которые
появляются из матричных элементов орбитальных моментов электронов
проводимости $(\mathbf{l})_{\mathbf{k}\mathbf{k}^{\prime }}$,
описывают анизотропное рассеяние электронов. Такие члены
соответствуют связи тока электронов проводимости во внешнем
электрическом поле и момента $J$ и дают поэтому аномальный эффект
Холла. Коэффициент Холла пропорционален $A(g-2)$, что
соответствует взаимодействию орбитальных моментов электронов с
локализованными орбитальными моментами $\mathbf{L}$. Эта картина
отличается от картины в $d$"~металлах, где аномальный эффект
Холла возникает из-за слабой спин-орбитальной связи. Для
$f$"~электронов она сильна (порядка~$1$~эВ), что позволяет
рассматривать только один $J$"~мультиплет, так что константа
спин-орбитальной связи не~будет явно входить в результаты.

Гамильтониан косвенного $f$--$f$~взаимодействия через электроны
проводимости получается во втором порядке по~$\mathcal{H}_{sf}$.
Основные вклады могут быть записаны в форме~\cite{389}:
\begin{multline}
\mathcal{H}_{ff}(\nu _1\nu
_2)=-I_1(g-1)^2(\mathbf{J}_1\mathbf{J}_2)
-I_2D_1(g-1)\left[
(\mathbf{J}_1\mathbf{J}_2)-3(\mbox{\boldmath$\rho
$}_{12}\mathbf{J} _1)(\mbox{\boldmath$\rho
$}_{12}\mathbf{J}_2)/\rho _{12}^2\right] -
\\
-I_3nD_3\left[ (\mathbf{J}_1\mathbf{J}_2)-3(\mbox{\boldmath$\rho
$}_{12}\mathbf{J} _2)^2/\rho _{12}^2\right],
 \label{eq:K.11}
\end{multline}
Наибольший член этого разложения, который
пропорционален~\mbox{$(g-1)^2$}, соответствует обычному обменному
взаимодействию между спинами согласно формуле де~Женна
(\ref{eq:B.20}). Зависимость $f$---$f$~обменного
параметра~$J_{\text{eff}}\sim (g-1)^2$ находится в хорошем
согласии с экспериментальными данными для парамагнитных температур
Кюри в ряду редкоземельных металлов. Орбитальные вклады
в~$f$--$f$~взаимодействие, пропорциональные~$D_1$ и~$D_2$,
исчезают при~$L=0$ и значительно меньше. Еще более малый член
чисто орбитального взаимодействия получается во втором порядке
по~$A_2$:
\begin{equation}
\mathcal{H}_{ff}^{\prime }(\nu _1\nu
_2)=-I_4(g-2)^2(\mathbf{J}_1\mathbf{J}
_2)=-I_4(\mathbf{L}_1\mathbf{L}_2).
 \label{eq:K.13}
\end{equation}

В отличие от спинового обмена, обменные взаимодействия в~(\ref{eq:K.11}), которые
определяются орбитальными моментами,
сразу становятся анизотропными после учета анизотропии кристалла. Вклады
анизотропного обмена в энергию магнитной анизотропии рассчитаны
в~\cite{389}. Полный результат для гексагональной плотно упакованной решетки с
параметрами $c$ и $a$ имеет вид
\[
\mathcal{E}_{\text{cr}} =(K_1^{\text{cr}}+K_1^{\text{exch}})\cos
^2\theta +\ldots ,
\]
\begin{equation}
K_1^{\text{cr}} =\alpha _JJ\left( J-\frac 12\right)
\frac{Z^{\text{eff}}e^2}a\frac{\langle r_f^2\rangle
}{a^2}1{,}2\left( \frac{c}a- \sqrt{\frac{8}3}\right), \,
K_1^{\text{exch}} \sim (g-1)D_1J^2I_{sf}^2N(E_{\text{F}}).
 \label{eq:4.120}
\end{equation}
Здесь $\alpha _J$~"--- параметр Стивенса; $Z^{\text{eff}}$~"---
эффективный заряд иона; $\langle r_f^2\rangle $~"--- среднее от
квадрата радиуса $f$"~оболочки. Выражения~(\ref{eq:4.120}) дают оценки порядка
величин $K_1^{\text{cr}}$ и $K_1^{\text{exch}}$. Так как для
тяжелых редкоземельных элементов $\alpha _J\sim 10^{-2}-10^{-3}$,
получаем $K_1^{\text{cr}}\sim 10^7-10^8$~эрг/см${}^3$. Тогда
$D_1\sim 10^{-2}$, так что $K_1^{\text{exch}}\sim
10^6-10^7$~эрг/см${}^3$.

Таким образом, магнитная анизотропия редкоземельных элементов по
величине на один или два порядка больше, чем у наиболее сильно
анизотропных гексагональных $d$"~магнетиков. Эта разница есть
следствие того факта, что для РЗ магнитная анизотропия
определяется электростатическими взаимодействиями кристаллического
поля или обменом анизотропного типа, а не~слабым спин-орбитальным
взаимодействием (как у $d$"~магнетиков).

Как показывает сравнение с экспериментом \cite{39}, вклад от кристаллического поля,
вероятно, доминирует, а анизотропный обмен вносит
только~$10-20\,\%$. Надежное экспериментальное определение
последнего имеет фундаментальный интерес для теории обменных
взаимодействий. В~отличие от одноионного механизма
кристаллического поля, анизотропный обмен приводит к двухионной
анизотропии, так что он может быть выделен на основе зависимости
от состава сплава.

Что касается знака магнитной анизотропии, здесь теория углового момента
дает точные предсказания. Знаки как $\alpha _J$, так и $D_1$ меняются
при переходе от конфигурации $f^3$~($f^{10})$ к конфигурации
$f^4$~($f^{11})$ в первой (второй) половине РЗ ряда, а также при
переходе от первой половины ко второй.
Этот  математический результат имеет ясный физический смысл.
Магнитная анизотропия связана с величиной и ориентацией
орбитальных компонент полных орбитальных моментов в электрическом
кристаллическом поле.

\begin{table}
\caption{Значения орбитального момента и тип магнитной анизотропии
для редкоземельных ионов. a~"--- легкая ось, p~"--- легкая
плоскость, c~"--- кубическая структура; радиоактивный прометий не
исследован, гадолиний не имеет орбитального момента}
\centering
\vskip 1em
\begin{tabular}{|l|c|ccc|ccc|ccc|ccc|c|}
\hline
R$^{3+}$ & Ce & Pr & Nd & Pm & Sm & Eu & Gd & Tb & Dy & Ho & Er & Tm & Yb \\ \hline
& f$^1$ & f$^2$ & f$^3$ & f$^4$ & f$^5$ & f$^6$ & f$^7$ & f$^8$ & f$^9$ & f$
^{10}$ & f$^{11}$ & f$^{12}$ & f$^{13}$ \\ \hline
& F & H & I & I & H & F & S & F & H & I & I & H & F \\
L & 3 & 5 & 6 & 6 & 5 & 3 & 0 & 3 & 5 & 6 & 6 & 5 & 3 \\ \hline
 & p & p & p & ? & c  &0 &  p  & p &p & a & a &c& a \\ \hline
\end{tabular}
\end{table}

Как видно из Таблицы 1,
помимо тривиальной электронно-дырочной симметрии в значениях $L$
между первой и второй половинами ряда, имеется также симметрия в
пределах каждой половины, связанная с заполнением орбитальных
квантовых состояний. Например, $f^1$- и $f^6$"~состояния имеют
одно и то~же значение $L=3$, и могло бы показаться, что
анизотропия должна также быть одинаковой. Однако следует принять во
внимание, что $L(f^1)$~"--- угловой момент одного электрона, тогда
как $L(f^6)$~"--- орбитальный момент дырки в сферической
конфигурации~$f^7$, характеризуемой величиной $L=0$. Таким
образом, анизотропия электрического заряда будет противоположной
для конфигураций~$f^1$ и~$f^6$.

\subsection{Эффект Кондо}

Эффект Кондо впервые
обсуждался в связи с проблемой минимума электросопротивления в
разбавленных сплавах переходных металлов \cite{552}. Даже в "<чистых">
образцах меди, золота и цинка наблюдалось увеличение сопротивления
при температурах ниже~$10-20$~К. Экспериментально установлено, что
это явление сильно связано с присутствием малого количества
($10^{-2}-10^{-3}\,\%$) примесей переходных металлов (Cr, Fe, Mn),
которые сохраняют магнитный момент в матрице простого металла.
Такой сильный эффект нельзя объяснить в простых одноэлектронных
приближениях для примесного электросопротивления. Кондо
показал, что в третьем порядке теории возмущений
$s$---$d$~обменное взаимодействие электронов проводимости с
локализованными моментами приводит к сингулярной поправке вида
$\ln{} T$ к сопротивлению вследствие многочастичных эффектов
(фермиевской статистики). После объединения с обычным
низкотемпературным вкладом $T^5$, вызванным электрон-фононным
рассеянием, эта поправка приводит к минимуму сопротивления.
Минимизируя выражение $\rho =Ac\ln{} T+BT^5$,
где $c$~"--- концентрация примесей, получаем $T_{\text{min}}\sim
c^{1/5}$, т.~е. очень слабую зависимость от $c$.

При очень низких температурах увеличение сопротивления подавляется
магнитным упорядочением примесей, которое обусловлено
дальнодействующим РККИ--взаимодействием (в упорядоченной фазе
ориентация спинов фиксируется и рассеяние становится
неэффективно). Для $d$"~примесей логарифмическая поправка третьего
порядка в большинстве случаев оказывается достаточной для того,
чтобы описать экспериментальные данные, а вклады более высоких
порядков теории возмущений малы вплоть до температуры магнитного
упорядочения моментов примесей. Вместе с тем, редкоземельные
примеси (например Ce, Yb, Sm, Tm в матрицах Y или Lа) могут
рассматриваться как изолированные до~$c\sim 1\,\%$; даже при
больших концентрациях взаимодействие между ними не~обязательно
приводит к обычному магнитному упорядочению, поскольку происходит
формирование "<плотных"> кондо-систем~\cite{545}. Тогда возникает
проблема точного учета многоэлектронных эффектов, обусловленных
$s$---$d(f)$~обменным взаимодействием при низких температурах. Это
и есть проблема Кондо, сыгравшая столь большую роль в
теоретической физике.

В дальнейшем были рассмотрены различные обобщения задачи Кондо на
случай взаимодействия электронов проводимости с различными
двухуровневыми системами, туннельными состояниями, сильно
ангармоническими фононами, зарядовыми степенями свободы (см.,
напр., \cite{574,442}). Формально такие системы описываются
псевдо-спиновыми гамильтонианами, так что теория возмущений
приводит к логарифмическим расходимостям.

Суммирование главных логарифмических членов в сопротивлении дает \cite{552}
\begin{equation}
\rho _{sd}=\rho _{sd}^{(0)}\left( 1+2I\rho \ln{} \frac WT\right)
^{-2}, \rho=N(E_F).
 \label{eq:6.10}
\end{equation}
В~случае "<ферромагнитного"> $s$---$d$~обмена $I>0$ это
"<паркетное"> приближение дает полное решение проблемы
Кондо. Однако в более важном случае $I<0$ (например,
для магнитных примесей в
благородных металлах, когда эффективный $s$---$d$~обмен имеет
гибридизационную природу) такое приближение
приводит к расходимости сопротивления при температуре
\begin{equation}
T_{\text{K}}=W\exp{} \frac{1}{2I\rho },
 \label{eq:6.11}
\end{equation}
которая называется температурой Кондо.
Эта величина совпадает с полюсом собственной энергии (\ref{Nag3}).

В~отличие от критической температуры ферромагнетика или
сверхпроводника, температура Кондо соответствует не~фазовому
переходу, а только характерному энергетическому масштабу
кроссовера между высоко- и низкотемпературными областями.
Рассмотрение области $T<T_{\text{K}}$~"--- очень трудная и
красивая математическая проблема. Случай $T\ll T_{\text{K}}$
исследован в рамках феноменологической теории
ферми-жидкости~\cite{553} и методов аналитической
ренормализационной группы~\cite{554,555}. Численное решение
получено Вильсоном с использованием метода ренормализационной
группы~\cite{556}. Наконец, в некоторых упрощающих приближениях
(которые сводят задачу к одному измерению) Андреем и Вигманом было
предложено точное решение однопримесной $s$---$d$~модели с
использованием подстановки Бете~\cite{557,558}.
Имеются также попытки получить аналитическое описание
режима сильной связи диаграммными методами \cite{Barnes}.

Оказывается, что при $T\rightarrow 0$ эффективное
(перенормированное) $s$---$d$~взаимодействие становится бесконечно
сильным, так что примесный магнитный момент полностью
компенсируется (экранируется) электронами проводимости. Строго
говоря, в обычной $s$---$d$~модели с нулевым орбитальным
моментом~(\ref{eq:G.2}) такая компенсация имеет место только для
$S=1/2$, а для произвольного~$S$ эффект Кондо приводит к
уменьшению примесного спина: $S \rightarrow S-1/2$ \cite{555}. Однако в
реальной ситуации вырожденных электронных зон число "<каналов
рассеяния"> для электронов проводимости достаточно, чтобы
обеспечить экранирование.
Для редкоземельных систем более правильно применять
модель Коблина-Шриффера (\ref{eq:6.39}), в которой имеем
\begin{equation}
T_{\text{K}}=W\exp{} \frac{1}{NI\rho },
\label{eq:6.22}
\end{equation}
Эффекты кристаллического поля могут приводить к ряду кроссоверов
(смен поведения) с понижением температуры и последовательным
снятием вырождения;  при этом меняется и выражение для температуры
Кондо.

Пройдя через температуру Кондо, сопротивление стремится при
$T\rightarrow 0$ к конечному унитарному пределу (который
соответствует максимально возможному фазовому сдвигу $\pi /2$),
причем поправки при низкой температуре пропорциональны
$(T/T_{\text{K}})^2$~\cite{552,558}:
\begin{equation}
\rho _{sd}=\frac{3}{\pi }(\rho v_{\text{F}}e)^{-2}\left(
1-\frac{\pi ^2T^2}{T_{\text{K}}^2}+O\left(
\frac{T}{T_{\text{K}}}\right) ^4\right).
 \label{eq:6.12}
\end{equation}

Удельная теплоемкость системы имеет максимум при $T\sim
T_{\text{K}}$ и ведет себя линейно при $T\rightarrow 0$:
\begin{equation}
C_{sd}(T)=\frac{\pi }{3}\frac{T}{T_{\text{K}}}\left( 1+O\left(
\frac{T}{T_{\text{K}}}\right) ^2\right),
 \label{eq:6.13}
\end{equation}
что напоминает обычное выражение для электронной
теплоемкости с заменой~$E_{\text{F}}\rightarrow \pi T_{\text{K}}$.

Магнитная энтропия при $T=0$, равная $\mathcal{S}(0)=R\ln{}
(2S+1)$, устраняется  вследствие экранирования магнитного момента,
а не магнитного упорядочения. Магнитная восприимчивость
\begin{equation}
\chi =\frac{(g\mu _{\text{B}})^2}{2\pi T_{\text{K}}}\left(
1-O\left( \frac{T^2}{T_{\text{K}}^2}\right) \right)
 \label{eq:6.14}
\end{equation}
демонстрирует паулиевское поведение (в отличие от закона
Кюри при $T>T_{\text{K}}$) и значительно усилена,
так~же как и теплоемкость. Данные результаты могут быть
описаны в терминах узкого многочастичного резонанса
Абрикосова"--~Сула на уровне Ферми с шириной порядка
$T_{\text{K}}$ и высотой порядка $1/T_{\text{K}}$, так что
$T_{\text{K}}$ играет роль эффективной температуры вырождения.
Таким образом, формируется новое фермижидкостное состояние, что
сопровождается большими многоэлектронными перенормировками.

Интерполяционную формулу для $\chi (T)$ можно записать в форме
закона Кюри"--~Вейсса с отрицательной парамагнитной температурой
Кюри $|\theta |\sim T_{\text{K}}$. В~этой связи отметим, что
разница между примесями переходных металлов, которые сохраняют
магнитный момент в данном образце, и немагнитными примесями имеет
по большому счету не~качественный, а количественный характер. Во
втором случае можно считать, что $T_{\text{K}}$ высока~"---
порядка $10^2-10^4$~К, что иногда выше, чем точка плавления (в
случае обычной паулиевской восприимчивости
$T_{\text{K}}\rightarrow E_{\text{F}}$). Подобные рассуждения
могут применяться к чистым веществам, где локальные магнитные
моменты при низких температурах не~существуют (хотя конкретные
теоретические модели могут быть весьма различны). Для усиленных
паулиевских парамагнетиков типа Pd, Pt, UAl${}_2$, где при высоких
температурах выполняется закон Кюри"--~Вейсса, вместо температуры
Кондо вводят так называемую температуру спиновых флуктуаций.

Альтернативное описание эффекта Кондо может быть дано в модели решетки Андерсона.
Пренебрегая спин-орбитальным взаимодействием, что разумно для переходных
металлов и их соединений, запишем ее гамильтониан в виде
\begin{equation}
\mathcal{H}=\mathcal{H}_0+\sum_{\mathbf{k}\sigma
}t_{\mathbf{k}}c_{\mathbf{k}\sigma }^{\dagger }c_{\mathbf{k}\sigma
}+\sum_{\mathbf{k}lm\sigma }(V_{\mathbf{k}lm}c_{\mathbf{k}\sigma
}^{\dagger }a_{\mathbf{k}lm\sigma }+\text{h.~c.}),
 \label{eq:N.1}
\end{equation}
где $\mathcal{H}_0$~"--- гамильтониан внутриузельного
взаимодействия между $d$"~-электронами. Для простоты будем
описываем состояния электронов проводимости плоскими волнами.
Используя разложение по сферическим гармоникам,
получаем для матричного элемента гибридизации
\begin{equation}
V_{\mathbf{k}lm}=i^lY_{lm}^{*}(\hat{\mathbf{k}})v_l(k), \,
v_l(k)=4\pi \int r^2R_l(r)v(r)j_l(kr)\,dr
 \label{eq:N.2}
\end{equation}
где $v(r)$~"--- сферически симметричный потенциал для данного узла.
В~пределе $jj$"~связи (соединения актинидов) нужно подставить
в~(\ref{eq:N.1}) $ lm\sigma \rightarrow j\mu $, где $j=l\pm
1/2$~"--- полный момент электронов, а $\mu$~"--- его проекция.

В~случае сильных корреляций для $d$"~-электронов удобно перейти к
представлению операторов Хаббарда, которое приводит
$\mathcal{H}_0$ к диагональному виду~(см. (\ref{eq:C.21})). Сохраняя два
низших терма $\Gamma _n=\{{SL\}}$, $\Gamma _{n-1}=\{{S^{\prime
}L^{\prime }\}}$ для конфигураций $d^n$ и $d^{n-1}$ следуя (\ref{eq:A.31}), определяя
новые электронные операторы проводимости
\begin{equation}
d_{\mathbf{k}lm\sigma }^{\dagger } =\sum_{\mu \mu ^{\prime
}MM^{\prime }}C_{S^{\prime }\mu ^{\prime },\frac 12 \sigma }^{S\mu
}C_{L^{\prime }M^{\prime },lm}^{LM}X_{\mathbf{k}}(SL\mu
M,S^{\prime }L^{\prime }\mu ^{\prime }M^{\prime }),
 \label{eq:N.4}
\end{equation}
\[
c_{\mathbf{k}lm\sigma }^{\dagger
}=i^lY_{lm}^{*}(\hat{\mathbf{k}})c_{\mathbf{k}\sigma }^{\dagger },
\]
представляем гамильтониан~(\ref{eq:N.1}) в форме
\begin{equation}
\mathcal{H}=\mathcal{H}_0+\sum_{\mathbf{k}\sigma }\left[
t_{\mathbf{k}}c_{\mathbf{k}\sigma }^{\dagger }c_{\mathbf{k}\sigma
}+\tilde{v}_l(k)(c_{\mathbf{k}lm\sigma }^{\dagger
}d_{\mathbf{k}lm\sigma }+\text{h.~c.})\right],
 \label{eq:N.5}
\end{equation}
где
\begin{equation}
\mathcal{H}_0 =\Delta \sum_{\mathbf{k}lm\sigma
}d_{\mathbf{k}lm\sigma }^{\dagger }d_{\mathbf{k}lm\sigma }+\const, \Delta =E_{SL}-E_{S^{\prime }L^{\prime }}
 \label{eq:N.6}
\end{equation}
Эффективные гибридизационные
параметры даются формулой
\begin{equation}
\tilde{v}_l(k)=n^{1/2}G_{S_{n-1}L_{n-1}}^{S_nL_n}v_l(k).
 \label{eq:N.7}
\end{equation}
Теперь обсудим редкоземельные системы. Вследствие сильного
кулоновского взаимодействия между $4f$"~электронами образование
$f$"~зон, содержащих $14$~электронных состояний, нереалистично.
Таким образом, нужно использовать модель с двумя
конфигурациями~$f^n$ и~$s(d)f^{n-1}$, что соответствует
делокализации одного электрона на атоме. В~схеме
Рассела"--~Саундерса можно ограничиться двумя низшими
мультиплетами $4f$"~иона: $\Gamma _n=SLJ$ и $\Gamma
_{n-1}=S^{\prime }L^{\prime }J^{\prime }$. Переходя
в~(\ref{eq:N.1}), (\ref{eq:A.31}) к $J$"~представлению
и суммируя произведения
коэффициентов Клебша"--~Гордана, получим
\begin{equation}
\mathcal{H}=\sum_{\mathbf{k}j\mu }\left[ \Delta f_{\mathbf{k}j\mu
}^{\dagger }f_{kj\mu }+t_{\mathbf{k}}c_{\mathbf{k}j\mu }^{\dagger
}c_{\mathbf{k}j\mu }+\tilde{v}_j(k)(c_{\mathbf{k}j\mu }^{\dagger
}f_{\mathbf{k}j\mu }+\text{h.~c.})\right].
 \label{eq:N.9}
\end{equation}
Здесь введены новые электронные операторы
\begin{equation}
f_{\mathbf{k}j\mu }^{\dagger } =\sum_{M_JM_{J^{\prime
}}}C_{J^{\prime }M_J^{\prime },j\mu
}^{JM_J}X_{\mathbf{k}}(SLJM_J,S^{\prime }L^{\prime }J^{\prime
}M_J^{\prime }),
c_{\mathbf{k}j\mu }^{\dagger }=i^l\sum_{m\sigma }C_{\frac 12
\sigma ,lm}^{j\mu }Y_{lm}^{*}(\hat{\mathbf{k}})c_{\mathbf{k}\sigma
}^{\dagger },
 \label{eq:N.10}
\end{equation}
а эффективные гибридизационные параметры выражены через
$9j$"~символы:
\begin{equation}
\tilde{v}_j(k)=
\left\{
\begin{matrix}
S           & L           & J           \\ %
S^{\prime } & L^{\prime } & J^{\prime } \\ %
1/2         & l           & j              %
\end{matrix}
\right\}
([j][j^{\prime }][L])^{1/2}G_{S^{\prime }L^{\prime
}}^{SL}v_l(k).
 \label{eq:N.11}
\end{equation}
где $[J] = (2J+1)$. Следовательно, гибридизационные эффекты в МЭ системах сильно
зависят от МЭ квантовых чисел $S$, $L$, $J$ и атомных
номеров~\cite{709}. Такая зависимость в редкоземельном ряде
подобна корреляции де~Женна для $s$---$f$~обменного параметра и
парамагнитной температуры Кюри. Экспериментальные исследования
этой зависимости, например спектроскопические данные, представляют
большой интерес.

Рассмотрим антикоммутаторную запаздывающую функцию Грина для
локализованных $d$"~-электронов~(\ref{eq:H.3}) в немагнитной фазе
модели~(\ref{eq:N.1}). Простейшее расцепление дает (ср.~\cite{710,563})
\begin{equation}
G_{\mathbf{k}lm}(E)=\left[ \Phi
(E)-\frac{|V_{\mathbf{k}lm}|^2}{E-t_{\mathbf{k}}}\right] ^{-1},
 \label{eq:N.14}
\end{equation}
где функция $\Phi $ определена в~(\ref{eq:H.4}). Соответствующий
энергетический спектр содержит систему подзон, разделенных
гибридизационными щелями (или псевдощелями в случае, когда
$V(\mathbf{k})$ исчезает для некоторых $\mathbf{k}$), которые
окружены пиками плотности состояний. В~модели с сильными
корреляциями~(\ref{eq:N.7}) имеем
\begin{equation}
E_{\mathbf{k}}^{1,2}=\frac 12(t_{\mathbf{k}}+\Delta )\pm \left[
\frac 14(t_{\mathbf{k}}-\Delta
)^2+|\tilde{V}_{\mathbf{k}lm}|^2\right],
 \label{eq:N.15}
\end{equation}
где
\begin{equation}
\tilde{V}_{\mathbf{k}lm}=i^lY_{lm}^{*}
(\hat{\mathbf{k}})\tilde{v}_l(k)\left\{ \frac{[S][L]}{2[l]}
(N_{SL}+N_{S^{\prime }L^{\prime }})\right\} ^{1/2}.
 \label{eq:N.16}
\end{equation}
Легко видеть, что ширина гибридизационной щели явно зависит от
многоэлектронных чисел заполнения (в частности, от положения
$d$"~уровня).

Приближение~(\ref{eq:N.14}) не~учитывает процессов с
переворотом спина, которые приводят к эффекту Кондо и могут
существенно изменять структуру электронного спектра вблизи уровня
Ферми. Чтобы учесть кондовские аномалии, следует выполнить более точные
вычисления функций Грина. Для краткости рассмотрим
модель~(\ref{eq:N.9}); в модели~(\ref{eq:N.5})
$[J]\rightarrow [S][L], \tilde{v}_j\rightarrow \tilde{v}_l$.
В уравнениях движения для функций
Грина $f$"~-электронов  при коммутировании $X$-операторов и расцеплении во втором порядке по гибридизации возникнут фермиевские функции распределения электронов проводимости.
Выполняя интегрирование, получим логарифмический вклад в собственную энергию
\begin{equation}
\Sigma _f(E)=2\rho
\sum_j\tilde{v}_j^2(k_{\text{F}})\frac{J-J^{\prime }}{[J^{\prime
}]}\ln{} \left| \frac WE\right|.
 \label{eq:N.22}
\end{equation}
Рассмотренным выше влиянием
гибридизационной щели можно пренебречь, если
она лежит намного ниже уровня Ферми. Тогда при $J>J^{\prime
}$ функция Грина  $f$"~-электронов имеет полюс
\begin{equation}
|\Delta ^{*}|=T_{\text{K}}\approx W\exp{} \left\{ -\left(
\frac{[J]}{[J^{\prime }]}-1\right) ^{-1}|\Delta |\left[ \rho
\sum_j\tilde{v}_j^2(k_{\text{F}})\right] ^{-1}\right\}.
 \label{eq:N.23}
\end{equation}
Обычный эффект Кондо соответствует полной компенсации магнитного
момента ($J^{\prime }=0$). При $J^{\prime }>J$
полюс~(\ref{eq:N.23}) отсутствует (режим сильной связи
не~возникает), так как рассматриваемая модель переходит в модель
Коблина"--~Шриффера с положительным обменным параметром.
При $J'=0$ результат~(\ref{eq:N.23}) отличается от результата теории
возмущений при высоких температурах~(\ref{eq:6.22}) только
единицей в знаменателе показателя степени. Такая разница типична
для вычисления температуры Кондо в вырожденной модели
Андерсона и объясняется тем, что использованный подход
оправдан, строго говоря, только в пределе больших~$N$.

\subsection{Свойства аномальных $f$-соединений}

Обсудим теперь поведение некоторых классов $4f$- и
$5f$"~соединений, которые имеют аномальные электронные свойства.
К~ним относятся так называемые решетки Кондо, системы с
промежуточной валентностью и тяжелыми фермионами. В чем-то похожие физические свойства проявляют и некоторые $d$"~системы, в частности, медь-кислородные высокотемпературные сверхпроводники, где имеют место сильные
эффекты корреляции в плоскостях CuO${}_2$.

Наиболее экзотические свойства характерны для соединений с
тяжелыми фермионами. Они обладают гигантскими значениями
эффективной электронной массы, что наиболее ярко проявляется в
огромном значении коэффициента при линейном члене в теплоемкости.
Несколько произвольно определяя тяжелофермионные системы, обычно
полагают граничное значение~$\gamma $
равным~$400$~мДж/(моль${}\cdot {}$К${}^2$). Кроме того,
наблюдаются большая парамагнитная восприимчивость при низких
температурах и большой коэффициент при $T^2$"~члене в
сопротивлении.

Аномальные редкоземельные и актинидные соединения обычно классифицируются
как концентрированные кондо-системы, или решетки Кондо, так как
образование низкотемпературного состояния Кондо дает наиболее
естественное объяснение их необычных свойств
(хотя возможны и другие механизмы, см., напр., \cite{367,607}).
Для большинства таких соединений имеется
$\ln{} T$"~вклад в сопротивление при высоких температурах, но они
имеют металлическое основное состояние с $\rho (T\rightarrow
0)\sim T^2$. Однако известны и примеры непроводящих решеток Кондо.
В~частности, система CeNiSn обладает при низких температурах
чрезвычайно малой энергетической щелью порядка нескольких
градусов, причем частичная замена Ni на Cu приводит к
тяжелофермионному металлическому поведению.

Картина изоляторной решетки Кондо используется иногда для
узкощелевых полупроводников с промежуточной валентностью
SmB${}_6$, SmS (золотая фаза)~\cite{545}.
Образование непроводящего состояния Кондо
может быть описано в терминах когерентного рассеяния Кондо, когда
резонанс Абрикосова"--~Сула трансформируется в узкую
многоэлектронную щель.

Помимо температуры Кондо, вводят иногда второй энергетический
масштаб~"--- температуру когерентности $T_{\text{coh}}$, которая
соответствует переходу к когерентному кондовскому рассеянию
различными узлами решетки. Она обычно мала по сравнению с
$T_{\text{K}}$. Именно картина образования когерентного состояния
позволяет объяснить экспериментальные данные по низкотемпературным
аномалиям термоэдс в системах с тяжелыми
фермионами~\cite{545}. С~уменьшением $T$ ниже
высокотемпературного экстремума $\alpha (T)$ часто
изменяет знак, снова имеет экстремум и линейно исчезает при
$T\rightarrow 0$. Такое поведение можно
объясить появлением псевдощели с изменением знака величины
$dN(E)/dE$ на уровне Ферми, которая определяет знак $\alpha (T)$.
Кроме того, формирование когерентного
состояния приводит к положительному магнитосопротивлению и резкому
отрицательному пику в коэффициенте Холла.

Описание кроссовера между когерентным и некогерентным режимом
проводилось в рамках модифицированной SU$(N)$"~модели
Андерсона (\ref{eq:6.38})~\cite{582}.
Температурная зависимость эффективного
параметра гибридизации была получена в форме
\[
V_{\text{eff}}^2\sim \langle b_i^{\dagger }b_i\rangle \sim \varphi
(T),
\]
\begin{equation}
\varphi (T)=\left( N+e^{-T_{\text{K}}/T}+1\right) ^{-1}=
\begin{cases}
1,      & T\ll T_{\text{coh}},                \\ %
O(1/N), & T_{\text{coh}}\ll T\ll T_{\text{K}}    %
\end{cases}
 \label{eq:6.42}
\end{equation}
с температурой когерентности $T_{\text{coh}}=T_{\text{K}}/\ln{}
N$.

Рассмотрим проявления эффекта Кондо для периодической решетки
локализованных $f$"~моментов в рамках $s$---$f$~обменной
модели~(\ref{eq:G.2}). Этот случай отличается от случая одиночной
кондовской примеси присутствием межузельных обменных
взаимодействий и, следовательно, спиновой динамики, которая
приводит к ослаблению обычных кондовских расходимостей, а также к
некоторым новым эффектам.
С этой целью можно обобщить результат ((\ref{Nag3}), учитывая в энергетических знаменателях частоты спиновых флуктуаций. Такое вычисление вклада
второго порядка в электронную собственную энергию дает~\cite{367}
\begin{equation}
\Sigma _{\mathbf{k}}^{(2)}(E)=I^2\sum_{\mathbf{q}}\int
K_{\mathbf{q}}(\omega )\left(
\frac{1-n_{\mathbf{k}+\mathbf{q}}}{E-t_{\mathbf{k}+\mathbf{q}}+\omega
}+\frac{n_{\mathbf{k}+\mathbf{q}}}{E-t_{\mathbf{k}+\mathbf{q}}-\omega
}\right)\,d\omega ,
 \label{eq:6.25}
\end{equation}
где $K_{\mathbf{q}}(\omega )$~"--- спектральная плотность
подсистемы локализованных спинов,
\begin{equation}
K_{\mathbf{q}}(\omega )=-\frac{1}{\pi }N_{\text{B}}(\omega )\Im{}
\chi _{\mathbf{q}\omega }, \,
\chi _{\mathbf{q}\omega }=\langle \!\langle
\mathbf{S}_{\mathbf{q}}|\mathbf{S}_{-\mathbf{q}}\rangle \!\rangle
_{\omega }, \, \Im{} \chi _{\mathbf{q}\omega }=-\Im{} \chi
_{\mathbf{q}-\omega }.
 \label{eq:6.28}
\end{equation}
Таким образом, при наличии спиновой динамики расходимости
кондовского типа в собственной энергии возникают уже во~втором
порядке. Формально они связаны с функцией Ферми:
\begin{equation}
\sum_{\mathbf{q}}\frac{n_{\mathbf{k}+\mathbf{q}}}
{E-t_{\mathbf{k}+\mathbf{q}}\pm \overline{\omega }}\simeq \rho
\ln{} \frac{W}{\max{} \{|E|,T,\overline{\omega }\}},
 \label{eq:6.27}
\end{equation}
где $\overline{\omega }$~"--- характерная частота спиновых
флуктуаций. В классическом пределе $\omega \ll T$ имеем
$K_{\mathbf{q}}(\omega )=K_{\mathbf{q}}(-\omega )=-\frac{1}{\pi
}(T/\omega) \Im{} \chi _{\mathbf{q}\omega }$,
так что члены с функциями Ферми сокращаются. Однако в квантовом
случае $\Sigma (E)$ резко изменяется в окрестности~$E_{\text{F}}$
с шириной~$\overline{\omega }$, что ведет к заметной
перенормировке вычета функции Грина~$Z$ и,
следовательно, электронной эффективной массы $m^{*}$ и удельной
теплоемкости. Эти перенормировки обращаются в нуль при $T\gg
\overline{\omega }$. В~частности, результат~(\ref{eq:6.25}) с
$I\rightarrow U$ дает спин-флуктуационную (парамагнонную)
перенормировку в модели Хаббарда~\cite{573}.

Выражения, подобные~(\ref{eq:6.25}), можно получить в других
ситуациях. Для
\begin{equation}
K_{\mathbf{q}}(\omega )\sim [1-f(\Delta _{\text{cf}})]\delta
(\omega +\Delta _{\text{cf}})+f(\Delta _{\text{cf}})\delta (\omega
-\Delta _{\text{cf}})
 \label{eq:6.29}
\end{equation}
формула~(\ref{eq:6.25}) описывает эффекты взаимодействия с
 возбуждениями в кристаллическом поле~\cite{263}, причем $\Delta
_{\text{cf}}$~"--- величина расщепления уровня~КП.
Спектральная плотность вида~(\ref{eq:6.29}) с $\Delta $, которое
слабо зависит от волнового вектора, соответствует локализованным
спиновым флуктуациям. Перенормировка $m^{*}$ вследствие таких
флуктуаций намного больше, чем перенормировка, даваемая мягкими
парамагнонами, из-за малости фазового объема флуктуаций в
последнем случае.

Таким образом, определение эффекта Кондо в системах с динамикой
нетривиально. Условие $Z\ll 1$, характеризующее решетки Кондо,
может быть удовлетворено не~только из-за обычного эффекта Кондо
(образование резонанса Абрикосова"--~Сула при $T<T_{\text{K}}$),
но также из-за взаимодействия с низкоэнергетическими спиновыми или
зарядовыми флуктуациями.

Результат $m^{*}\sim 1/\overline{\omega }$, который следует
из~(\ref{eq:6.25}), (\ref{eq:6.27}), не~меняет свой вид при учете
членов более высокого порядка, даже для произвольно
малой~$\overline{\omega }$. Данная проблема исследована
в~\cite{574} для простой модели, описывающей взаимодействие с
локальными возбуждениями двухуровневой системы.  Отметим, что все
такие особенности исчезают при $\overline{\omega }\rightarrow 0$ из-за
множителей типа~$\tanh{} (\overline{\omega }/2T)$ и параметр обрезания
для них есть $\overline{\omega }$, а не~ширина полосы $W$.

Рассмотрим теперь "<истинные"> кондовские расходимости,
соответствующие другой последовательности сингулярных членов,
которая описывает процессы с переворотом спина и начинается
с~третьего порядка по $s$---$f$~параметру. Эти расходимости
не~исчезают при отсутствии динамики и действительно дают при
$E\rightarrow 0$ множители $\ln{} (W/\max{} \{\overline{\omega
},T\})$. С~учетом спиновой динамики соответствующий вклад в мнимую
часть собственной энергии имеет вид
\begin{equation}
\Im{} \Sigma _{\mathbf{k}}^{(3)}(E)=2\pi I^3\rho (E)\int
\sum_{\mathbf{q}}K_{\mathbf{q}}(\omega
)\frac{n_{\mathbf{k}+\mathbf{q}}}
{E-t_{\mathbf{k}+\mathbf{q}}-\omega }\,d\omega
 \label{eq:6.31}
\end{equation}
(вещественная часть сингулярного вклада отсутствует в силу компенсации средними $
m_{\bf k}$, ср. с (\ref{Nag3})).
Величина~(\ref{eq:6.31}) определяет затухание одночастичных
состояний и, следовательно, скорость релаксации $\tau ^{-1}(E)$.
Видно, что спиновая
динамика приводит к размытию логарифмического вклада в
сопротивление. Использование, например, простого диффузионного
приближения для спектральной плотности
\begin{equation}
K_{\mathbf{q}}(\omega )=\frac{S(S+1)}{\pi
}\frac{D_{\text{s}}\mathbf{q}^2}{\omega
+(D_{\text{s}}\mathbf{q}^2)^2},
 \label{eq:6.32}
\end{equation}
где $D_{\text{s}}$~"--- коэффициент спиновой диффузии, дает
\begin{equation}
\delta \tau ^{-1}(E)=4\pi I^3\rho ^2S(S+1)\ln{}
\frac{E^2+\overline{\omega }^2}{W^2},
 \label{eq:6.33}
\end{equation}
где $\overline{\omega }=4D_{\text{s}}k_{\text{F}}^2$. Таким
образом, в сопротивлении
$
\ln{} T\rightarrow \frac{1}{2}\ln{} (T^2+\overline{\omega
})\approx \ln{} (T+a\overline{\omega }),\quad a\sim 1.$

Теперь обсудим термоэдс $\alpha (T)$ в решетках Кондо. При
достаточно высоких (по сравнению с $T_{\text{K}}$) температурах
$\alpha (T)$ обычно велика и имеет экстремум (максимум при $\alpha
>0$, минимум при $\alpha <0$). Большие кондовские вклады в $\alpha
(T)$ соответствуют аномальному нечетному вкладу в $\tau
^{-1}(E)$~\cite{552}, который должен возникать, в силу
аналитических свойств $\Sigma (E)$, из логарифмической особенности
в Re$\Sigma (E)$~\cite{367}. Хотя такая особенность отсутствует в
обычной проблеме Кондо, она имеется в случае рассеивающего
потенциала $V$, который ведет к появлению комплексных множителей
\begin{equation}
1+V\sum_{\mathbf{k}}(E-t_{\mathbf{k}}+i0)^{-1},
 \label{eq:6.35}
\end{equation}
которые "<перемешивают"> $\Im{} \Sigma $ и $\Re{} \Sigma $ в
некогерентном режиме. Спиновая динамика ведет к заменам
\begin{equation}
\ln{} \frac{|E|}{W}\rightarrow \frac{1}{2}\ln{}
\frac{E^2+\overline{\omega }^2}{W^2},\qquad \sign{} E\rightarrow
\frac{2}{\pi }\arctan{} \frac{E}{\overline{\omega }},
 \label{eq:6.36}
\end{equation}
в $\Im{} \Sigma $ and $\Re{} \Sigma $ соответственно, и аномальный
вклад к $\alpha (T)$ имеет вид
\begin{equation}
\alpha (T)\sim \frac{I^3V}{e\rho (T)}\int
\frac{E}{T}\frac{\partial f(E)}{\partial E}\arctan{}
\frac{E}{\overline{\omega }}\,dE\sim \frac{I^3V}{e\rho
(T)}\frac{T}{\max{} \{T,\overline{\omega }\}}.
 \label{eq:6.37}
\end{equation}
Следовательно, величина $\overline{\omega }$ играет роль
характерного флуктуирующего магнитного поля, которое введено
в~\cite{552}, чтобы описать термоэдс разбавленных кондовских систем.

Теперь обсудим проблему магнитного упорядочения в решетках Кондо.
Долгое время считалось, что конкуренция межузельного обменного
РККИ-взаимодействия и эффекта Кондо должна привести к формированию
или обычного магнитного упорядочения с большими атомными магнитными
моментами (как в чистых редкоземельных металлах), или немагнитного
состояния Кондо с подавленными магнитными моментами. Однако затем
экспериментальные исследования убедительно продемонстрировали, что
магнитное упорядочение и выраженные спиновые флуктуации весьма
широко распространены среди систем с тяжелыми фермионами и других
аномальных $4f$- и $5f$"~соединений, которые обычно
рассматриваются как концентрированные кондо-системы.

Класс "<кондовских"> магнетиков характеризуют следующие
особенности~\cite{601,II}:
\begin{enumerate}
    \item Логарифмическая температурная зависимость удельного
сопротивления при $T>T_{\text{K}}$, присущая кондо"~системам.
    \item Малое значение магнитной энтропии в точке
упорядочения по сравнению со значением $R\ln{} (2S+1)$, которое
соответствует обычным магнетикам с локализованными моментами. Это
явление связано с подавлением магнитной теплоемкости вследствие
эффекта Кондо (экранирования моментов): лишь малая часть изменения
энтропии связана с дальним магнитным порядком.
    \item Упорядоченный магнитный момент $M_{\text{s}}$ мал по
сравнению с высокотемпературным моментом $\mu _{\text{eff}}$,
найденным из постоянной Кюри. Последний имеет, как правило,
нормальное значение, близкое к соответствующему значению для
редкоземельного иона (например $\mu _{\text{eff}}\simeq 2{,}5\mu
_{\text{B}}$ для иона Ce${}^{3+}$). Такое поведение напоминает
слабые коллективизированные магнетики.
    \item Парамагнитная точка Кюри $\theta $, как правило,
отрицательна (даже для ферромагнетиков) и заметно превышает по
абсолютной величине температуру магнитного упорядочения, что
обязано большому одноузельному кондовскому вкладу в парамагнитную
восприимчивость ($\chi (T=0)\sim 1/T_{\text{K}}$). Наиболее яркий
пример здесь~"--- кондовский ферромагнетик CeRh${}_3$B${}_2$
с~$T_{\text{C}}=115$~К, $\theta =-370$~К~\cite{284} и небольшим
$\gamma =16$~мДж/(моль${}\cdot {}$К${}^2)$.
\end{enumerate}

Существуют многочисленные примеры систем, где кондовские аномалии
в термодинамических и кинетических свойствах сосуществуют с
магнитным упорядочением, а момент насыщения $M_{\text{s}}$ имеет
величину порядка магнетона Бора. Это ферромагнетики CePdSb,
CeSi${}_x$, Sm${}_3$Sb${}_4$, Ce${}_4$Bi${}_3$, NpAl${}_2$,
антиферромагнетики CeAl${}_2$, TmS, CeB${}_6$, UAgCu${}_4$;
экспериментальные данные и библиография представлены в
\cite{II}.

Что касается "<классических"> систем с тяжелыми фермионами, здесь
положение более сложное. Существуют недвусмысленные свидетельства
антиферромагнетизма в UCd${}_{11}$ и U${}_2$Zn${}_{17}$ с тем~же
порядком величины $M_{\text{s}}$~\cite{507}. Для соединений
UPt${}_3$ и URu${}_2$Si${}_2$, \linebreak $M_{\text{s}}\simeq
2-3\cdot 10^{-2}\mu _{\text{B}}$. Признаки антиферромагнитного
упорядочения с очень малым $M_{\text{s}}$ также наблюдалось для
CeAl${}_3$, UBe${}_{13}$, CeCu${}_2$Si${}_2$, CeCu${}_6$ (впрочем,
ряд данных для этих систем подвергались сомнению, см. также обзор
\cite{Vojta1}).

Вообще, типичная особенность тяжелофермионных магнетиков~"---
высокая чувствительность $M_{\text{s}}$ к внешним параметрам,
таким, как давление и легирование малым количеством примесей.
Например, UBe${}_{13}$ становится антиферромагнитным с заметным
$M_{\text{s}}$ под давлением $P>23$~кбар; напротив, CeAl${}_3$
становится парамагнитным под давлением выше $P=3$~кбар. Момент в
UPt${}_3$ увеличивается до значений порядка одного $\mu
_{\text{B}}$ при замене~$5\,\%$ Pt на Pd или $5\,\%$ U на Th. Ряд
систем с тяжелыми фермионами претерпевают метамагнитные переходы в
слабых магнитных полях с резким увеличением магнитного момента.
Здесь характерно название недавнего обзора П. Коулмена
\cite{Coleman1}: "<Тяжелые фермионы. Электроны на грани
магнетизма">.

Относительные роли
эффекта Кондо и межузельного РККИ-взаимодействия задаются
величинами двух энергетических масштабов: температуры Кондо
$T_{\text{K}}=W\exp{} (1/2I\rho )$, которая определяет кроссовер
между режимом свободных моментов и областью сильной связи, и
$T_{\text{RKKY}} \sim I^2\rho $. Последняя величина имеет
порядок температуры магнитного упорядочения $T_{\text{M}}$ в
отсутствие эффекта Кондо. Отношение~$T_{\text{K}}/T_{\text{M}}$
может изменяться в зависимости от внешних параметров и состава
системы при легировании.

В~немагнитном случае $T_{\text{RKKY}}\sim \overline{\omega }$, где
$\overline{\omega }$~"--- характерная частота cпиновых флуктуаций.
Для большинства обсуждаемых соединений
$T_{\text{K}}>T_{\text{RKKY}}$. Однако существуют также аномальные
магнетики, содержащие церий и уран, с $T_{\text{K}}\ll
T_{\text{N}}$, например CeAl${}_2$Ga${}_2$, UAgCu${}_4$. Этот случай
близок к обычным редкоземельным магнетикам, где эффект Кондо почти
полностью подавлен магнитным упорядочением.

Для описания формирования состояния кондовского магнетика
рассмотрим поправки теории возмущений к магнитным характеристикам
с учетом спиновой динамики. Вычисление магнитной
восприимчивости~\cite{367} приводит к результату
\begin{equation}
\chi =\frac{S(S+1)}{3T}(1-4I^2L), \,
L=\frac{1}{S(S+1)}\sum_{\mathbf{p}\mathbf{q}}\int
K_{\mathbf{p}-\mathbf{q}}(\omega )
\frac{n_{\mathbf{p}}(1-n_{\mathbf{q}})}
{(t_{\mathbf{q}}-t_{\mathbf{p}}-\omega )^2}\,d\omega
 \label{eq:6.83}
\end{equation}
где
спиновая спектральная плотность определена в~(\ref{eq:6.28}).
Из простой оценки интеграла в~(\ref{eq:6.83}) следует
\begin{equation}
\chi =\frac{S(S+1)}{3T}\left( 1-2I^2\rho ^2 \ln{}
\frac{W^2}{T^2+\overline{\omega }^2}\right) ,
 \label{eq:6.85}
\end{equation}
где величина в скобках описывает подавление эффективного момента.

Кондовские поправки к магнитному моменту в ферро- и
антиферромагнитном состояниях получаются с использованием
стандартного спин-волнового результата
\begin{equation}
\delta \overline{S}=-\sum_{\mathbf{q}}\langle b_{\mathbf{q}}^{\dagger
}b_{\mathbf{q}}^{}\rangle
 \label{eq:6.86}
\end{equation}
подстановкой поправки к числам заполнения магнонов при нулевой
температуре, обусловленной их затуханием вследствие рассеяния на
электронах проводимости. В случае ферромагнетика имеем
\begin{equation}
\delta \langle b_{\mathbf{q}}^{\dagger }b_{\mathbf{q}}\rangle
=2I^2S\sum_{\mathbf{k}}\frac{n_{\mathbf{k}\downarrow }
(1-n_{\mathbf{k}-\mathbf{q}\uparrow })}{(t_{\mathbf{k}\downarrow }
-t_{\mathbf{k}-\mathbf{q}\uparrow }-\omega _{\mathbf{q}})^2},
 \label{eq:6.87}
\end{equation}
Интегрирование как для ферромагнетика, так и антиферромагнетика дает
\begin{equation}
\delta \overline{S}/S=-2I^2\rho ^2\ln{} \frac{W}{\overline{\omega }}.
 \label{eq:6.89}
\end{equation}
Эти поправки к моменту в основном состоянии возникают в любых
проводящих магнетиках, включая чистые $4f$"~металлы. Однако в
последнем случае они должны быть малы (порядка $10^{-2}$).

В~целях получения самосогласованной картины для магнетика с
заметными кондовскими перенормировками нужно вычислить
поправки к характерным частотам спиновых флуктуаций
$\overline{\omega }$. В~парамагнитной фазе
оценка из поправки второго порядка к динамической восприимчивости
дает~\cite{608}:
\begin{equation}
\omega _{\mathbf{q}}^2=\frac{4}{3}S(S+1)\sum_{\mathbf{p}}
(J_{\mathbf{q}-\mathbf{p}}-J_{\mathbf{p}})^2 [1-4I^2L(1-\alpha
_{\mathbf{q}})].
 \label{eq:6.92}
\end{equation}
Здесь величина $L$ определена в~(\ref{eq:6.83}),
\begin{equation}
\alpha _{\mathbf{q}}=\sum_{\mathbf{R}}J_{\mathbf{R}}^2\left(
\frac{\sin{} k_{\text{F}}R}{k_{\text{F}}R}\right) ^2[1-\cos{}
\mathbf{q}\mathbf{R}]\Biggm/ \sum_{\mathbf{R}}
J_{\mathbf{R}}^2[1-\cos{} \mathbf{q}\mathbf{R}].
 \label{eq:6.93}
\end{equation}
Поскольку $0<\alpha _{\mathbf{q}}<1$, эффект Кондо приводит к
уменьшению зависимости $\overline{\omega }(T)$ при понижении
температуры. В~приближении ближайших соседей (с периодом решетки
$d$) для $J(\mathbf{R})$ значение~$\alpha $ не~зависит
от~$\mathbf{q}$:
\begin{equation}
\alpha _{\mathbf{q}}=\alpha =\left( \frac{\sin{}
k_{\text{F}}d}{k_{\text{F}}d}\right) ^2.
 \label{eq:6.94}
\end{equation}
Вычисление поправок к частоте спиновых волн в ферромагнитной и
антиферромагнитной фазе вследствие магнон-магнонного
взаимодействия, а также электрон-магнонного рассеяния также приводит к результату
\cite{612}
\begin{equation}
\delta \omega _{\mathbf{q}}/\omega _{\mathbf{q}}=-4I^2\rho
^2a\ln{} \frac {W}{\overline{\omega }},
 \label{eq:6.95}
\end{equation}
где множитель $a$ зависит от типа магнитного упорядочения.

Приведенные результаты теории возмущений дают возможность
качественного описания состояния кондо-решетки как магнетика с
малым магнитным моментом. Предположим, что мы понижаем
температуру, стартуя с парамагнитного состояния. При этом
магнитный момент "<компенсируется">, но, в отличие от
однопримесной ситуации, степень компенсации определяется
$(T^2+\overline{\omega }^2)^{1/2}$ вместо~$T$. В~то~же время сама
$\overline{\omega }$ уменьшается согласно~(\ref{eq:6.92}). Этот
процесс не~может быть описан аналитически в рамках теории
возмущений. Впрочем, если иметь в виду образование универсального
энергетического масштаба порядка $T_{\text{K}}$, то нужно выбрать
$\overline{\omega }\sim T_{\text{K}}$ при $T<T_{\text{K}}$.
Последний факт подтверждается большим числом экспериментальных
данных относительно квазиупругого нейтронного рассеяния в
кондо-системах, которые показывают, что при низких температурах
типичная ширина центрального пика $\Gamma \sim \overline{\omega }$
имеет тот~же самый порядок величины, что и фермиевская температура
вырождения, определенная из термодинамических и кинетических
свойств, т.~е. $T_{\text{K}}$. Следовательно, процесс компенсации
магнитного момента завершается где-то на границе области сильной
связи и приводит к состоянию с конечным (хотя, возможно, и малым)
моментом насыщения~$M_{\text{s}}$.

Количественное рассмотрение проблемы магнетизма решеток Кондо
может быть выполнено в рамках подхода ренормгруппы в
простейшей форме андерсоновского "<скейлинга для бедных"> (poor
man scaling)~\cite{554}. Результаты теории
возмущений позволяют записать ренормгрупповые уравнения для
эффективного $s$---$f$~параметра и $\overline{\omega
}$~\cite{612}, что достигается рассмотрением интегралов по
$\mathbf{k}$ с фермиевскими функциями в кондовских поправках к
электронной собственной энергии~(см. разд. 3.1) и
частоте спиновых флуктуаций. Чтобы построить процедуру скейлинга,
нужно выделить вклады от энергетического слоя $C<E<C+\delta C$
около уровня Ферми $E_{\text{F}}=0$. Например, в случае
ферромагнетика из эффективного расщепления в электронном спектре
\begin{equation}
2I_{\text{eff}}S=2IS-\left[ \Sigma _{\mathbf{k}\uparrow
}^{\text{FM}}(E_{\text{F}})-\Sigma _{\mathbf{k}\downarrow
}^{\text{FM}}(E_{\text{F}})\right] _{k=k_{\text{F}}}
 \label{eq:6.96}
\end{equation}
находим, используя~(\ref{eq:G.34}),
\begin{equation}
\delta
I_{\text{eff}}=I^2\sum_{C<t_{\mathbf{k}+\mathbf{q}}<C+\delta
C}\left( \frac {1}{t_{\mathbf{k}+\mathbf{q}}+\omega _{\mathbf{q}}}
+\frac{1}{t_{\mathbf{k}+\mathbf{q}}-\omega _{\mathbf{q}}}\right)=
\frac{\rho I^2}{\overline{\omega }}\delta C\ln{} \left| \frac
{C-\overline{\omega }}{C+\overline{\omega }}\right|,
 \label{eq:6.97}
\end{equation}
где $\overline{\omega }=4Dk_{\text{F}}^2$,
$D$~"--- спин-волновая жесткость. Вводя безразмерные константы
связи $g=-2I\rho, g_{\text{eff}}(C)=-2I_{\text{eff}}(C)\rho$,
в однопетлевом приближении получаем систему уравнений ренормгруппы вида
\begin{equation}
\partial g_{\text{eff}}(C)/\partial C=-\Phi , \qquad
\partial\ln{} \overline{\omega }_{\text{eff}}(C)/\partial C=a\Phi /2, \qquad
\partial\ln{} \bar{S}_{\text{eff}}(C)/\partial C=\Phi /2,
 \label{eq:6.99}
\end{equation}
где
\begin{equation}
\Phi =\Phi (C,\overline{\omega }_{\text{eff}}(C))
=[g_{\text{eff}}^2(C)/C]\phi (\overline{\omega
}_{\text{eff}}(C)/C).
 \label{eq:6.100}
\end{equation}
Скейлинговая функция для пара-, ферро- и антиферромагнитных фаз
равна:
\begin{equation}
\phi (x)=
\begin{cases}
x^{-1}\arctan{} x,                               & \text{PM}, \\ %
\frac{1}{2x}\ln{} \left| \frac{1+x}{1-x}\right|, & \text{FM}, \\ %
-x^{-2}\ln|1-x^2|,                               & \text{AFM}.   %
\end{cases}
 \label{eq:6.101}
\end{equation}
Как показывают результаты исследования
уравнений~(\ref{eq:6.99})---(\ref{eq:6.101})~\cite{612},
в~зависимости от соотношения между однопримесной температурой
Кондо и затравочной частотой спиновых флуктуаций возможны три
режима при $I<0$:
\begin{enumerate}
    \item Режим сильной связи, где $g_{\text{eff}}$ расходится
при некотором $C$. Он грубо определен условием
$\overline{\omega }<T_{\text{K}}=W\exp{} (-1/g)$.
Здесь $I_{\text{eff}}(C\rightarrow 0)=\infty $, так что все
электроны проводимости связаны в синглетные состояния и спиновая
динамика подавлена.
    \item Режим "<кондовского"> магнетика с заметной, но
неполной компенсацией магнитных моментов, который реализуется в
интервале $T_{\text{K}}<\overline{\omega }<AT_{\text{K}}$
($A$~"--- числовой множитель порядка единицы), соответствующем
малому интервалу $\delta g\sim g^2$. В~этом интервале
перенормированные значения магнитного момента и частоты спиновых
флуктуаций $S_{\text{eff}}(0)$ и $\overline{\omega
}_{\text{eff}}(0)$ увеличиваются от нуля  почти до затравочных
значений.
    \item Режим "<обычных"> магнетиков с малыми
логарифмическими поправками к моменту основного состояния
(см.~(\ref{eq:6.89})), возникающий при $\overline{\omega
}>AT_{\text{K}}$.
\end{enumerate}

Высокая чувствительность магнитного состояния к внешним факторам,
которая обсуждалась выше, объясняется тем, что в случае~2
магнитный момент сильно меняется при малых вариациях затравочного
параметра взаимодействия.

Критическое значение $g_c$ существенно зависит от типа магнитного
упорядочения и структуры магнонного спектра (например, наличия в
нем щели), размерности пространства и др. Таким образом, критерий
Дониаха $g_c\simeq 0.4$ \cite{Don}, который был получен для
простой одномерной модели, вряд ли может быть применимым к
реальным системам. В рамках первопринципных расчетов эти проблемы
обсуждаются в недавней работе \cite{Don1}.

Разумеется, при рассмотрении квантового магнитного фазового
перехода требуется более точный учет магнитных флуктуаций. Недавно
для анализа фазовой диаграммы двумерной антиферромагнитной
решетки Кондо было использовано $\epsilon$-разложение в методе
ренормгруппы с использованием нелинейной сигма-модели \cite{2Drg}.
Важную роль также может играть изменение топологии поверхности Ферми
\cite{Si}.

Описанный механизм формирования магнитного состояния с малым $M_{\text{s}}$
внешне радикально отличается от обычного механизма для слабых
коллективизированных ферромагнетиков, которые, как предполагается,
находятся в непосредственной близости к неустойчивости Стонера.
Вспомним, однако, что и энергетический спектр новых фермиевских
квазичастиц, и эффективное взаимодействие между ними претерпевают
сильные перенормировки. Поэтому практически очевидна
неприменимость критерия Стонера с затравочными параметрами для
кондовских магнетиков.

Так как существует непрерывный переход между
сильнокоррелированными кондо-решетками и "<стандартными">
системами коллективизированных электронов (в частности, обычные
паулиевские парамагнетики могут рассматриваться как системы с
высоким $T_{\text{K}}$~"--- порядка энергии Ферми), возникает
вопрос относительно роли, которую многоэлектронные эффекты играют
в "<классических"> слабых зонных магнетиках, подобных ZrZn${}_2$.
Может оказаться, что близость основного состояния к точке
стонеровской неустойчивости, т.~е. малость $M_{\text{s}}$, в
последних системах возникает из-за перенормировок параметра
взаимодействия и химпотенциала, а не~из-за случайных затравочных
значений $N(E_{\text{F}})$. Действительно, в такую случайность
трудно поверить, поскольку отклонение от граничного условия
Стонера $UN(E_{\text{F}})=1$ является крайне малым.

Отметим, что в реальной ситуация необходим учет особенностей
ван-Хова, которые как правило имеются в таких системах вблизи
$E_{\text{F}}$ (в случае гладкой плотности состояний критерию
Стонера не удовлетворить!). Эти особенности могут существенно повлиять на
структуру теории возмущений (ср.\cite{IKI}).
При этом скейлинговое поведение будет определяться не только близостью уровня Ферми ($C \rightarrow 0$), но и расстоянием до особенности ван-Хова;
в определенных ситуациях возможно ослабление зависимости константы связи от ее затравочного значения.
Отметим, что при наличии особенности плотности состояний вблизи $E_{\text{F}}$  однопримесный эффект Кондо  качественно видоизменяется \cite{gogolin}; в случае решетки такие особенности усиливают подавление момента и частот спиновой динамики \cite{kondovh}.

В~рассмотренном контексте было бы интересно описать слабые коллективизированные
магнетики не~с зонной точки зрения, а с точки зрения локальных
магнитных моментов, которые почти скомпенсированы. Поскольку
сейчас уже стало обычным совместно рассматривать решетки Кондо
и зонные магнетики \cite{Ohkawa,Vojta} и трактовать UPt${}_3$, CeSi${}_x$ и
CeRh${}_3$B${}_2$ как слабые коллективизированные магнетики (см.,
напр., \cite{613}),  то второй подход кажется уже не~менее
естественным, чем первый.

Специальное приближение среднего поля для
основного состояния кондо-решетки в режиме сильной связи~\cite{711}
использует псевдофермионное
представление для операторов  локализованных спинов $S=1/2$
\begin{equation}
\mathbf{S}_i=\frac 12\sum_{\sigma \sigma ^{\prime }}f_{i\sigma
}^{\dagger }\mbox{\boldmath$\sigma $}_{\sigma \sigma ^{\prime
}}f_{i\sigma ^{\prime }}
 \label{eq:O.1}
\end{equation}
с вспомогательным условием
$f_{i\uparrow }^{\dagger }f_{i\uparrow }+f_{i\downarrow }^{\dagger
}f_{i\downarrow }=1$. Используя приближение перевальной точки для интеграла по
траекториям, описывающего спин-фермионную взаимодействующую
систему~\cite{711}, можно свести гамильтониан $s$---$f$~обменного
взаимодействия к эффективной гибридизационной модели:
\begin{equation}
-I\sum_{\sigma \sigma ^{\prime }}c_{i\sigma }^{\dagger }c_{i\sigma ^{\prime }
}\left( \mbox{\boldmath$\sigma $}_{\sigma \sigma ^{\prime
}}\mathbf{S}_i-\frac 12\delta _{\sigma \sigma ^{\prime }}\right)
\rightarrow f_i^{\dagger }V_ic_i+c_i^{\dagger }V_i^{\dagger
}f_i-\frac 1{2I}\Sp{} (V_iV_i^{\dagger }),
 \label{eq:O.2}
\end{equation}
где введены векторные обозначения
$f_i^{\dagger }=(f_{i\uparrow }^{\dagger },\,f_{i\downarrow
}^{\dagger }),\, c_i^{\dagger }=(c_{i\uparrow }^{\dagger},
c_{i\downarrow }^{\dagger })$, а
$V$~"--- эффективная матрица гибридизации, определяемая из условия
минимума свободной энергии.
Коулмен и Андрей~\cite{711}
рассмотрели формирование состояния спиновой жидкости в двумерной
ситуации.
Соответствующий энергетический спектр содержит узкие пики
плотности состояний вследствие псевдофермионного вклада.
Следует подчеркнуть, что $f$"~-псевдофермионы в
рассматриваемой ситуации сами становятся коллективизированными.
Делокализация $f$"~-электронов в системах с тяжелыми фермионами,
которая подтверждена наблюдением больших электронных масс в
экспериментах по эффекту де~Гааза"--~ван~Альфена, нетривиальна в
$s$---$f$~обменной модели (в отличие от модели Андерсона с
$f$"~состояниями вблизи уровня Ферми, где имеется затравочная
$s$---$f$~гибридизация). Эта делокализация аналогична появлению
фермиевской ветви возбуждения в теории резонирующих валентных
связей~(РВС) для высокотемпературных сверхпроводников (см.
раздел 2.1).

Более простой случай ферро- и антиферромагнитного упорядочения
рассмотрен в работе~\cite{608}. При этом в случае ферромагнетика
были получены "<полуметаллические"> решения, типичные для плотности
состояний с резкими пиками (в этом случае энергетически выгодна
раздвижка спиновых подзон, позволяющая убрать пики с уровня Ферми).
Разумеется, эта задача требует дальнейшего исследования.

В 1990-е годы был открыт еще одни класс $f$"~-электронных систем:
для ряда соединений и сплавов на основе церия и урана было
обнаружено нефермижидкостное поведение~\cite{646,646a,507nfl}. Оно
проявляется в необычном поведении электронной теплоемкости вида $T
\ln T$ или $T^{1-\lambda}$, аномальных степенных температурных
Зависимостях магнитной восприимчивости $ T^\zeta $ с $\zeta<1$,
сопротивления $T^{\mu}$ с $\mu<2$ и~т.~д.
Часто нефермижидкостное поведение возникает на границе магнитного упорядочения.
Для его объяснения был
предложен ряд механизмов: сингулярное распределение температур Кондо
в неоднородных системах, сильные спиновые флуктуации вблизи квантового
фазового перехода, механизм точек Гриффитса и др.
\cite{646a}. Как показано в работе~\cite{731}, достаточно широкий интервал
нефермижикостного поведения возникает в решетках Кондо при учете затухания
спиновых возбуждений либо в случае
спиновой динамики с особой спектральной функцией; эти механизмы
могут иметь отношение и к описанию температурно-индуцированных моментов
в слабых зонных магнетиках.

\section*{Заключение}

Полярная и $s-d(f)$--обменная модель были сформулированы еще в
первой половине ХХ века. Тем не менее, они продолжают успешно
работать в физике твердого тела, не только ложась в основу все
более сложных и красивых теоретических концепций, но и описывая
новые физические явления, открытые экспериментаторами.

Многоэлектронные модели оказываются очень полезными с точки зрения
микроскопического описания эффектов сильных корреляций. Такие
эффекты являются особенно яркими для некоторых $d$-- и
$f$--соединений. В случае узких зон (сильное кулоновское
взаимодействие) корреляции приводят к радикальной перестройке
электронного спектра "--- формированию хаббардовских подзон.
Удобный инструмент для описания атомной статистики возбуждений в
данной ситуации "--- метод многоэлектронных хаббардовских
операторов (для орбитально вырожденных систем – в сочетании с формализмом
углового момента). Ряд проблем здесь еще далеки от решения
(например, совмещение зонной и атомной статистики, построение
убедительных интерполяционных описаний).

С другой стороны, даже слабое взаимодействие
между локализованными и коллективизированными электронами может
привести к перестройке электронного спектра при низких
температурах вследствие особенностей резонансного рассеяния в
многочастичных системах (эффект Кондо).

Во многих случаях эффекты корреляции сводятся к перенормировке
электронного спектра и плотности состояний (например формирование
щели гибридизации или резонанса Абрикосова--Сула в системах с
промежуточной валентностью и кондовских системах), так что
электронные свойства можно рассчитать феноменологическим способом
с использованием результатов одноэлектронной теории. В то же время
в некоторых ситуациях состояние многоэлектронной системы не
описывается в рамках обычной квазичастичной картины (например,
спектр имеет существенно некогерентный характер).

Спектр возбуждений сильно коррелированых систем часто описывается
в терминах вспомогательных ферми- и бозе-операторов, которые
соответствуют квазичастицам с экзотическими свойствами
(нейтральные фермионы, заряженные бозоны и т. д.). В последнее
время такие идеи широко применялись в связи с необычными спектрами
высокотемпературных сверхпроводников и систем с тяжелыми
фермионами. Подобные концепции существенно обогащают и изменяют
(иногда радикально) классические представления теории твердого
тела и магнетизма.

В своей лекции "<Многочастичная физика: незаконченная революция">
\cite{end} П. Коулмен говорит о трех эрах физики конденсированного
состояния: первых успехах после открытия квантовой механики
(свободные фермионы), многочастичной физике середины ХХ века
(начало изучения коллективных явлений) и современной эре физики
сильнокоррелированнной материи (“экзотические” системы). В своем
бурном развитии последний этап тесно связан с другими передовыми
отраслями человеческого знания~"--- физикой ядра и элементарных
частиц, космологией, квантовыми технологиями, даже биологией
\cite{bio}.

Автор благодарен за обсуждения своим коллегам и соавторам работ,
использованных в данном обзоре: А.О. Анохину, А.В. Зарубину, П.А. Игошеву, А.А.
Катанину, М.И. Кацнельсону. Работа выполнена при частичной поддержке программой
Президиума РАН “Квантовая физика конденсированного состояния”.

\end{document}